\documentclass[12pt]{article}

\usepackage{color,soul}

\usepackage{amssymb}
\usepackage{amsmath} 
\usepackage{amsthm}
\usepackage{lineno}
\usepackage{longtable}
\usepackage{subfig}  
\usepackage{float}
\usepackage{enumitem}

\usepackage{lineno}
\usepackage{amsmath} 
\usepackage{subfig}  
\usepackage{float}
\usepackage{enumitem}
\usepackage{array}
\newcolumntype{P}[1]{>{\centering\arraybackslash}p{#1}}
\usepackage[utf8]{inputenc}
\usepackage{booktabs}
\usepackage{ltablex}
\usepackage{hyperref}

\usepackage{multirow}
\usepackage{makecell}
\usepackage{graphicx}

\usepackage{hyperref}

\usepackage{blindtext}
\usepackage{authblk}

\title{Optimization and uncertainty quantification model for time-continuous geothermal energy extraction undergoing re-injection}

\author{Hussein Hoteit\footnote{Corresponding author: hussein.hoteit@kaust.edu.sa} , Xupeng  He, Bicheng Yan,  Volker Vahrenkam}

\affil[]{Physical Science and Engineering Division, King Abdullah University of Science and Technology, Thuwal, Saudi Arabia}

\date{\today}

\begin{document}

\maketitle
  

\begin{abstract}
Geothermal field modeling is often associated with uncertainties related to the subsurface static properties and the dynamics of fluid flow and heat transfer. Uncertainty quantification using simulations is a useful tool to design optimum field-development alternatives and to guide decision-making. The optimization process includes assessments of multiple time-dependent flow mechanisms, which are functions of operational parameters subject to subsurface uncertainties. This process requires careful determination of the parameter ranges, dependencies, and their probabilistic distribution functions. This study presents a new approach to assess time-dependent predictions of thermal recovery and produced-enthalpy rates, including uncertainty quantification and optimization. We use time-continuous and multi-objective uncertainty quantification for geothermal recovery, undergoing a water re-injection scheme. The ranges of operational and uncertainty parameters are determined from a collected database, including 135 geothermal fields worldwide. The uncertainty calculation is conducted non-intrusively, based on a workflow that couples low-fidelity models with Monte Carlo analysis. Full-physics reservoir simulations are used to construct and verify the low-fidelity models. The sampling process is performed with Design of Experiments, enhanced with space-filling, and combined with analysis of covariance to capture parameter dependencies. The predicted thermal recovery and produced-enthalpy rates are then evaluated as functions of the significant uncertainty parameters based on dimensionless groups. The workflow is applied for various geothermal fields to assess their optimum well-spacing in their well configuration. This approach offers an efficient and robust workflow for time-continuous uncertainty quantification and global sensitivity analysis applied for geothermal field modeling and optimization.

\end{abstract}




\section{Introduction}
The increasing global demand for clean, sustainable, and environmentally-friendly energy is driving accelerated growth for the renewable energy market worldwide. The contribution of geothermal energy was about 4\% of renewable energy in 2018 \cite{BP2019}, whose role is to complement the other sources of energy for electricity generation, water desalination, cooling, greenhouse farming, among others \cite{IEA2019}.

Geothermal resources manifest in the subsurface under different temperature conditions ranging from low to high enthalpy. The temperature ranges determine the applicability for electricity generation or other direct use. Resources with temperatures between $150\,^{\circ}$C - $300\,^{\circ}$C are suitable for electricity generation, while the ones with temperatures below $100\,^{\circ}$C are suitable for direct use \cite{Sanyal2005}. Geothermal resources are typically classified into three types of reservoir rocks: volcanic, sedimentary, metamorphic, and mix of them \cite{Moeck2014}. The rock hydraulic and thermal properties govern the flow behavior and the efficiency of thermal extraction. One of the main challenges of geothermal energy is related to its low efficiency that is typically below 16\% \cite{Zarrouk2014}. This low efficiency makes geothermal energy less competitive than other renewable energy sources, such as solar and wind. The efficiency of geothermal recovery is a function of complex parameters, including subsurface static reservoir properties (such as permeability, porosity, thickness, and depth), thermal properties (thermal conductivity and temperature), and energy extraction method (production/injection rate and well spacing) \cite{JULIUSSON201380,Zarrouk2014,Santoso2019,BIRDSELL2021116658}. Other technologies proposed improving efficiency and economics by  including the circulation of carbon dioxide or combination with natural gas production \cite{ADAMS2014409,ADAMS2015365,EZEKIEL2020115012}.\\
Subsurface properties are often poorly characterized and therefore associated with significant uncertainties, leading to inaccurate predictions and optimization \cite{YOON2014165,PANDEY201717,Ansari2017,Schulte2020,JULIUSSON2021102091}. Therefore, optimizing field recovery performance under various time-dependent uncertainties is crucial to improve efficiency. \cite{JULIUSSON201380} used tracer and flow-rate data to maximize the net present value of production from fractured geothermal reservoirs.  Various multi-objective-optimization approaches have been proposed to address uncertainties in the subsurface and in operational conditions  \cite{PASQUIER2019176,KESHAVARZZADEH2020101861,PARK2021102050}. \cite{SCHULTE2020101792} applied multi-objective particle swarm optimization with experimental-design proxy models to optimize the performance of low enthalpy geothermal reservoirs, taking into account the uncertainty in the reservoir heterogeneity. \cite{PATTERSON2020101906} used an analytical approach to optimize geothermal production from fractured reservoirs, considering the reservoir structural uncertainty. \cite{SALINAS2021102089} proposed an optimization approach based on dynamic mesh optimization to address the challenges in fixed simulation grids. 
{This issue of poor characterization of subsurface properties has also been reported for geothermal resources in  hot dry rock (HDR), where  rigorous  analytic and numerical modeling tools are crucial to capture uncertainties and optimize the development strategies \cite{GONG20201339,GUO2020113981}}.

There are three essential elements in performing uncertainty quantification and global sensitivity analysis: 1) accurate interpolation of time-dependent responses of thermal recovery and produced-enthalpy rates, 2) efficient evaluation and propagation of multivariables uncertainties with broad ranges of possibilities, and 3) identify the significant parameters and determine their interdependencies. Performing full-physics simulations to capture all uncertainties is computationally impractical.  A non-intrusive approach, based on low-fidelity models, is developed to efficiently perform uncertainty quantification \cite{Ng2014}. Low-fidelity modeling is a reduced-order process used to project the physical space onto a low-dimensional manifold. This model can be generated using different approaches, such as neural network, multivariate polynomial, k-nearest-neighbors, and decision tree \cite{Antoulas2001,Rozza2008,Shao2017,Swischuk2019a,cmes.2021.016619}. The construction of a low-fidelity model follows the "training and validation" process in  machine learning, which treats the high-fidelity model as a black-box \cite{Swischuk2019,santoso2019application,he2020application,en14227495,he2021}. With rigorous training and validation, the low-fidelity model is expected to capture the governing physics as in the high-fidelity model but with significantly less computational effort. The training and validation process usually involves the use of Design of Experiments (DoE) for sampling \cite{Friedmann2003,LeMaitre2010,Damsleth2011,Ansari2017,Rajabi2020,santoso2020investigation,Hoteit2021623004}. When the output is time-dependent, the response is often built by several models constructed at discrete time intervals \cite{Chinesta2004}. This approach is frequently utilized in petroleum engineering applications to perform resource assessment and decision-making in field development \cite{Friedmann2003,Damsleth2011}. In geothermal applications, \cite{Ansari2017} used it for uncertainty quantification for thermal recovery factor of the injection scheme in a tilted geothermal reservoir. \cite{Quinao2018} conducted uncertainty quantification for power capacity in the Ngatamariki geothermal field in New Zealand. \cite{Pratama2020} presented an uncertainty quantification for resource assessment in Atadei geothermal field, Indonesia. However, this approach is computationally intensive due to the construction of a separate  model  for each time point. The selection of the optimum number and  positions of the time points is not systematic as it depends on the  smoothness of the response surface in time and space.

Another challenge in developing a global sensitivity analysis process is to identify the significant  parameters and their dependencies. Within a geothermal system, there are various complex interacting parameters that can  contribute to the performance of the recovery \cite{Pandey2017,Daniilidis2020,Pratama2020}. \cite{Fernandez2017} deployed Sobol-sensitivity indices, based on   variance decomposition, to perform global sensitivity analyses for a geothermal heat exchanger. However, when dealing with time-continuous or nested functions, the method becomes inaccurate with the presence of correlated  inputs \cite{Caniou2012}.  

In this work, we propose a new workflow to efficiently and accurately perform uncertainty quantification and global sensitivity analysis applicable for time-continuous and multi-objective functions. The proposed approach captures the time-continuous profile of the response surface.  The concept is to first capture the overall time response with a parameterized empirical model. The reduced-order model is then used to predict the parameters of the empirical model. To reduce the number of runs of the high-fidelity model, we utilize DoE to sample the inputs and conduct feature selection to exclude insignificant inputs. The feature selection is based on a regression-derived sensitivity measure. Secondly, the uncertainty propagation is then performed for the nested functions using Monte Carlo. Finally, we use ANCOVA  (Analysis of Covariance)  \cite{Caniou2012} to perform global sensitivity analysis for the time-continuous low-fidelity models. To our knowledge, this workflow is applied in geothermal applications for the first time. 

The workflow produces time-continuous uncertainty response to hinder  inaccurate interpolations that occur when the traditional time-independent models are used to capture the overall time response. With this approach, the number of high-fidelity simulations needed to construct the time-continuous low-fidelity models can be reduced. The global sensitivity analysis provides accurate identification of significant parameters (heavy-hitters) that can be  strongly correlated. We apply the workflow for a geothermal reservoir undergoing re-injection with a doublet arrangement. The workflow provides predictions of time-dependent thermal recovery and produced-enthalpy rate under subsurface uncertainties, including permeability, porosity, rate, heterogeneity, temperature, well spacing, and thermal conductivity. The optimization model is then applied to 25 geothermal fields to estimate the optimum well spacing. The results are compared  to their actual well configurations.  

The outline of this paper is the following: In section 2, we discuss the ranges of subsurface uncertainties based on a database including 135 geothermal fields. In section 3,  the problem of geothermal production undergoing re-injection is reviewed, followed by the proposed workflow and the time-continuous parametrization in section 4. In section 5,  the reservoir simulation model, the physical phenomena, and the approximation functions are presented. In section 6, we present the uncertainty quantification results, followed by sensitivity analysis to show the effect of different designs and sample numbers towards the training accuracy, predictability, feature selection results, uncertainty propagation, and global sensitivity analysis. Finally, a discussion and conclusions are outlined. 

\section{Variability in subsurface uncertainties}
\begin{figure}[h!]
\centering
\includegraphics[width=\textwidth]{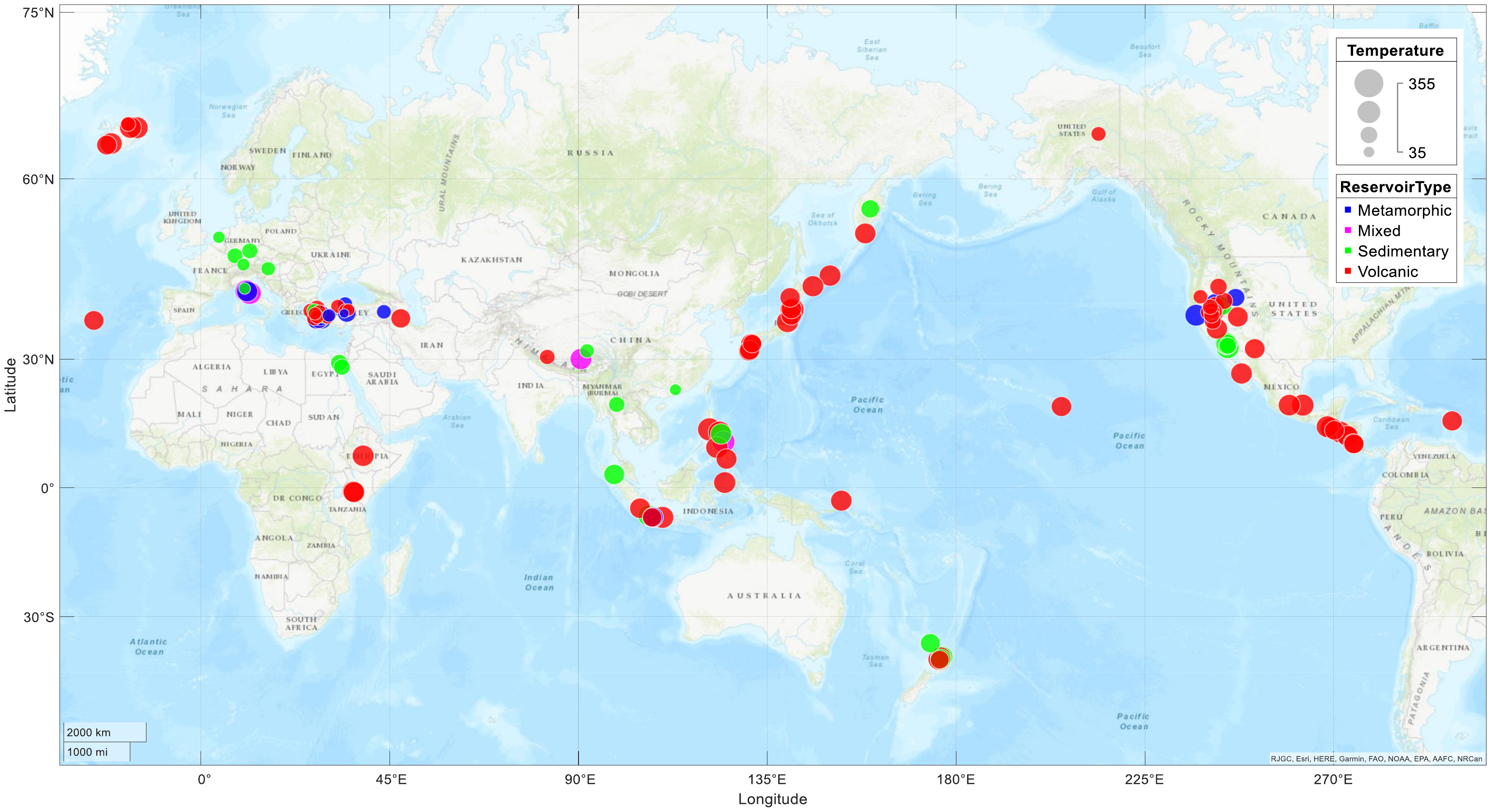}
\caption{Map showing the locations of 135 geothermal fields. The size of the bubbles reflects the temperature range, while the color indicates the reservoir rock type.}
\label{figure1}
\end{figure}

Rock formation, depositional setting, and diagenetic and tectonic overprints create wide variability in subsurface geothermal reservoirs, resulting in substantial variability in hydraulic and thermal properties \cite{Moeck2014}. We compiled data from 135 hydrothermal fields worldwide, covering various types of reservoir rocks, including sedimentary, volcanic, and metamorphic. The distribution of the geothermal fields, as appears in Figure \ref{figure1}, shows that most of the  fields are located around the Ring of Fire (Indonesia, Japan, Philippines, Russia, Tibet, New Zealand, western part of USA, and some Latin American countries), Turkey, western Europe, and some African countries. The high-temperature fields are mostly around the Ring of Fire, while the low-temperatures are mostly in Europe. The temperatures of the geothermal fields vary between 80 and 242$\,^\circ$C, are utilized for various applications, including electricity generation and heating. 
\begin{figure}[h!]
\centering
\begin{minipage}{0.45\linewidth}
\centering
\includegraphics[scale=0.4]{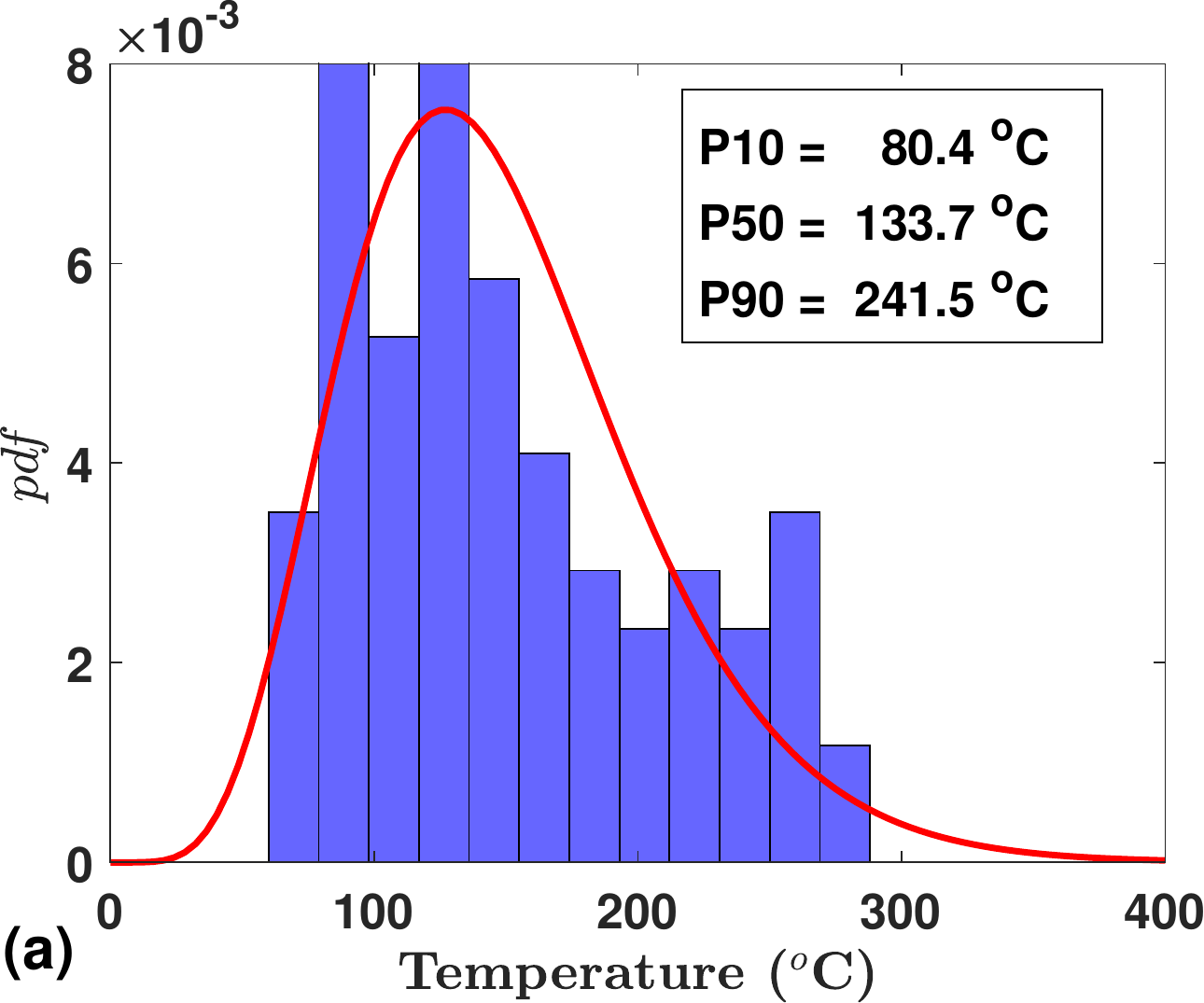}
\end{minipage}
\begin{minipage}{0.45\linewidth}
\centering
\includegraphics[scale=0.4]{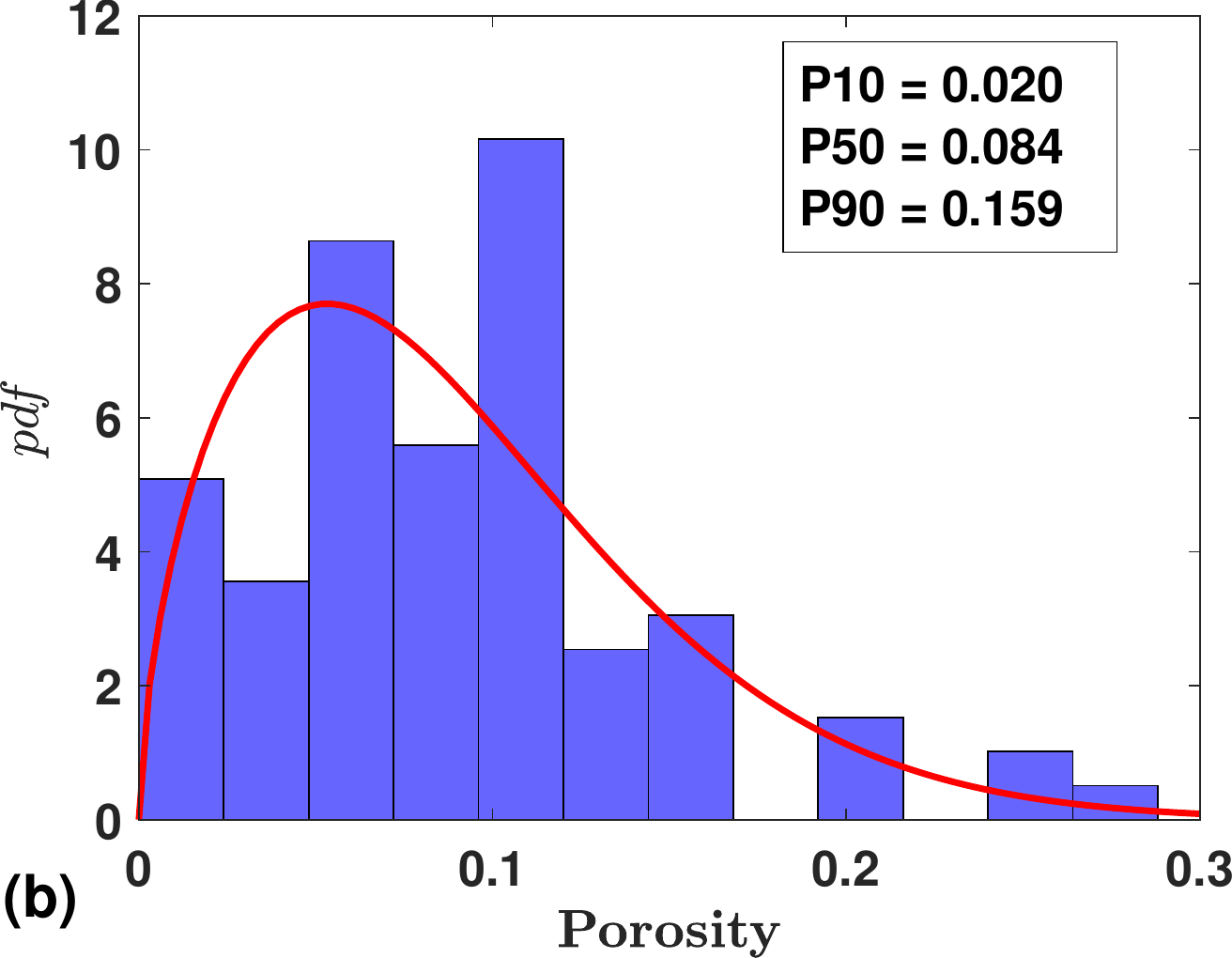}
\end{minipage}
\\
\centering
\begin{minipage}{0.45\linewidth}
\centering
\includegraphics[scale=0.4]{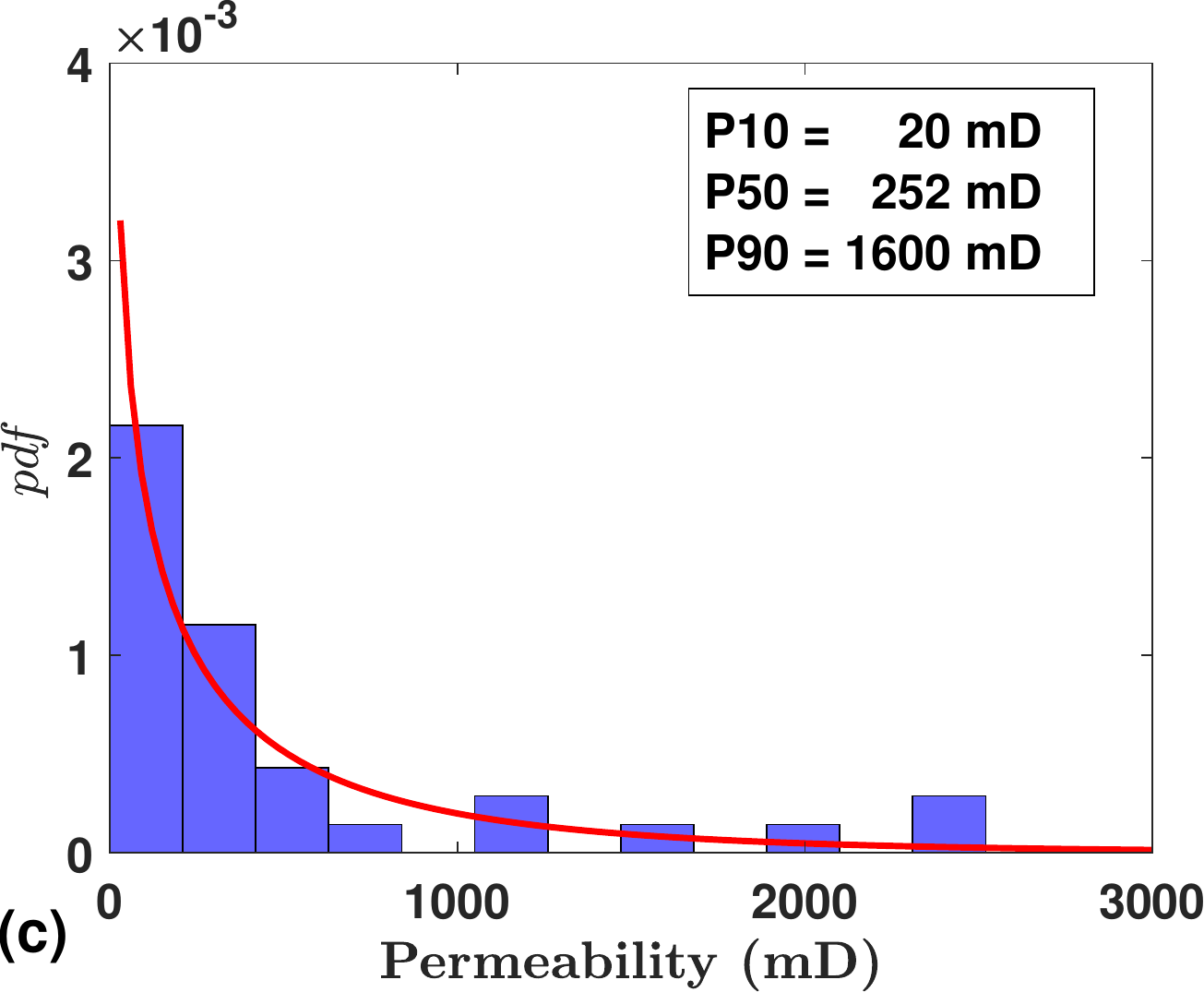}
\end{minipage}
\begin{minipage}{0.45\linewidth}
\centering
\includegraphics[scale=0.4]{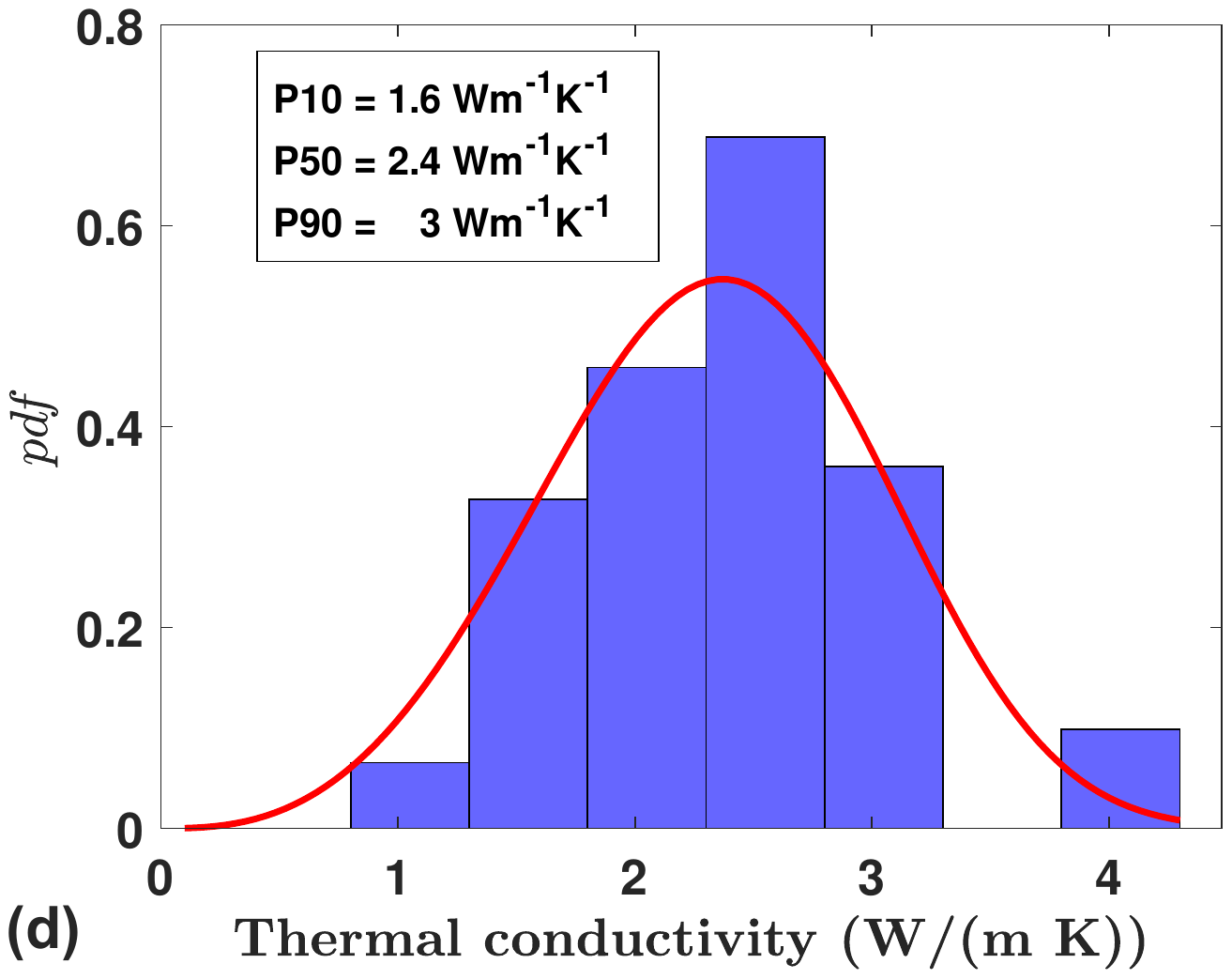}
\end{minipage}
\\
\centering
\begin{minipage}{0.45\linewidth}
\centering
\includegraphics[scale=0.4]{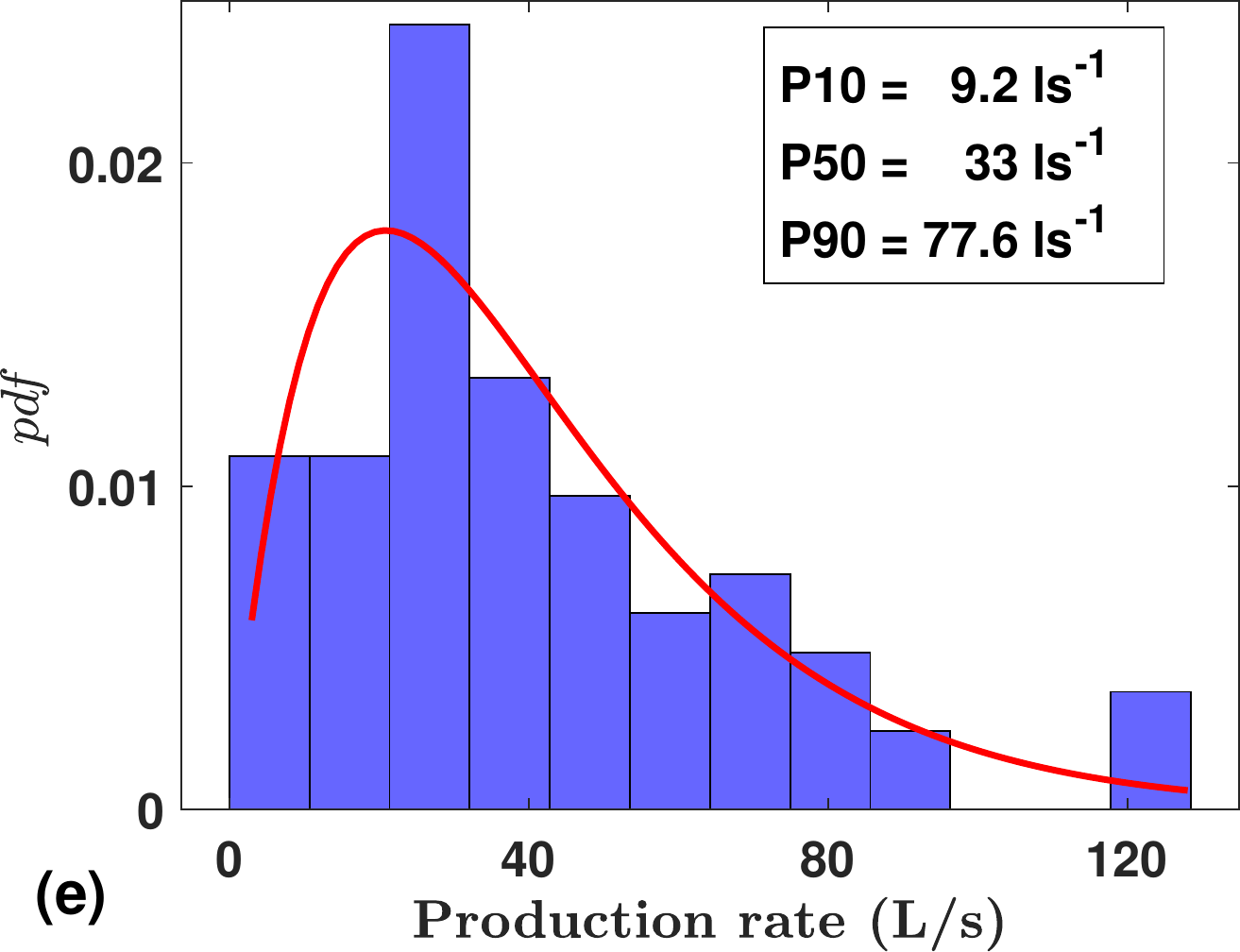}
\end{minipage}
\begin{minipage}{0.45\linewidth}
\centering
\includegraphics[scale=0.4]{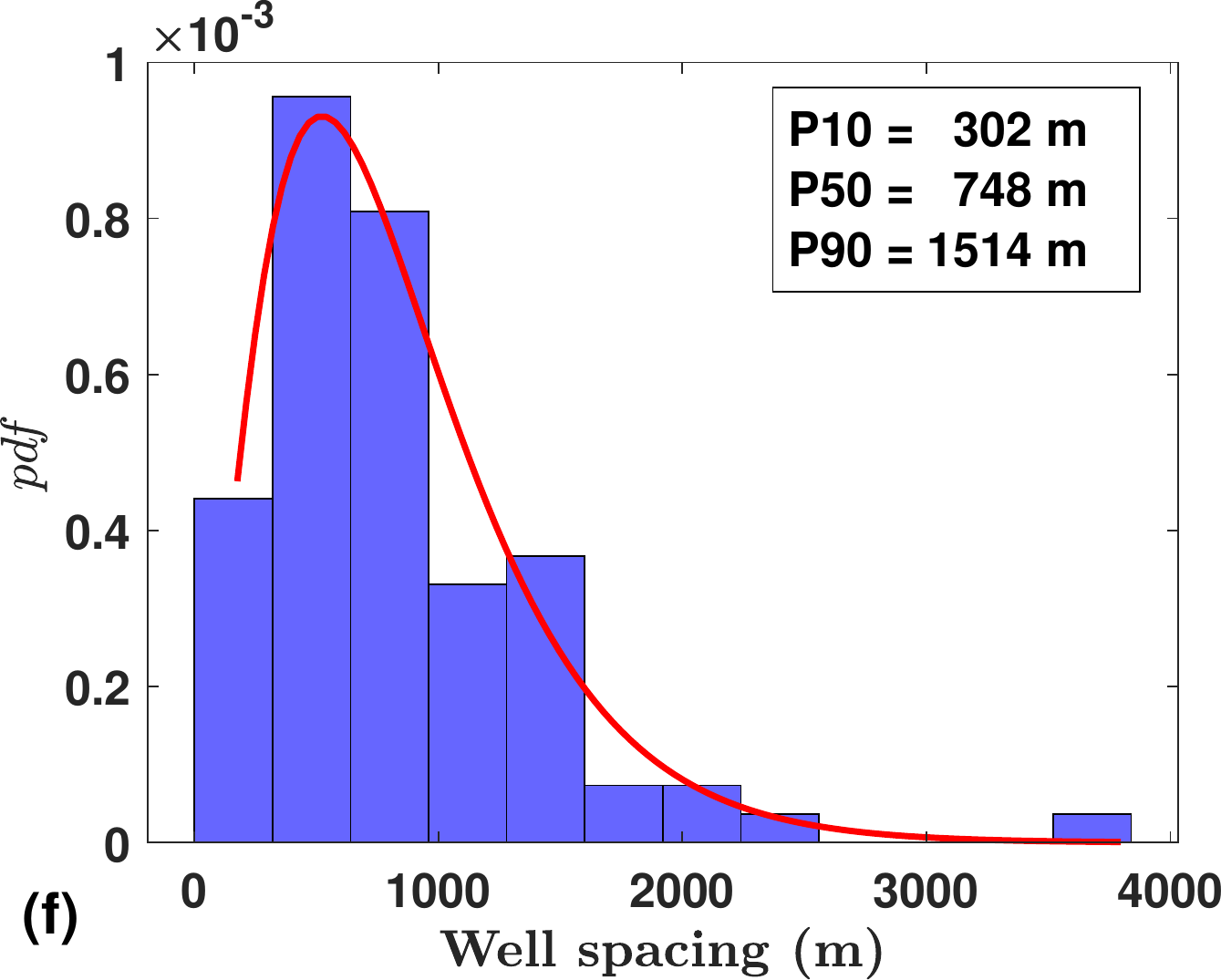}
\end{minipage}
\caption{Analog data gathered from 135 geothermal fields reflecting the ranges of several parameters, including (a) temperature, (b) porosity, (c) permeability, (d) thermal conductivity, (e) production rate per well, and (f) well spacing. The data is adopted from \cite{Santoso2019}.}
\label{figure2}
\end{figure}
Six critical parameters are identified from the database to be addressed in the uncertainty quantification study, including temperature, permeability, porosity, production rate, well spacing, and formation thermal conductivity. The distributions of the six parameters are depicted in Figure \ref{figure2}, which show that the permeability from various fields varies between 20 and 1600 $mD$, reflecting a wide range of variability due to the inherent heterogeneity of geothermal reservoirs. The porosity varies between 0.02 and 0.16. The low-porosity ranges correspond to volcanic and metamorphic reservoirs, while high-porosity ranges correspond to sedimentary reservoirs. The thermal conductivity is within the range of  1.6 - 3 $W/(m\cdot K)$, reflecting the variations among basaltic, sandstone, and metamorphic rocks. The 10$^{th}$ (P10) and 90$^{th}$ (P90) percentiles of the  average production rate per well varies within 9.2 - 77.6 $l/s$. The operating rate depends on the permeability and the amount of thermal energy of the reservoirs. Most fields are developed under doublet arrangement with injection and production wells with a well spacing between 302 and 1514 $m$.   
\section{Problem formulation}
The heat extraction from a geothermal reservoir with water re-injection scheme is governed by three fundamental equations: mass, momentum, and energy conservation equations \cite{Bird2006,Grant2011}. The mass conservation in 3D space is expressed by:
\begin{equation}\label{equation1}
\frac{\partial }{\partial t}\left( \phi {{\rho }_{\alpha }}{{S}_{\alpha }} \right)+\nabla .\left( {{\rho }_{\alpha }}{{u}_{\alpha }} \right)={{Q}_{\alpha }}\hspace{20pt}\alpha =\text{l,v},
\end{equation}
where $t$ is time, $\phi$ is the porosity of the formation, $\rho$ is the phase density, $S$ is the fluid phase saturation in the formation, $u$ is the phase superficial velocity, and $Q$ is the phase sink/source term. The subscription $\alpha$ denotes the phase where $l$ is for liquid, and $v$ is for vapor. The number of mass conservation equations follows the number of phases, $N$. 
\\~\\
The superficial velocity for each phase is modeled using the extended Darcy’s law, that is,
\begin{equation}\label{equation3}
{{u}_{\alpha }}=-\frac{\mathbf{k}{{k}_{r\alpha }}}{{{\mu }_{\alpha }}}\nabla {{\Phi }_{\alpha }},
\end{equation}
where $\mathbf{k}$ is the absolute permeability of the reservoir, ${k}_{r}$ is the phase relative permeability, $\mu$ is the phase dynamic viscosity, and $\Phi$ is the phase potential, which is given by:
\begin{equation}\label{equation4}
{{\Phi }_{\alpha }}={{P}_{\alpha }}-{{\rho }_{\alpha }}gz,
\end{equation}
where $P_\alpha$ is the phase pressure, $g$ is the gravitational acceleration, and $z$ is the phase hydrostatic height.\\
The energy equation describes the dynamics of the temperature in the reservoir, that is,
\begin{equation}\label{equation5}
\begin{split}
\frac{\partial }{\partial t}\left( \phi \sum\limits_{i\left( \alpha  \right)}^{N}{{{\rho }_{i}}{{S}_{i}}{{U}_{i}}}+\left( 1-\phi  \right){{\rho }_{r}}{{U}_{r}} \right)+\nabla .\left( \sum\limits_{i\left( \alpha  \right)}^{N}{{{\rho }_{i}}{{u}_{i}}{{H}_{i}}} \right) \\
-\nabla .\left( \left( \phi \sum\limits_{i\left( \alpha  \right)}^{N}{{{S}_{i}}{{\sigma }_{i}}}+\left( 1-\phi  \right){{\sigma }_{r}} \right)\nabla T \right)=W,
\end{split}
\end{equation}  
\begin{equation}\label{equation6}
{{H}_{\alpha }}={{U}_{\alpha }}+{{P}_{\alpha }}\overline{{{V}_{\alpha }}},
\end{equation}
\begin{equation}\label{equation7}
{{U}_{\alpha }}={{C}_{\alpha }}\left( T-{{T}_{ref}} \right),
\end{equation}
\begin{equation}\label{equation8}
{{U}_{r}}={{C}_{r}}\left( T-{{T}_{ref}} \right),
\end{equation} 
where $U$ is the internal energy of a phase or rock, $H$ is the enthalpy of a phase or rock, $\overline{V}$ is the volume per unit mass of a phase, $C$ is specific heat capacity per unit mass of a phase or rock, $T$ is the temperature of the reservoir system, ${T}_{ref}$ is the reference temperature, typically corresponds to the fluid injection temperature, $\sigma$ is the thermal conductivity of a phase or rock, and $W$ is the energy source/sink term. The subscription $r$ denotes the rock.
The density of the fluid is a function of pressure and temperature, which can be calculated with an Equation of State \cite{Grant2011}. The unknowns correspond to the saturation, pressure, and temperature/energy. In this work, the thermal recovery equations are solved implicitly using CMG-STARS simulator \cite{CMG2016}. The boundary conditions, corresponding to the heat loss/gain to the overburden and underburden, are also considered.          

\section{Proposed workflow}
\subsection{Dimensionless variables}
The main objective of the study is to evaluate the thermal recovery and produced-enthalpy rate under uncertainties. To be physically scalable for different  static and dynamic  conditions of a reservoir,  several dimensionless groups are considered. These  dimensionless groups  correspond to the operational time, thermal recovery factor, and enthalpy production factor, defined by:       
\begin{equation}\label{equation9}
{{t}_{D}}=\frac{{{Q}_{\alpha }}t}{{{V}_{p}}},
\end{equation}
where $V_{p}$ is the reservoir's pore-volume. The dimensionless time represents the cumulative volume of injected water into the reservoir at reservoir conditions relative to the reservoir’s pore-volume. We use the thermal recovery factor and enthalpy production factor to quantify the energy production efficiency.\\
 The thermal recovery factor ($F_{TR}$) reflects the change in reservoir energy over time due to the extraction of energy and interaction with the re-cycled cold water relative to the original energy in place, that is, 
\begin{equation}\label{equation10}
{{F}_{TR}}\left( {{t}_{D}} \right)=\frac{U\left( {{t}_{D}}=0 \right)-U\left( {{t}_{D}} \right)}{U\left( {{t}_{D}}=0 \right)}.
\end{equation}
The enthalpy production factor ($F_{EP}$) measures the decline rate (i.e., sustainability) of  heat production. It is defined as the ratio of the produced enthalpy at a given time and the initial enthalpy, that is, 
\begin{equation}\label{equation11}
{{F}_{EP}}\left( {{t}_{D}} \right)=\frac{\dot{H}\left( {{t}_{D}} \right)}{\dot{H}\left( {{t}_{D}}=0 \right)},
\end{equation}
where $\dot{H}$ is the enthalpy production rate during the extraction process. Initially, this indicator will be unity, and then it  declines as a result of reservoir cooling and cold front breakthrough.
The performance indicators $F_{EP}$ and  $F_{TR}$ are used as the objective functions to maximize  in our uncertainty quantification and optimization.
\subsection{Non-intrusive uncertainty quantification}
A non-intrusive uncertainty quantification consists of two essential steps: 1) constructing low-fidelity (surrogate) model to represent the high-fidelity (physics-based) model, and 2) propagating uncertainty with Monte Carlo  using  the low-fidelity model. The low-fidelity model is expressed in spectral expansion as follows: 
\begin{equation}\label{equation12}
f\left( \xi  \right)=\sum\limits_{\alpha \in A }{{{f}_{\alpha }}{{\Psi }_{\alpha }}\left( \xi        \right)},
\end{equation}
where $f$ represents the quantity of interest (QoI), $\Psi _{\alpha}\left( \xi \right)$ is an orthogonal basis function, and $\xi$ is the variable of interest. The symbol $\alpha$ denotes a multi-index scheme, defined by: 
\begin{equation}\label{equation13}
A = \left( {{\mathbf{\alpha }}^{\mathbf{1}}},{{\mathbf{\alpha }}^{\mathbf{2}}},...,{{\mathbf{\alpha }}^{{M+1}}} \right) ,
\end{equation}
where, $({M+1})$ denotes the number of basis functions, including the zeroth-order. For multiple inputs, the indexing for $\alpha$ becomes $\alpha _{i}^{k}$, where $i$ is the index for the variables, $k$  is the index for the basis function, and $\alpha$ returns the polynomial order used for a certain variable and certain basis number. The truncation of the basis functions follows the total order scheme \cite{LeMaitre2010}, expressed as:
\begin{equation}\label{equation14}
|\mathbf{\alpha }|=\sum\limits_{i=1}^{d}{\alpha _{i}^{k}\le p},
\end{equation}
where  $d$ is the total number of input parameters, and   $p$ is the highest order within the polynomial basis. The size of $\alpha$, corresponding  to the number of basis functions within this scheme, is given by:
\begin{equation}\label{equation15}
\left( M+1 \right)=\frac{\left( d+p \right)!}{d!p!}.
\end{equation}
For instance, a case with two input variables and a polynomial of order 2 has 6 basis functions, including the zeroth-order. The expansion is written as follows:
\begin{equation}\label{equation16}
\begin{split}
  & f\left( \mathbf{\xi } \right)={{f}_{0}}{{\Psi }_{0}}\left( {{\xi }_{1}} \right){{\Psi }_{0}}\left( {{\xi }_{2}} \right)+{{f}_{1}}{{\Psi }_{1}}\left( {{\xi }_{1}} \right)+{{f}_{2}}{{\Psi }_{1}}\left( {{\xi }_{2}} \right) \\ 
 & \text{           }+{{f}_{3}}{{\Psi }_{2}}\left( {{\xi }_{1}} \right)+{{f}_{4}}{{\Psi }_{2}}\left( {{\xi }_{2}} \right)+{{f}_{5}}{{\Psi }_{1}}\left( {{\xi }_{1}} \right){{\Psi }_{1}}\left( {{\xi }_{2}} \right)
 \end{split}
\end{equation}
The unknowns are the coefficients, ${{f}_{k}}$, of the basis functions. 
The coefficients in Eq. \ref{equation16} can be obtained through projection using Gauss quadrature or Least Square minimization \cite{LeMaitre2010}. In this study, we use the Least Square minimization method, which seeks  to minimize the residual error, defined as:
\begin{equation}\label{equation17}
\mathbf{f}=\underset{{{f}_{k}}\in \mathbb{R}}{\mathop{\arg \min }}\,\frac{1}{{{N}_{s}}}\sum\limits_{i=1}^{{{N}_{s}}}{{{\left( f_{i}^{true}-\sum\limits_{\alpha \in A }{{{f}_{\alpha }}{{\Psi }_{\alpha }}\left( \xi  \right)} \right)}^{2}}}
,\end{equation}
where $N_{s}$ is number of observations. Superscription $true$ denotes the observation data.      
\\
For uncertainty propagation, Monte Carlo  is applied to perform sampling from the  input ranges. The outputs are then calculated through the low-fidelity model. These outputs form a distribution from which the P10 (10\% probability), P50 (50\% probability), and P90 (90\% probability) are deduced.
\subsection{Modeling of time-continuous response}
We propose a new method to model the time-continuous responses of thermal recovery and enthalpy production,  based on a nested function approach. With this approach, a time-dependent objective function, $y$, is represented by an empirical function as follows:
\begin{equation}\label{equation18}
y=g\left( {{w}_{1}}\left( \xi  \right),\ldots ,{{w}_{{{N}_{w}}}}\left( \xi  \right),t \right),
\end{equation}
where $g$ is a given regression function, designed  to fit $y$ , and $w_{i, i=1,\dots,N_w}$ are the corresponding fitting parameters to be determined. The number of coefficients, ${{N}_{w}}$, is preferred to be small but high enough to capture the desired accuracy of approximation. We apply the spectral polynomial expansion to approximate the time-independent coefficients $w_i$, rather than the time-dependent objective function, $y$, which is commonly used in the literature. \\
In this work, the thermal recovery factor is described with a two-parameter function, expressed by:
\begin{equation}\label{equation19}
{{F}_{TR}}\left( {{t}_{D}} , \xi\right)=\frac{{{w}_{1}}\left( \xi  \right){{t}_{D}}}{1+{{w}_{2}}\left( \xi  \right){{t}_{D}}}.
\end{equation}
While the enthalpy production function is described with a three-parameter function: 
\begin{equation}\label{equation20}
{{F}_{EP}}\left( {{t}_{D}} , \xi\right)={{w}_{3}}\left( \xi  \right)+\frac{1-{{w}_{3}}\left( \xi  \right)}{1+{{\left( \frac{{{t}_{D}}}{{{w}_{4}}\left( \xi  \right)} \right)}^{{{w}_{5}}\left( \xi  \right)}}}.
\end{equation}
The regression functions in Eqs. \ref{equation19} and \ref{equation20} were selected based on their suitability to capture the profiles of the physical functions with the least number of parameters. The  parameters $w_{i, i=1,\dots,N_w}$ are determined from the low fidelity model given in Eq. \ref{equation12} .
\subsection{Uncertainty quantification and optimization workflow}
The proposed workflow to develop time-continuous, regression-based uncertainty quantification consists of five steps, as illustrated in  Figure \ref{figure3}. The workflow consists of selection and identification of the significant uncertainty parameters, building the proxy models, and perform sensitivities, which are described in more detail, as follows:  
\begin{figure}[h!]
\centering
\includegraphics[width=0.9\textwidth]{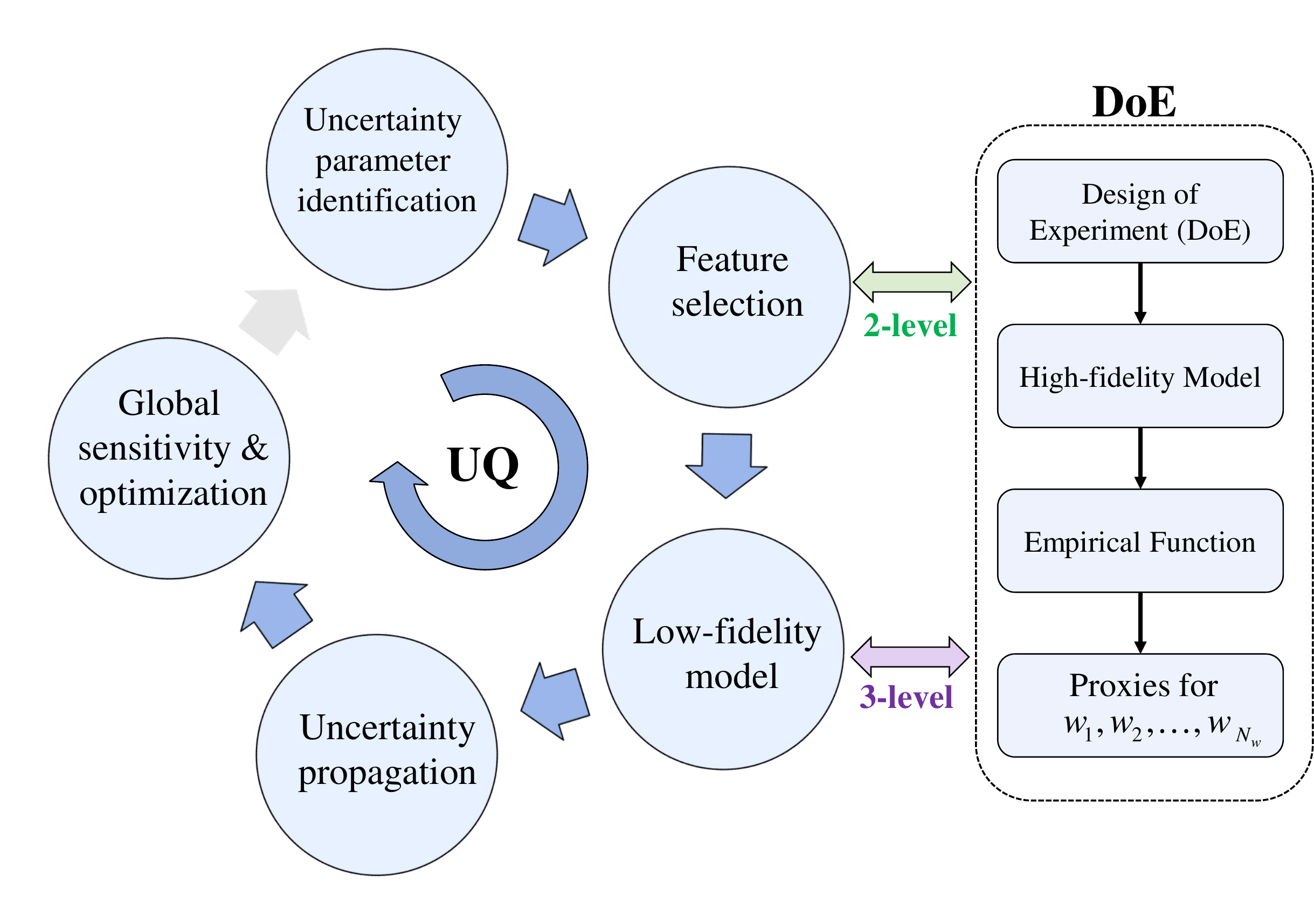}
\caption{Proposed Uncertainty Quantification (UQ) workflow to develop time-continuous, regression-based uncertainty quantification. The processes within the dashed line consist of fitting, training, and testing of the low-fidelity model from several realizations of the high-fidelity model. The parameters within the processes are sampled with either 2-level (for feature selection) or 3-level (for final response) DoE. The final response describes time-continuous objective functions, $F_{TR}$ and $F_{EP}$.  Monte Carlo simulations  are conducted for uncertainty propagation. The global sensitivity analysis uses ANCOVA technique.}
\label{figure3}
\end{figure}
\begin{itemize}
\item[]
\textbf{Step 1 - Identification of uncertainty parameters:}   uncertainty parameters are identified based on the impact of their range of variability (upper and lower bounds) on the response functions, corresponding to the thermal recovery and enthalpy production factors. Therefore, the parameter ranges affect the universality of the reduced-order model and the sensitivity analysis. We opted to group selected variables to accommodate dependency among them. This step is important to ensure physical consistency of the low-fidelity model.
\item[]
\textbf{Step 2 - Selection of significant parameters:} in this step, a 2-level DoE, such as folded Plackett-Burman, 2-level Full Factorial, or 2-level Fractional Factorial, is used to perform the screening process to select the most significant parameters in the objective functions. The concept is to measure the sensitivity of the time-continuous coefficients $w_i$, representing the regression functions of the responses, to the input parameters (modifiers).  Since the 2-level DoE adopt  first-order polynomials as the basis for the low-fidelity model, the sensitivity measure is also of the first-order \cite{Homma1996}. This can be inferred directly from the regression coefficients. The reduced-order model from a 2-level DoE can be expressed by:
\begin{equation}\label{equation21}
f\left( \xi  \right)={{f}_{0}}+\sum\limits_{i=1}^{d}{{{f}_{i}}{{\Psi }_{1}}\left( {{\xi }_{i}} \right)}.
\end{equation}
\\
The variance  of $f$ is defined by:
\begin{equation}\label{equation22}
Var\left[ f \right]=\sum\limits_{i=1}^{d}{{{f}_{i}}^{2}Var\left[ {{\xi }_{i}} \right]}
\end{equation}
The sensitivity is of the first-order, that is,
\begin{equation}\label{equation23}
{{S}_{i}}=\frac{{{f}_{i}}^{2}Var\left[ {{\xi }_{i}} \right]}{Var\left[ f \right]},\hspace{20pt}i=1,...,d,
\end{equation}
with,
\begin{equation}\label{equation24}
\sum\limits_{i=1}^{d}{{{S}_{i}}}=1.
\end{equation}
By conducting a hypothesis testing of 95\% confidence based on the sensitivity indices,   insignificant input parameters can be eliminated.
\item[]
\textbf{Step 3 - Low-fidelity model:} after selecting the significant parameters, a more accurate 3-level DoE, such as D-Optimal, Taguchi, Latin Hypercube, or Box-Behken, is used to sample the variables. The responses corresponding to the DoE samples are obtained from the high-fidelity simulation model. The input-output samples are then used to train the low-fidelity model with a Least-Square scheme, using the Least Angle Regression (LARS) algorithm in UQLab \cite{Marelli2014}.  Variable degrees of the polynomial-basis-function were investigated using  two metrics  to evaluate the low-fidelity model: 1) $R^{2}$ for measuring the accuracy of the training, and 2) $Q^{2}$ for measuring the predictability of the model, such that,
\begin{equation}\label{equation25}
{{R}^{2}}=\frac{\sum\limits_{i=1,i\in T}^{{{N}_{s}}}{{{\left( f_{i}^{true}-\sum\limits_{\alpha \in A }{{{f}_{\alpha }}{{\Psi }_{\alpha }}\left( \xi  \right)} \right)}^{2}}}}{\sum\limits_{i=1,i\in T}^{{{N}_{s}}}{{{\left( f_{i}^{true}-\frac{1}{{{N}_{s}}}\sum\limits_{k=1,k\in T}^{{{N}_{s}}}{f_{k}^{true}} \right)}^{2}}}}, 
\end{equation}
The difference between   $R^{2}$ and $Q^{2}$   is in the data sets used for their evaluation, where  ${R}^{2}$ is evaluated using the training data $T$, and ${Q}^{2}$ is evaluated using the testing data $D$. The testing set contains sampling from the space of  modifiers, which are different from the training set. 
\item[]
\textbf{Step 4 - Uncertainty propagation:}  Monte Carlo  simulations are applied to the low-fidelity model to capture uncertainty propagation. The sampled realizations are collected in the form of   distributions from which the probabilistic forecast of percentiles, P10, P50, and P90 are  evaluated.     
\item[]
\textbf{Step 5 - Global sensitivity analysis using ANCOVA:} this step aims to evaluate the global sensitivity  of the objective functions, including the time-dependent thermal-recovery  and enthalpy-production factors. Since the objective functions are in the form of nested functions (see Eqs. \ref{equation20} and \ref{equation21}), direct evaluation of sensitivity analysis (Eq. \ref{equation22} to \ref{equation24}) cannot be performed \cite{Caniou2012}. The commonly used ANOVA (Analysis of Variance) or Sobol sensitivity indices may miss-predict the sensitivity value
due to the complex dependency  among   variables.  To overcome this issue, we utilized ANCOVA indices, which provide triplet sensitivity values: individual (uncorrelative) sensitivity, interactive, and correlative sensitivity \cite{Caniou2012}. The uncorrelative sensitivity represents the uncorrelated structural contribution of a parameter, which is expressed as:
\begin{equation}\label{equation27}
{{S}_{i}}^{U}=\frac{Var\left( {{f}_{i}}\left( {{\xi }_{i}} \right) \right)}{Var\left[ f \right]},\hspace{20pt}i=1,\dots,d.
\end{equation}
Both interactive and correlative sensitivity denote the correlated contribution among parameters. The interactive sensitivity indices are given by: 
\begin{equation}\label{equation28}
\begin{split}
{{S}_{i}}^{I}=\frac{Cov\left[ {{f}_{i}}\left( {{\xi }_{i}} \right),\sum\nolimits_{\begin{matrix}
   _{\mathbf{V}\subset \left\{ 1,...,{{N}_{r}} \right\}}  \\
   _{\left\{ i \right\}\in \mathbf{V}}  \\
\end{matrix}}{{{f}_{\mathbf{V}}}\left( {{\xi }_{\mathbf{V}}} \right)} \right]}{Var\left[ f \right]}, \\
i=1,\dots,d.
\end{split}
\end{equation}
While the correlative sensitivity indices are defined by:
\begin{equation}\label{equation29}
\begin{split}
{{S}_{i}}^{C}=\frac{Cov\left[ {{f}_{i}}\left( {{\xi }_{i}} \right),\sum\nolimits_{\begin{matrix}
   _{\mathbf{W}\subset \left\{ 1,...,{{N}_{r}} \right\}}  \\
   _{\left\{ i \right\}\notin \mathbf{W}}  \\
\end{matrix}}{{{f}_{\mathbf{W}}}\left( {{\xi }_{\mathbf{W}}} \right)} \right]}{Var\left[ f \right]}, \\
i=1,\dots,d,
\end{split}
\end{equation}
In the above equations,  $N_{r}$ is the number of Monte Carlo realizations, and ${{\xi }_{\mathbf{V}}}$ and ${{\xi }_{\mathbf{W}}}$ are independent realizations used to approximate the covariance and variance  \cite{Caniou2012}. The total contribution of uncorrelative ($U$), interactive ($I$), and correlative ($C$) sensitivity is of the  first-order, which is expressed by:
\begin{equation}\label{equation30}
{{S}_{i}}={{S}_{i}}^{U}+{{S}_{i}}^{I}+{{S}_{i}}^{C}.
\end{equation}
This step is critical to identify the governing factors that impact the objective functions. The ANCOVA calculations were performed using UQLab  \cite{Marelli2014}.
\end{itemize}
\section{Simulation model}
\subsection{Simulation domain}
\begin{figure}[h!]
\centering
\includegraphics[scale=0.35]{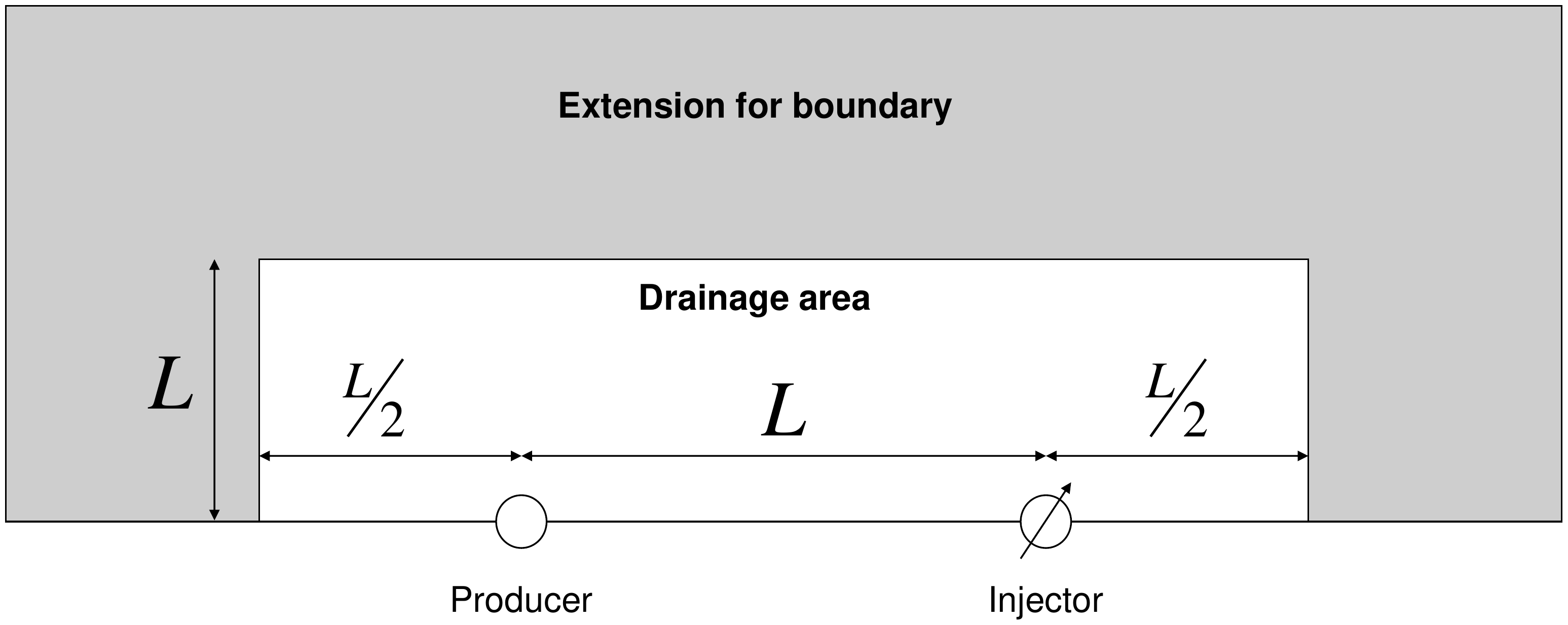}
\caption{Illustration of the 3D mechanistic model with two wells (top view). The shaded zone is used to account for the thermal boundary conditions. }
\label{figure4}
\end{figure}

\begin{figure}[h!]
\centering
\begin{minipage}[b]{0.47\linewidth}
\centering
\includegraphics[scale=0.47]{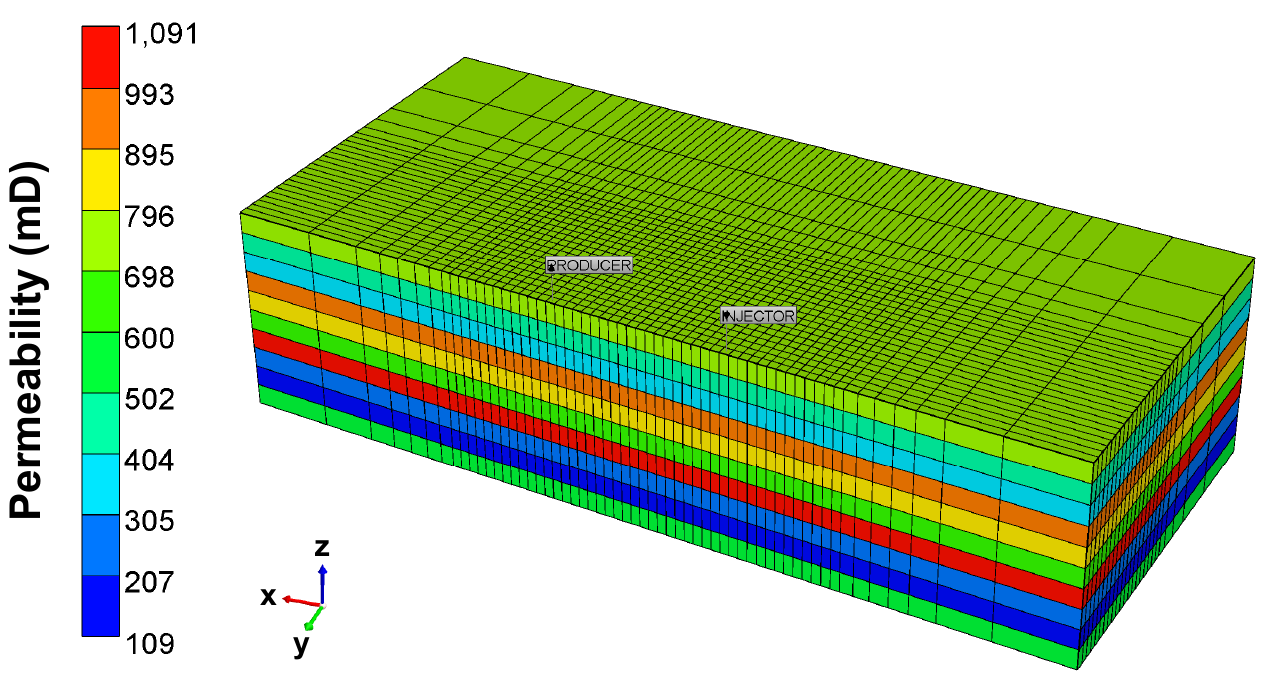}
{(a)}
\end{minipage}
\hspace{0.05cm}
\begin{minipage}[b]{0.47\linewidth}
\centering
\includegraphics[scale=0.47]{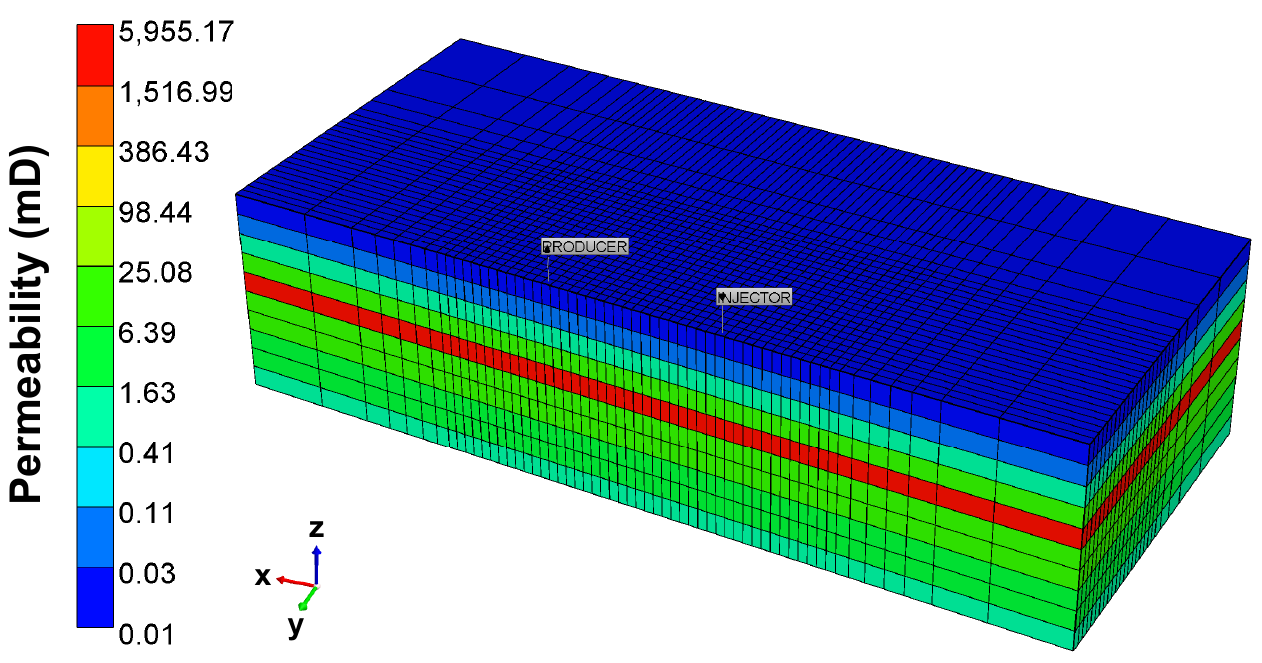}
{(b)}  
\end{minipage}
\caption{Two realizations of the model with vertical permeability distribution corresponding to  Dykstra-Parson coefficients,  ${{V}_{DP}}=0.56$ (a), and ${{V}_{DP}}=0.94$ (b).}
\label{figure5}
\end{figure}

\begin{figure}[h!]
\centering
\begin{minipage}[b]{0.48\linewidth}
\centering
\includegraphics[scale=0.48]{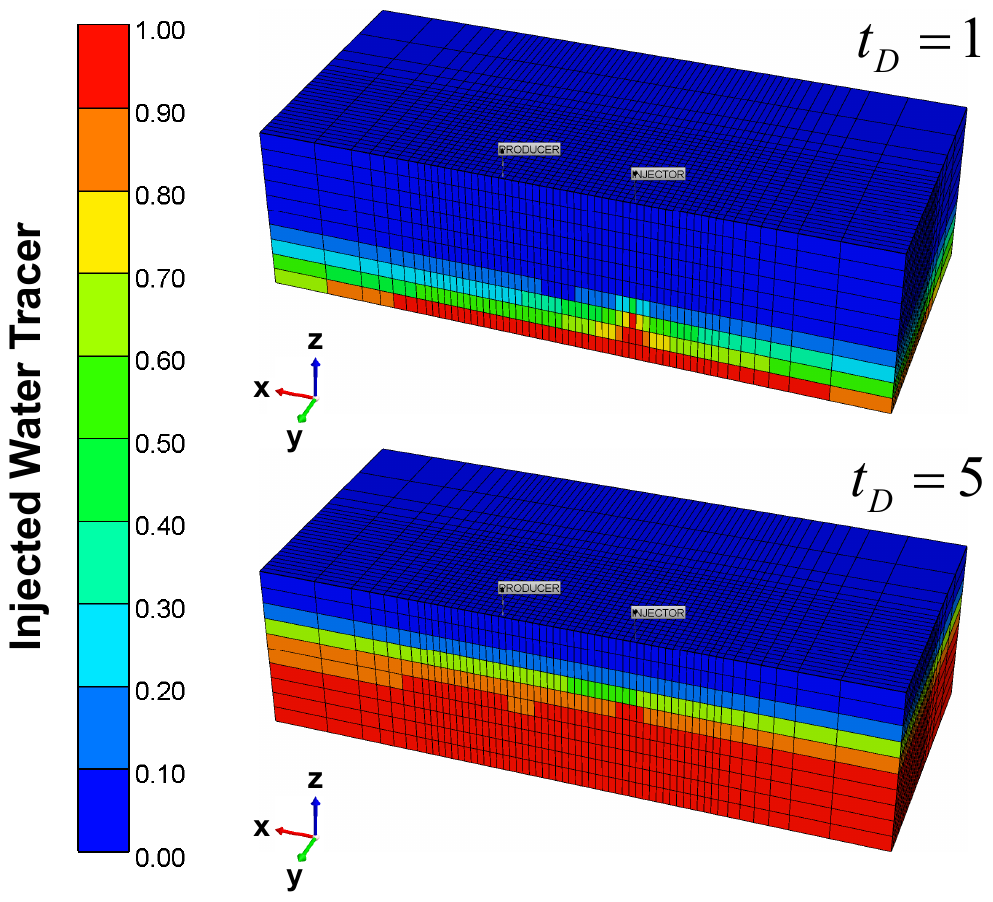}
{(a)}
\end{minipage}
\begin{minipage}[b]{0.48\linewidth}
\centering
\includegraphics[scale=0.48]{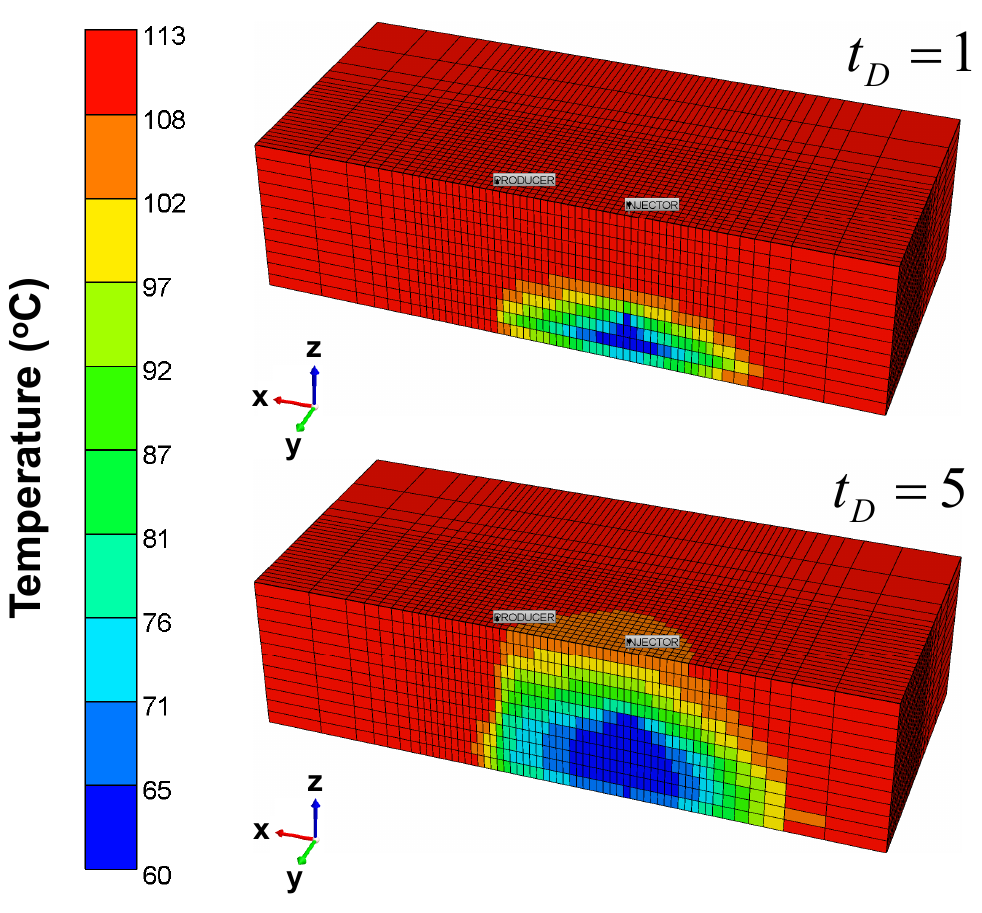}
{(b)}
\end{minipage}
\\
\begin{minipage}[b]{0.48\linewidth}
\centering
\includegraphics[scale=0.48]{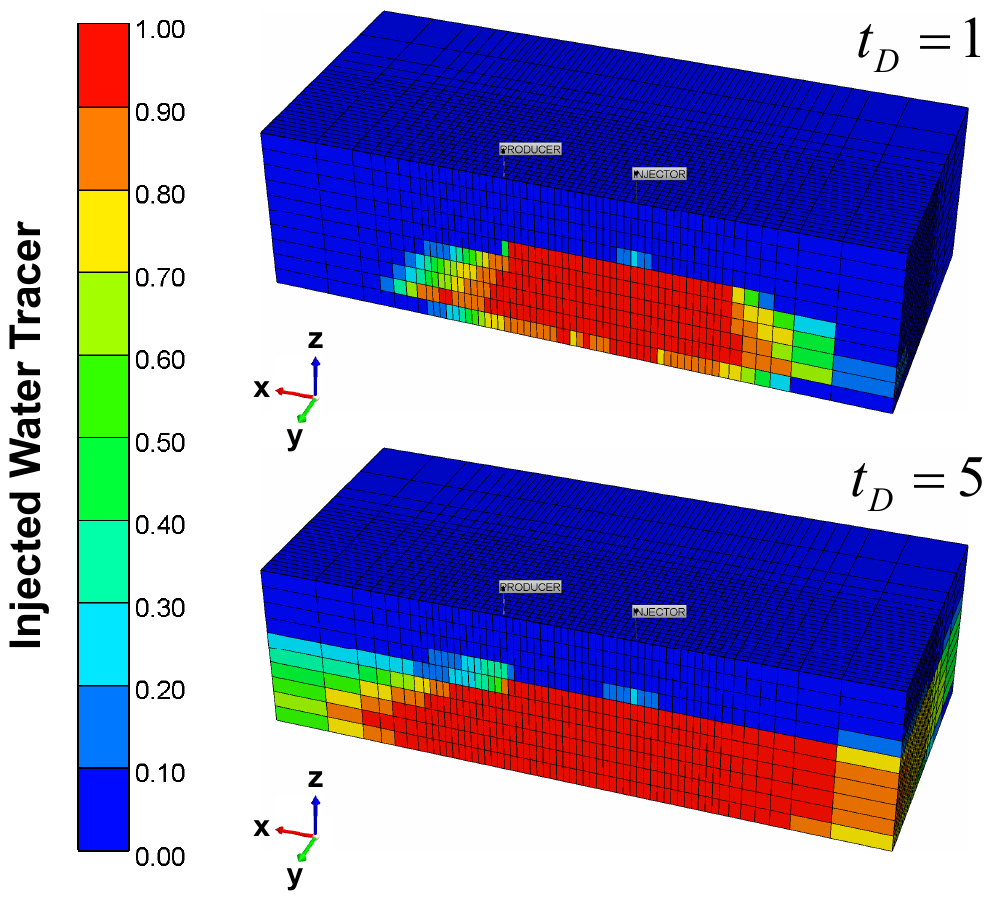}
{(c)}
\end{minipage}
\begin{minipage}[b]{0.48\linewidth}
\centering
\includegraphics[scale=0.48]{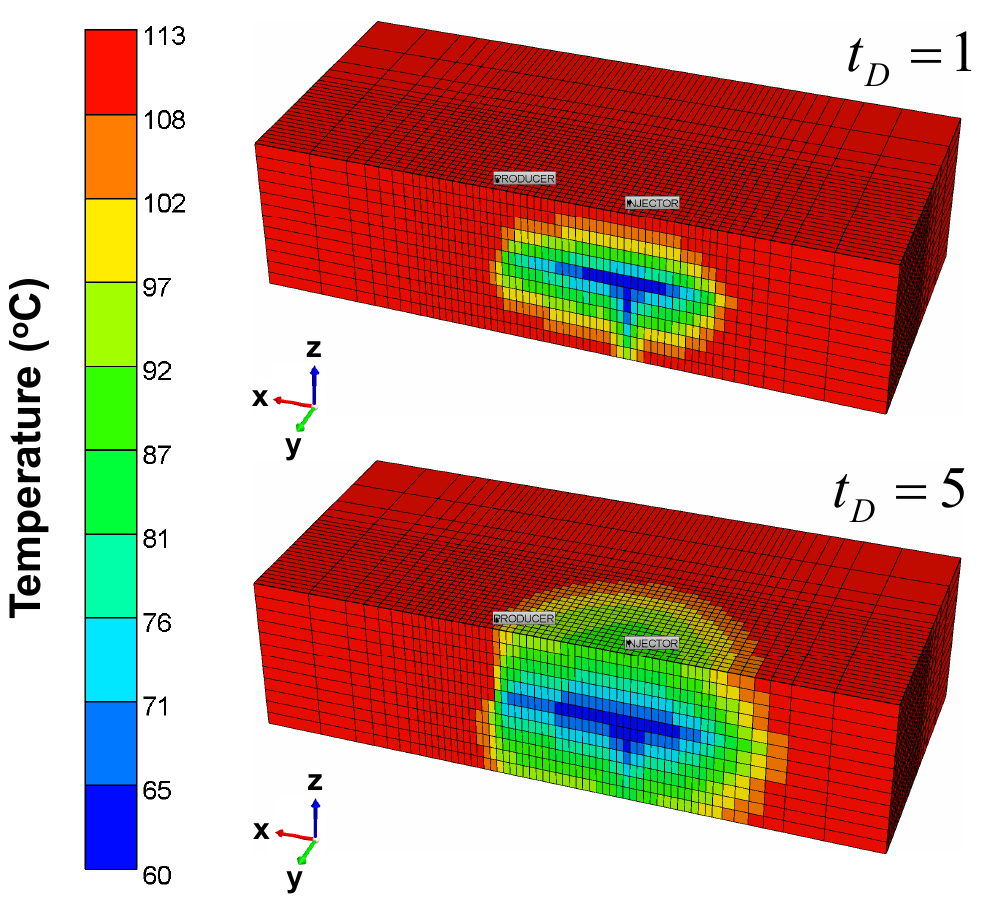}
{(d)}
\end{minipage}
\caption{Simulation results for the propagation of injected water-front, captured by a tracer (a, and c) and temperature profiles (b, and d) at different dimensionless times $t_D=1$ and $t_D=5$, respectively. The maps in (a) and (b) are for a low-heterogeneous case with $V_{DP}=0.56$, and the rest are for a  high-heterogeneous case with $V_{DP}=0.94$.}
\label{figure6}
\end{figure}

To assess the impact of subsurface uncertainties on the thermal recovery and enthalpy production, we consider a 3D mechanistic model with one vertical injector and one vertical producer, as illustrated in Figure \ref{figure4}. This setup follows the doublet arrangement, which is commonly implemented for production-injection recovery schemes in geothermal development \cite{Sanyal2005a}. 
\\
The model consists of 12 geological layers with different permeabilities. The  top and bottom layers in the 3D model are impermeable for fluid flow. However, they are conductive for heat to capture heat loss or gain to the overburden and the underburden. The model resembles a layered-cake configuration where each layer is assumed homogeneous. Besides the six uncertainty parameters discussed in Section 2, we add a parameter to reflect the vertical heterogeneity to account for multi-layer cross-flow behavior. The measure of the contrast in vertical permeability is quantified by the Dykstra-Parson coefficient \cite{Su2020}, denoted by  $V_{DP}$, and defined by: 
\begin{equation}\label{equation31}
{{V}_{DP}}=\frac{{{k}_{0.5}}-{{k}_{0.16}}}{{{k}_{0.5}}},
\end{equation}
where the subscripts $0.16$ and $0.5$  denote, respectively, the 16\% and 50\% probabilities from the Cumulative Distribution Function (CDF) of the permeability. In this work, $V_{DP}$ varies between 0.56, representing an almost homogeneous case, to 0.94 representing a very heterogeneous case. Examples of the realizations are shown in Figure \ref{figure5}. 
\\
Because of the symmetry across the vertical plane connecting the two wells, only one side of the model is considered in the simulations. To account for the lateral boundary conditions, we extended the model and used a pore-volume multiplier on the boundary grid blocks to mimic infinite-acting regime \cite{Sanyal2005a}. The drainage volume used to calculate the recovery factor and energy in place is assumed to be relative to the well spacing $L$, as shown in Figure \ref{figure4}. The production rate is a function of permeability multiplied by the reservoir pay thickness (that is, $kh$) and varies between 3 $l/s$ to 128 $l/s$, following the collected data ranges in Figure \ref{figure2}. The porosity is considered as a function of permeability. The 3D model is discretized into a 58$\times$25$\times$12 irregular Cartesian grid. This grid resolution was verified and found to be adequate for this study. The model is set to re-inject the wastewater at a constant temperature of 60$\,^\circ$C. The initial reservoir pressure is set at 15000 kPa. In Figure \ref{figure6}, we show the results for a typical simulation case  corresponding to 46.3 $l/s$ production rate, 600 $mD$ permeability, 113$\,^{\circ}$C initial temperature, 0.1 porosity, 2.4 $W/(m \cdot K)$ thermal conductivity, 700 $m$ well-spacing, and various Dykstra-Parson coefficients with values at 0.56 and 0.94, respectively. 
\\
A typical flow behavior occurring in the reservoir during water recycling is gravity-dominated. Since all  layers are hydraulically connected, the injected cold water tends to flow downward because of buoyancy.  Figure \ref{figure6} also  shows that the thermal breakthrough is slower than the water-front breakthrough due to heat conduction from the reservoir and the surrounding formation.
\begin{figure}[h!]
\centering
\begin{minipage}[b]{0.45\linewidth}
\includegraphics[scale=0.45]{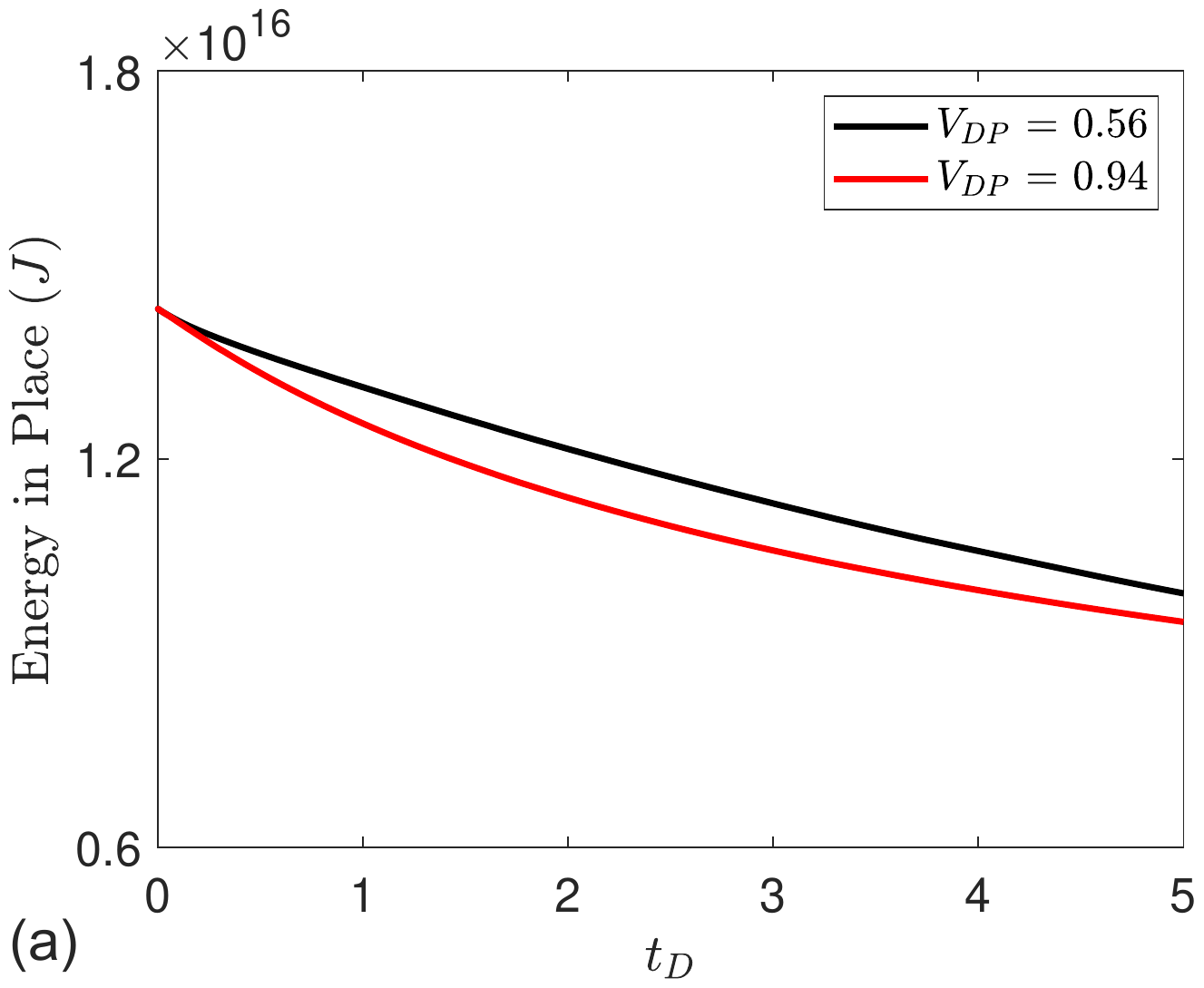}
\end{minipage}
\hspace{0.5cm}
\begin{minipage}[b]{0.45\linewidth}
\centering
\includegraphics[scale=0.45]{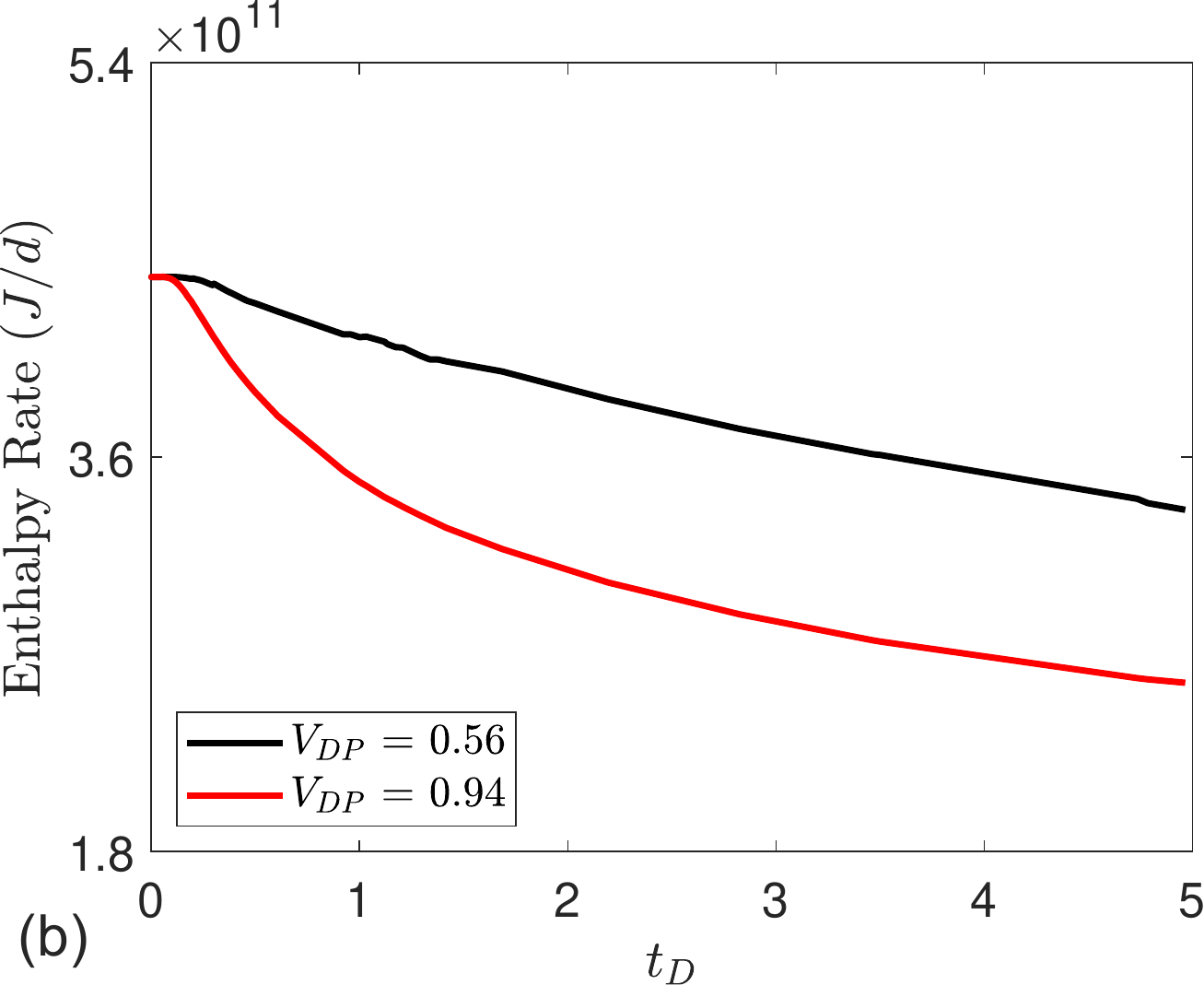}
\end{minipage}
\caption{Simulation results for two scenarios with different heterogeneity  showing the reservoir energy decline versus the dimensionless time (a), and  enthalpy production rate (b).}
\label{figure7}
\end{figure}
\\
Figure \ref{figure7}a and \ref{figure7}b show, respectively, the decline in the energy in place and the enthalpy production rate as a function of the dimensionless time $t_D$ for two different Dykstra Parson coefficients. The energy in place reflects the change in temperature, pressure, and mass at the reservoir conditions. In this case, water is re-injected, and therefore, no significant change in  the mass of the fluid occurs in the reservoir. Since the reservoir pressure is almost constant, the change in the internal reservoir energy is mainly influenced by reservoir cooling. The energy decline can be simply calculated from the temperature decline and vice-versa using Eqs. \ref{equation6} to \ref{equation8}. In the case of significant heterogeneity, the average temperature and energy in place decline faster, which is due to the presence of thief layers that can lead to a faster breakthrough, resulting in lower efficiency \cite{Ganguly2017}.  Note that thief layers (or high conduit fractures) serve to maintain high injectivity/productivity. However, water breakthrough along thief layers is a key factor in diminishing the efficiency of heat recovery, similar to hydrocarbon production from oilfields and water flow in fractured media \cite{HE2021103984,KOOHBOR2020103602,HOTEIT2008891} . Therefore, well spacing between the injectors and producers should be optimized to avoid premature water breakthrough and maximize thermal recovery \cite{Sommer2013,BIRDSELL2021116658}.\\
Figure \ref{figure7}b shows the enthalpy production rates versus the dimensionless time. The produced enthalpy rate is constant at constant production mass rate and temperature. The enthalpy decline is a result of the cold water-front breakthrough, which leads to a drop in the production fluid temperature. This concept is illustrated in Figure \ref{figure6}, which shows  the thermal front propagation from the injector to the producer at different times.  Reservoir  heterogeneity   causes early breakthrough, which decreases the temperature at the producer leading to a faster decline in enthalpy production rate. These simulation cases represent typical behaviors of the simulation model, which is used in the optimization process. 
\subsection{Capturing the physics with the low-fidelity model}
\begin{figure}[h!]
\centering
\begin{minipage}[b]{0.45\linewidth}
\centering
\includegraphics[scale=0.47]{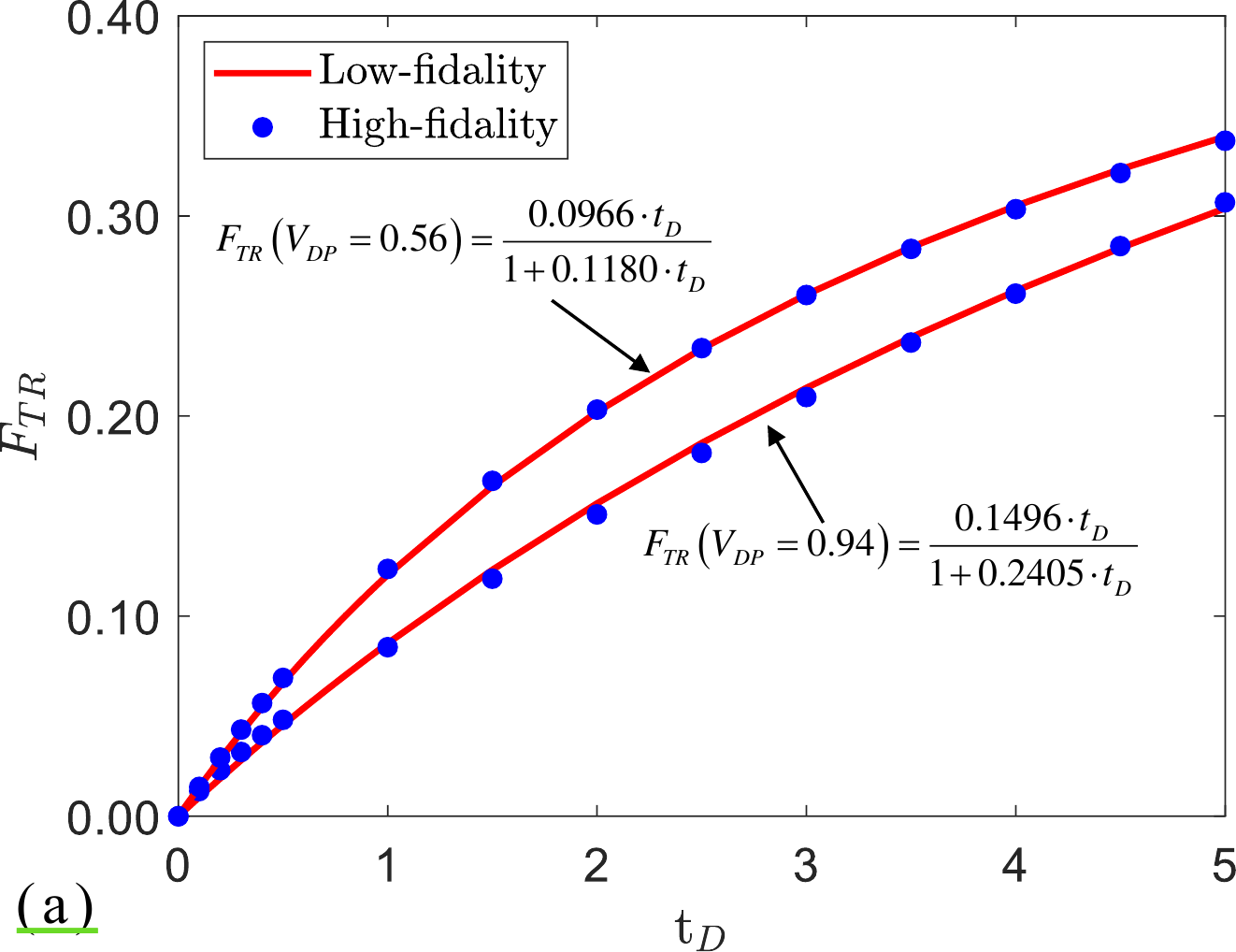}\\
\end{minipage}
\hspace{0.3cm}
\begin{minipage}[b]{0.45\linewidth}
\centering
\includegraphics[scale=0.47]{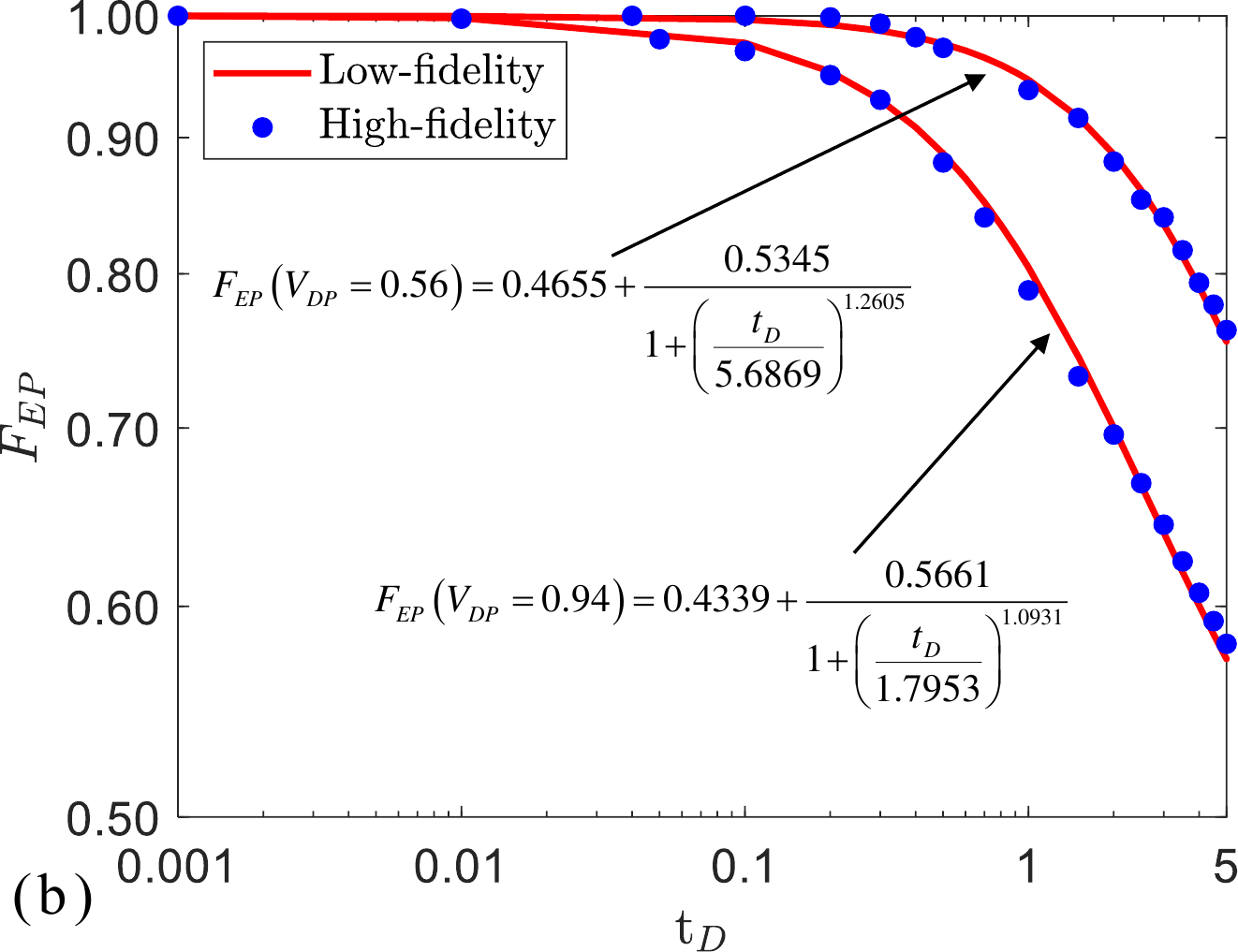}\\
\end{minipage}
\caption{Fitting the physics-based (high-fidelity) solutions for two test cases with different heterogeneities by the low-fidelity models for (a) thermal recovery factor $F_{TR}$,  and (b) enthalpy production factor $F_{EP}$.}
\label{figure8}
\end{figure}

We demonstrate the applicability of the  low-fidelity models in capturing the physics-based solutions. For a given set of simulation results, regression is performed to determine the fitting parameters $\omega_{i,i=1,...,5}$, corresponding to the analytical expressions of the thermal recovery factor and enthalpy production factors, given in Eqs. \ref{equation19} and \ref{equation20}. The low-fidelity models for $F_{TR}$ and  $F_{EP}$ can adequately fit the change in thermal recovery and enthalpy production rate versus time, as demonstrated in Figure \ref{figure8} for two test cases with different  heterogeneities.  In the workflow, the fitting coefficients $\omega_{i,i=1,...,5}$ are automatically determined as functions of the uncertainty parameters, including permeability, heterogeneity,  well spacing, and temperature. 
\section{Results}
\begin{figure}[h!]
\begin{minipage}[b]{0.45\linewidth}
\centering
\includegraphics[scale=0.45]{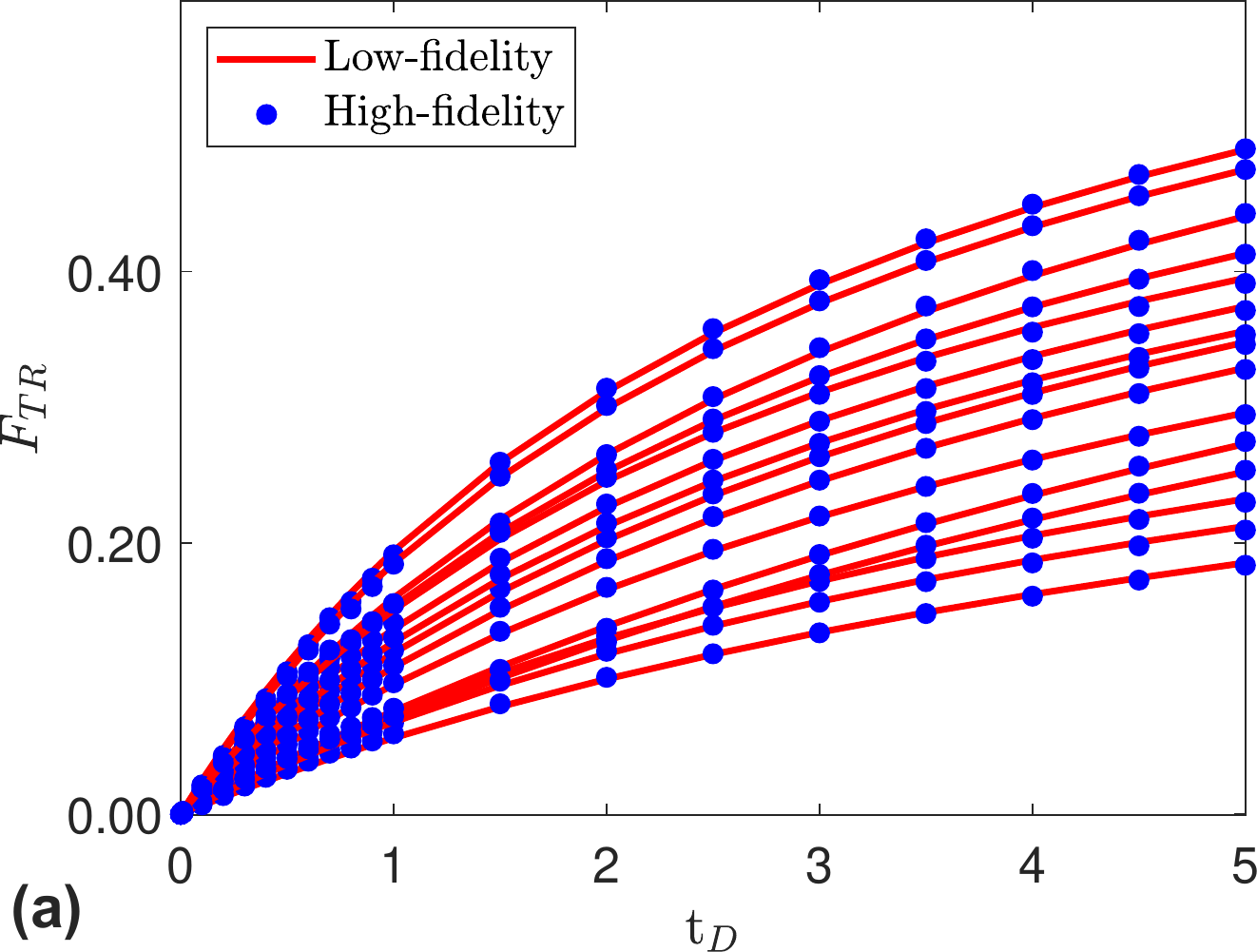}
\end{minipage}
\hspace{0.5cm}
\begin{minipage}[b]{0.45\linewidth}
\centering
\includegraphics[scale=0.45]{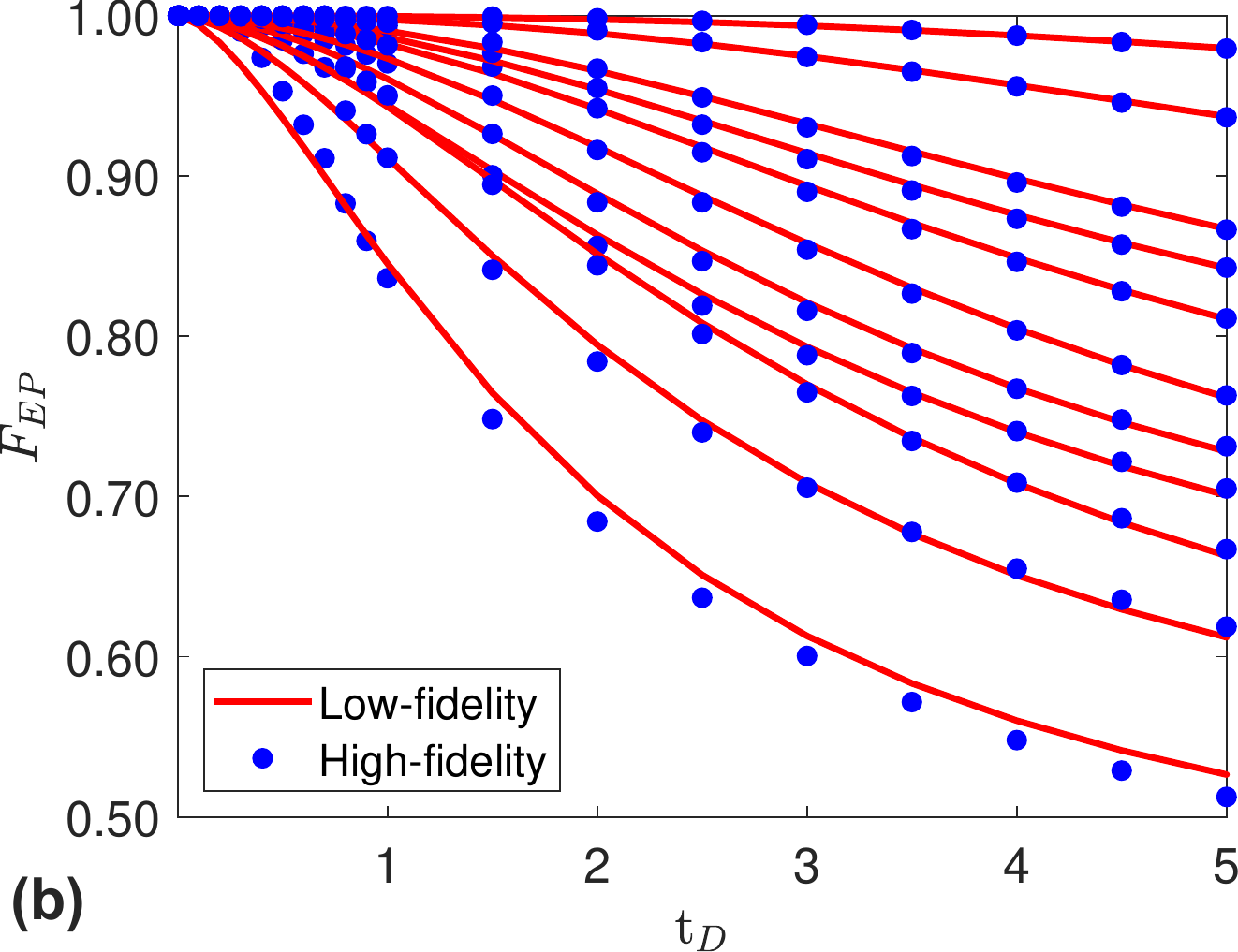}
\end{minipage}
\caption{ Results showing the reproduction of the high-fidelity model versus time by the low-fidelity model for several cases used in the sensitivity analysis by the low-fidelity model for (a) $F_{TR}$ and (b) $F_{EP}$.}
\label{figure10}
\end{figure}

\begin{figure}[h!]
\begin{minipage}[b]{0.45\linewidth}
\centering
\includegraphics[scale=0.45]{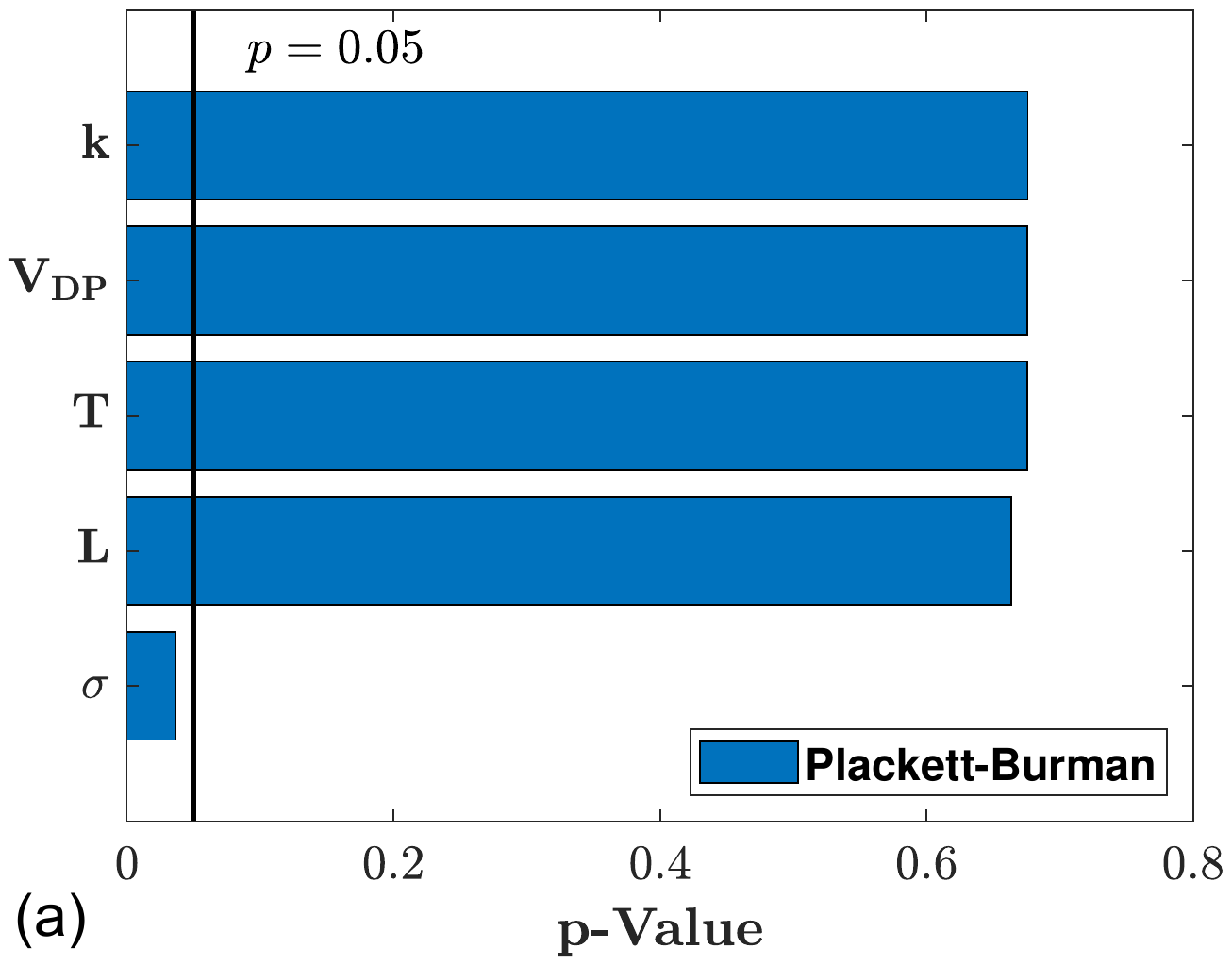}
\end{minipage}
\hspace{0.5cm}
\begin{minipage}[b]{0.45\linewidth}
\centering
\includegraphics[scale=0.45]{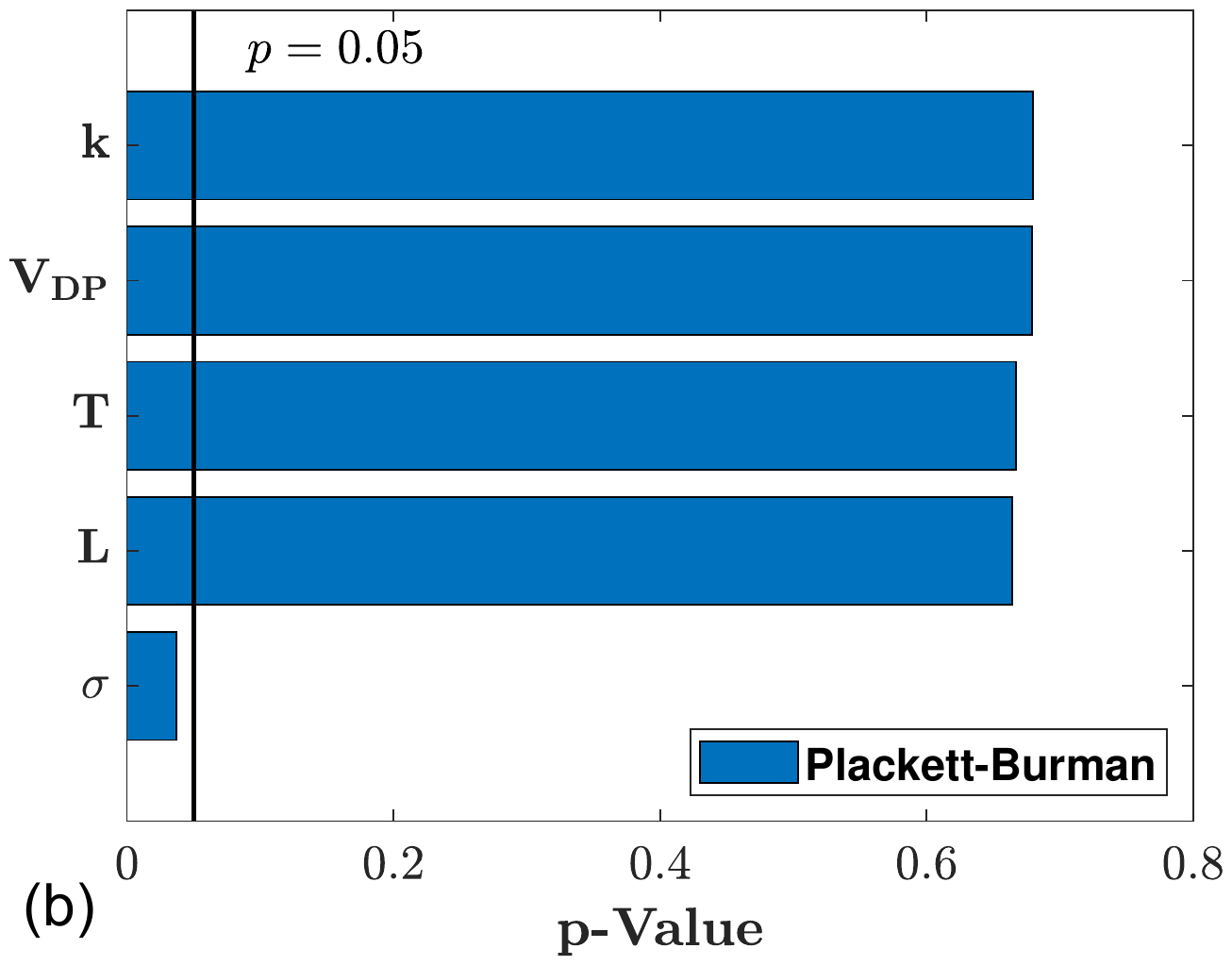}
\end{minipage}
\\
\vspace{0.5cm}
\begin{minipage}[b]{0.45\linewidth}
\centering
\includegraphics[scale=0.45]{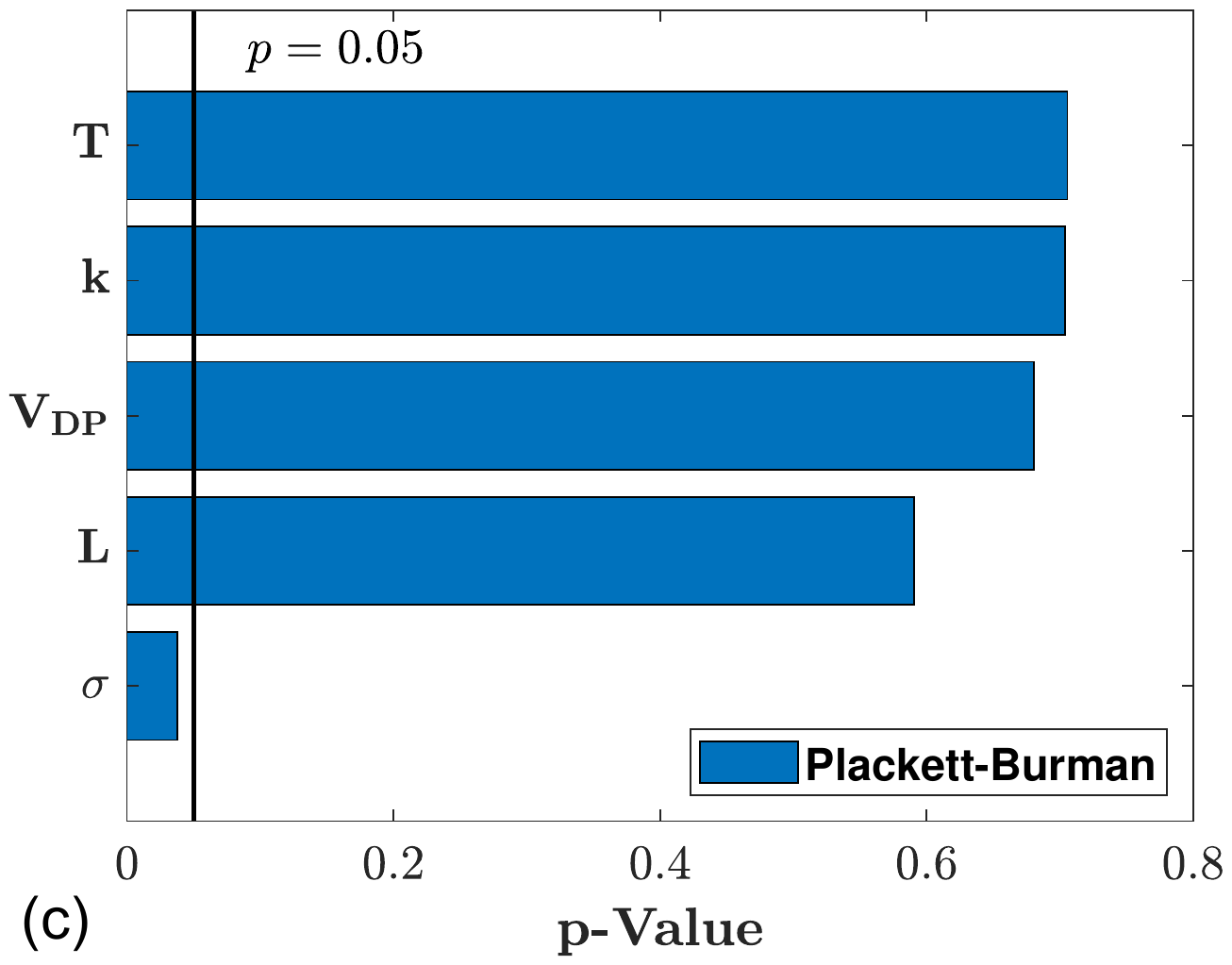}
\end{minipage}
\hspace{0.5cm}
\begin{minipage}[b]{0.45\linewidth}
\centering
\includegraphics[scale=0.45]{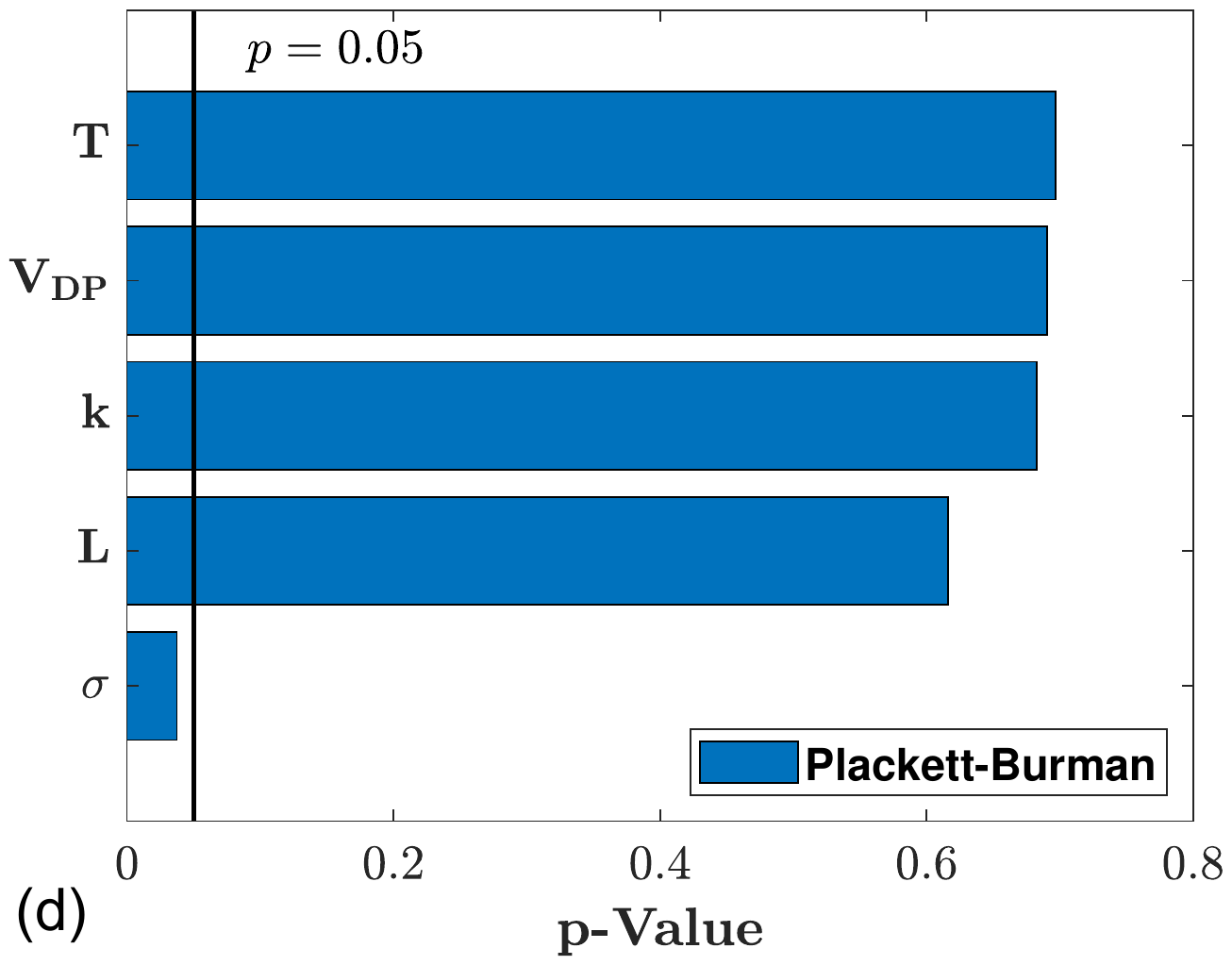}
\end{minipage}
\caption{Feature selection results from folded Placket-Burman design showing the uncertainty parameters to the fitting coefficients  (a) $w_{1}$ , (b) $w_{2}$ , (c) $w_{3}$ , (d) $w_{4}$; the behavior for $w_{5}$ is similar (figure not shown). The black line denotes a cut-off for 0.05 probability, reflecting 95\% confidence.}
\label{figure9}
\end{figure}

\begin{figure}[h!]
\begin{minipage}[b]{0.45\linewidth}
\centering
\includegraphics[scale=0.45]{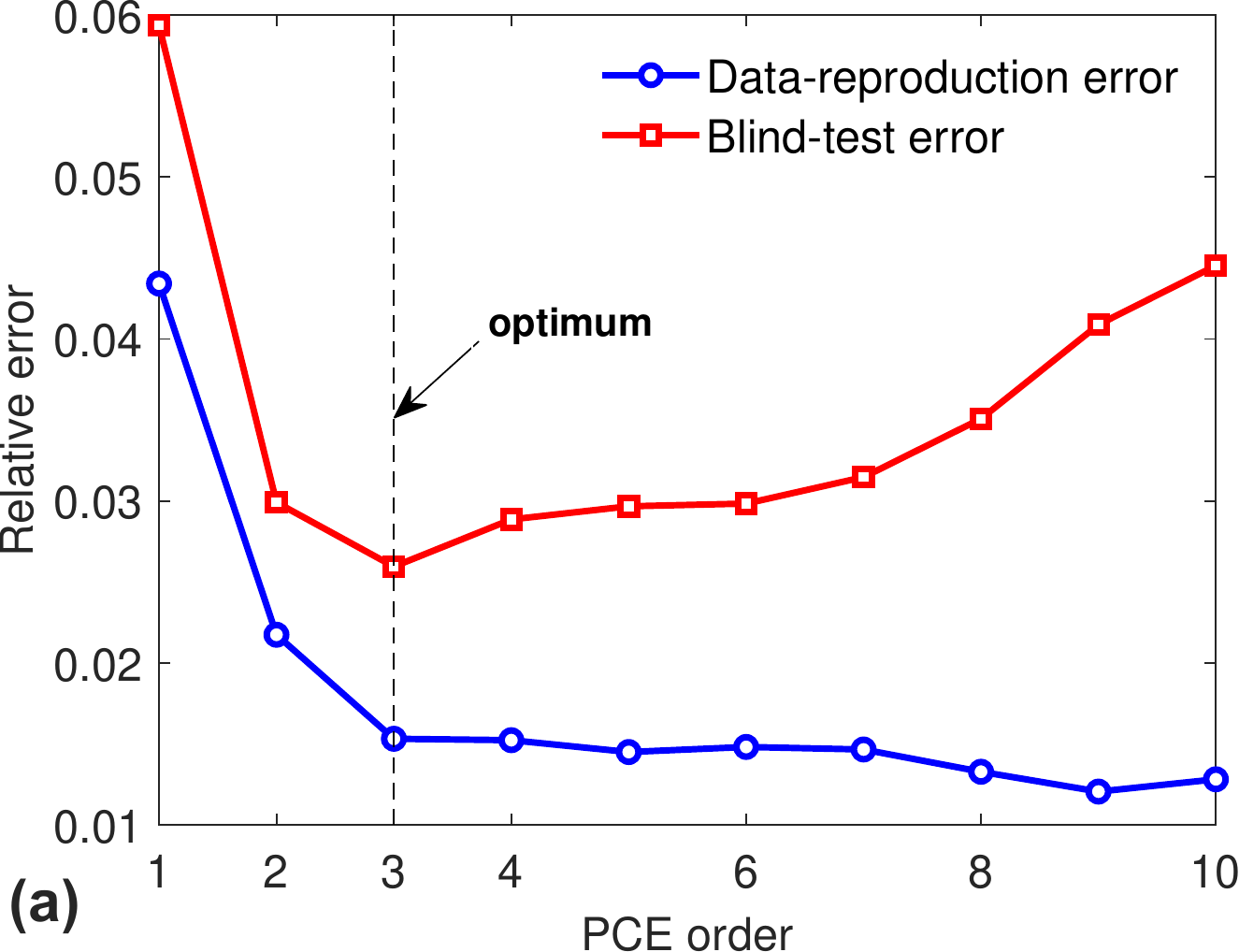}
\end{minipage}
\hspace{0.5cm}
\begin{minipage}[b]{0.45\linewidth}
\centering
\includegraphics[scale=0.45]{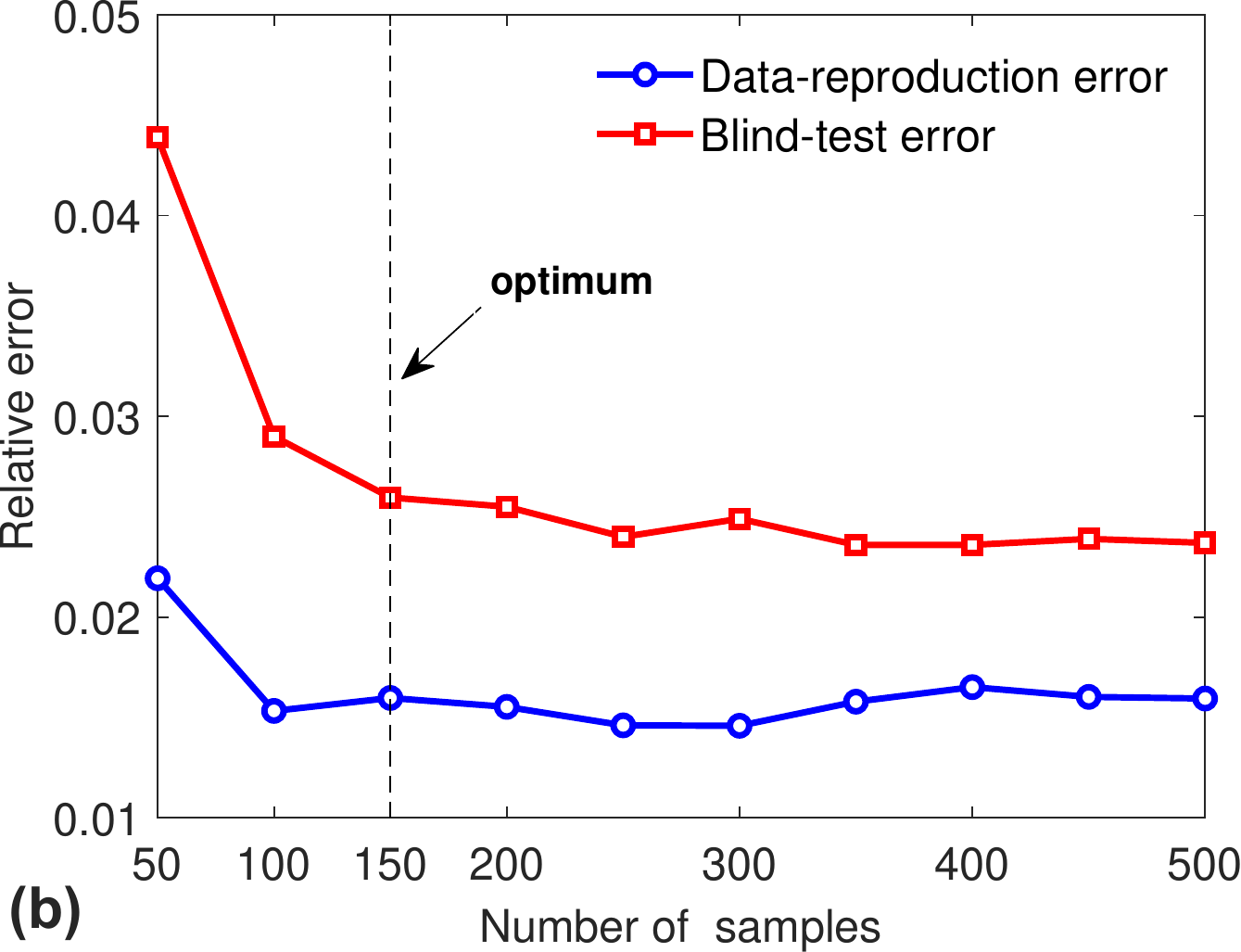}
\end{minipage}
\caption{Accuracy of the low-fidelity model as a function of the order of PCE in reproducing the training set (blue line) and in predicting the blind test set (red line) are shown in the left plot(a). The plot on the right (b)  shows the accuracy behavior of the low-fidelity model as a function of the size of the design.}
\label{figuren1}
\end{figure}

\begin{figure}[h!]
\begin{minipage}[b]{0.48\linewidth}
\centering
\includegraphics[scale=0.48]{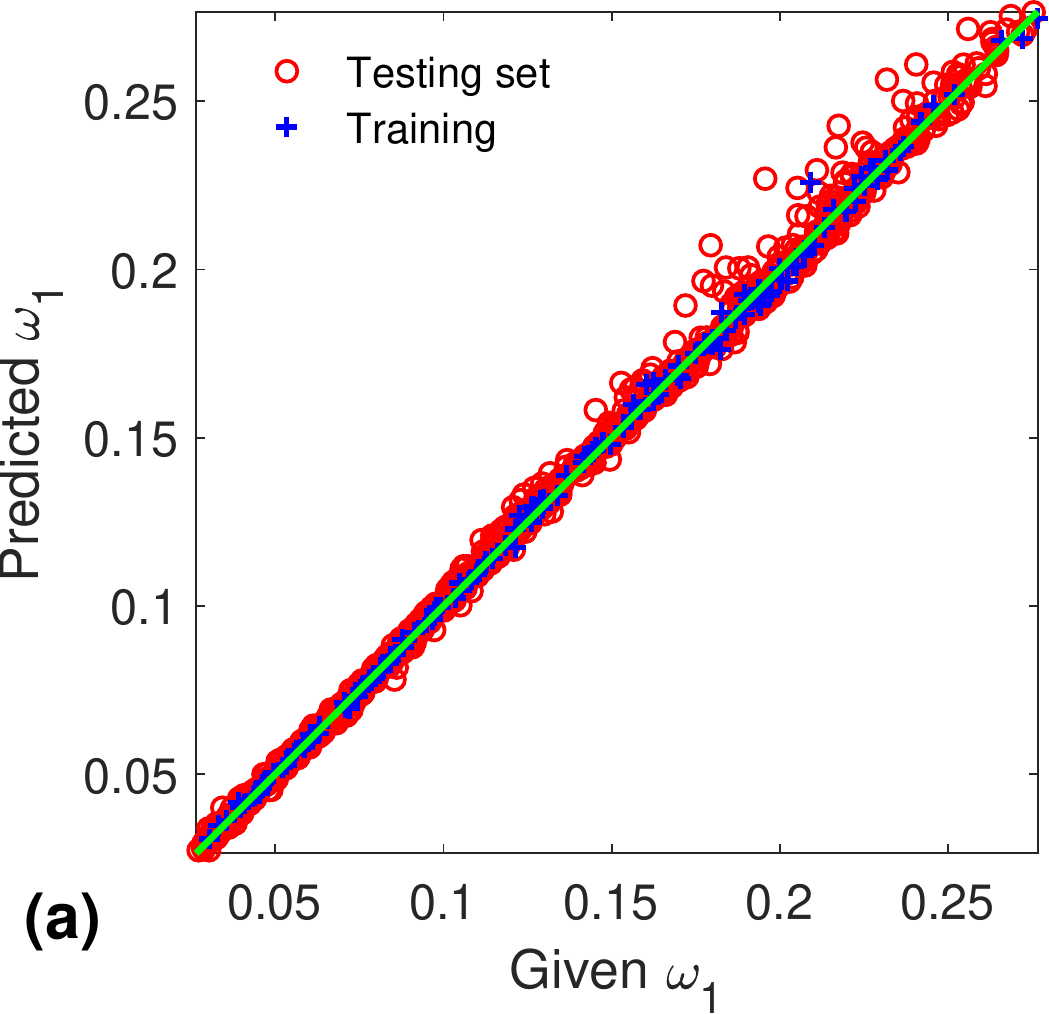}
\end{minipage}
\hspace{0.1cm}
\begin{minipage}[b]{0.48\linewidth}
\centering
\includegraphics[scale=0.48]{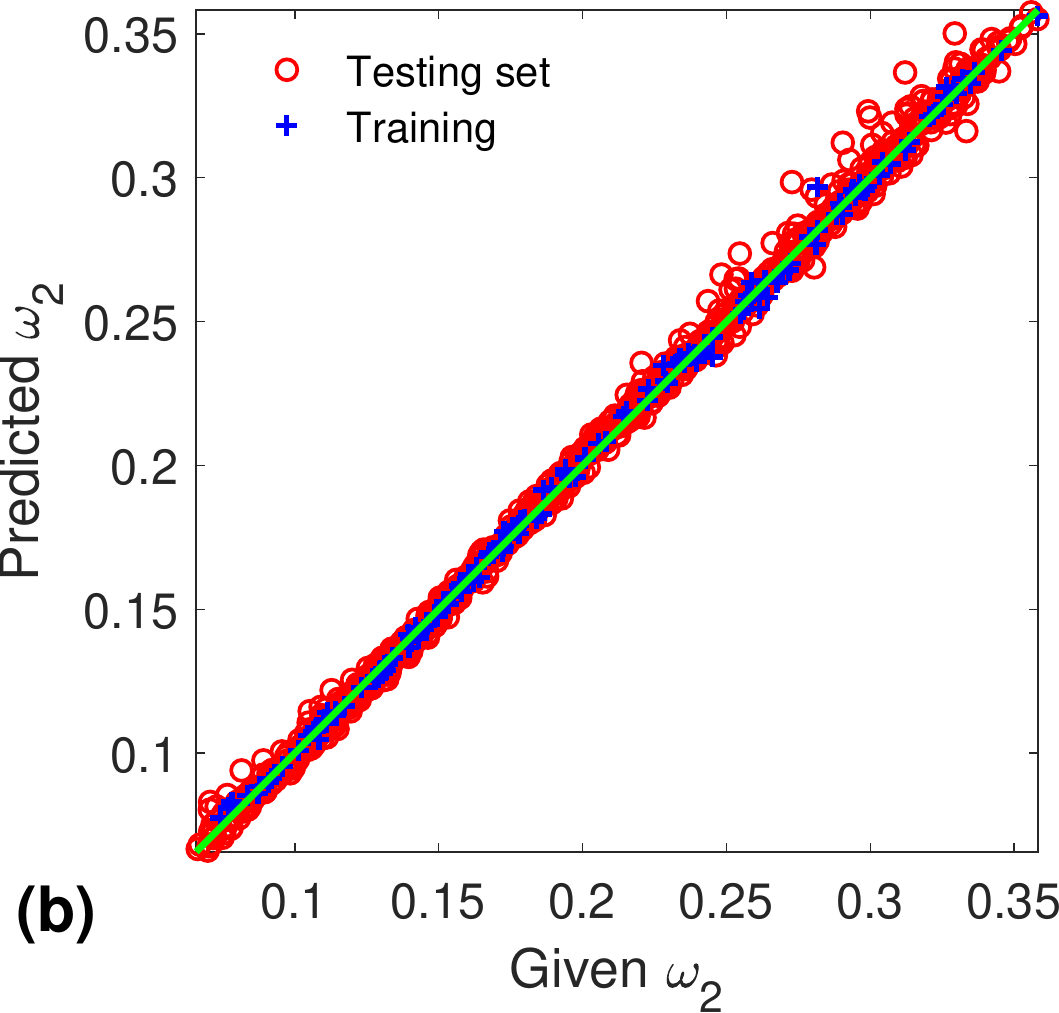}
\end{minipage}
\\
\begin{minipage}[b]{0.48\linewidth}
\centering
\includegraphics[scale=0.48]{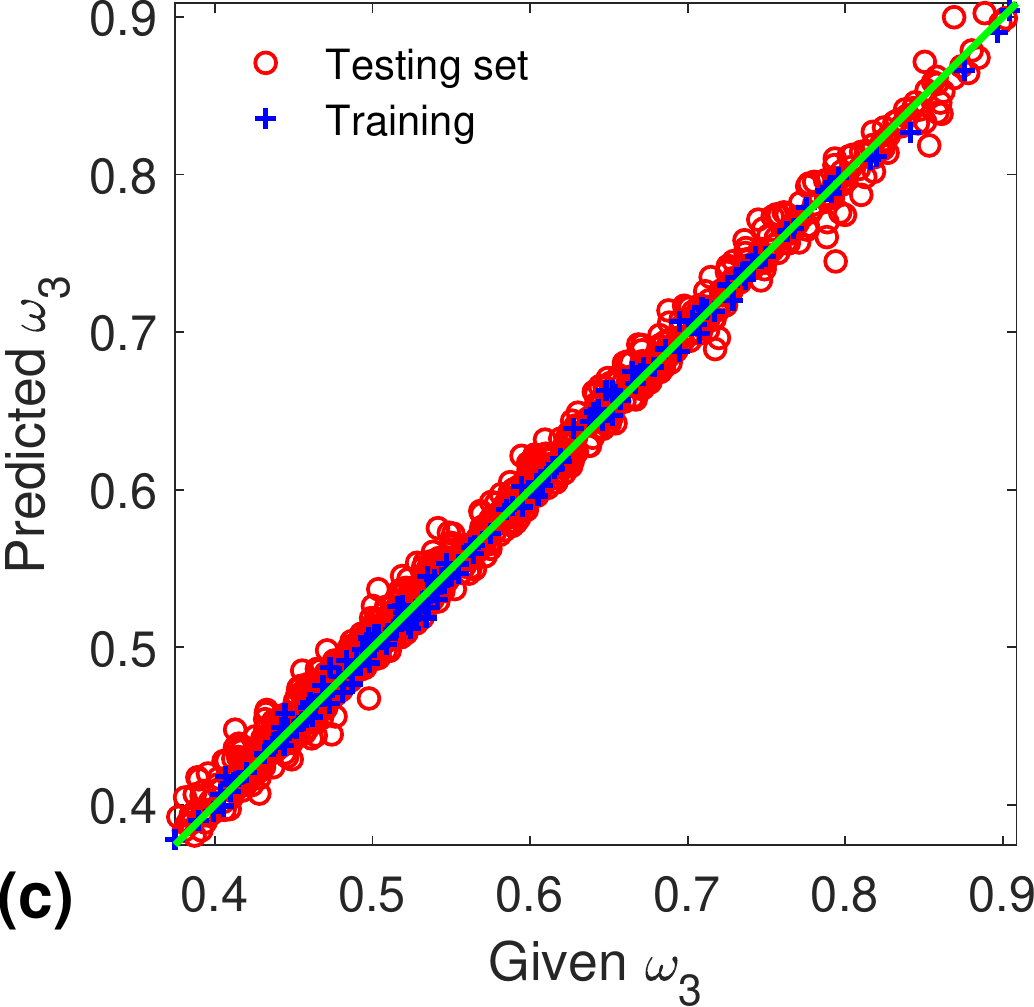}
\end{minipage}
\hspace{0.1cm}
\begin{minipage}[b]{0.48\linewidth}
\centering
\includegraphics[scale=0.48]{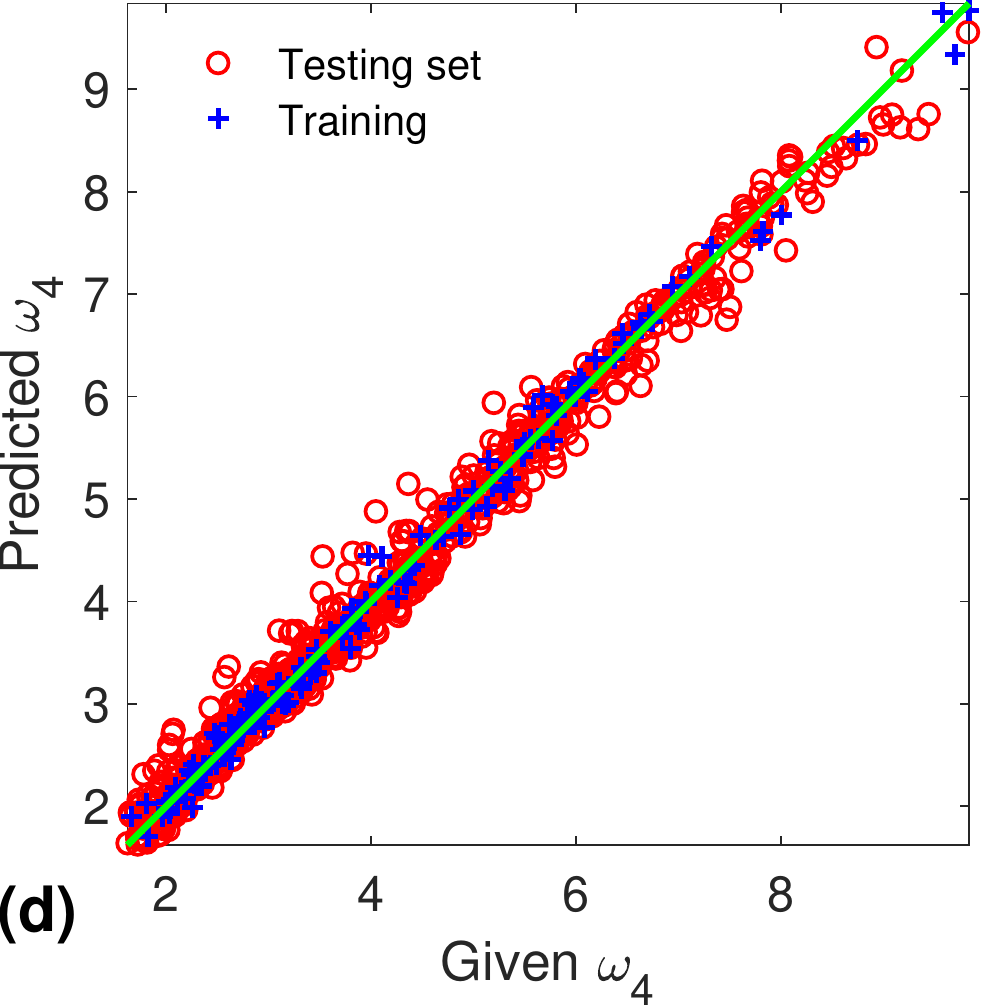}
\end{minipage}
\caption{Training accuracy and predictability of the low-fidelity models for the time-independent fitting coefficients including (a) $w_{1}$, (b) $w_{2}$, (c) $w_{3}$, (d) $w_{4}$. Similar behavior was obtained for $w_{5}$; figure is not shown.}
\label{figure11}
\end{figure}

\begin{figure}[h!]
\centering
\begin{minipage}[b]{0.45\linewidth}
\centering
\includegraphics[scale=0.35]{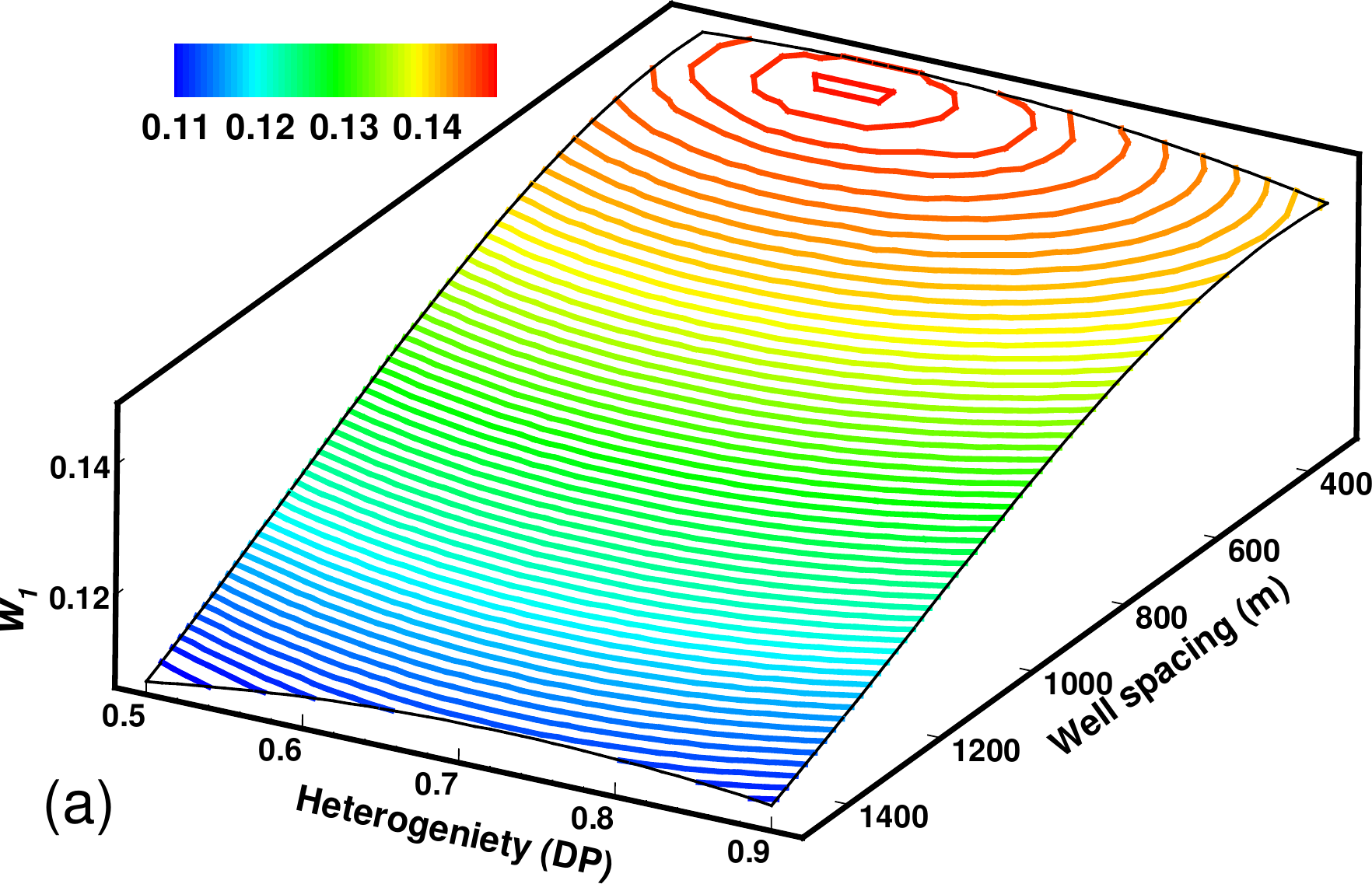}
\end{minipage}
\hspace{0.1cm}
\begin{minipage}[b]{0.45\linewidth}
\centering
\includegraphics[scale=0.35]{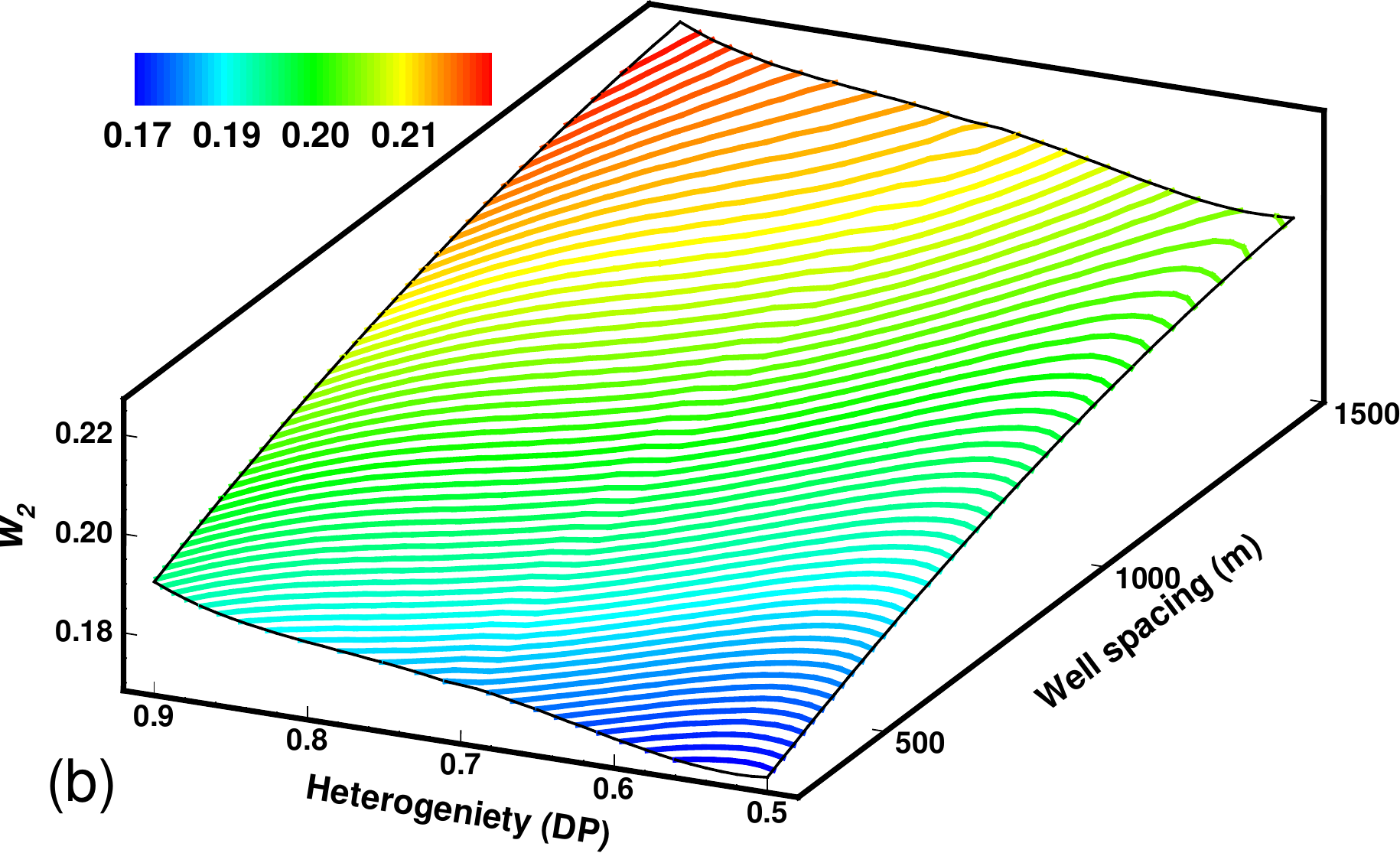}
\end{minipage}
\\~\\
\begin{minipage}[b]{0.45\linewidth}
\centering
\includegraphics[scale=0.33]{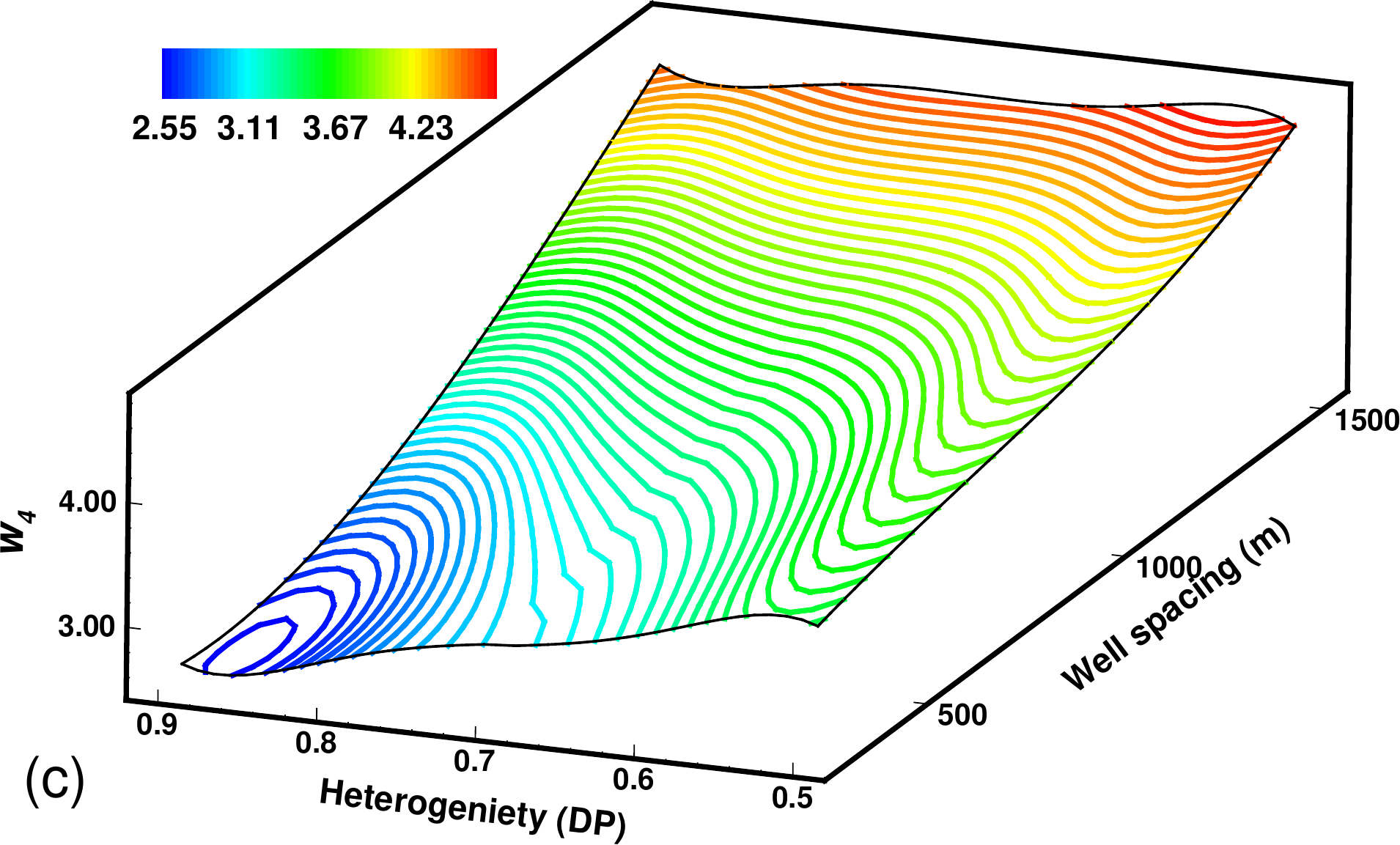}
\end{minipage}
\hspace{0.1cm}
\begin{minipage}[b]{0.45\linewidth}
\centering
\includegraphics[scale=0.33]{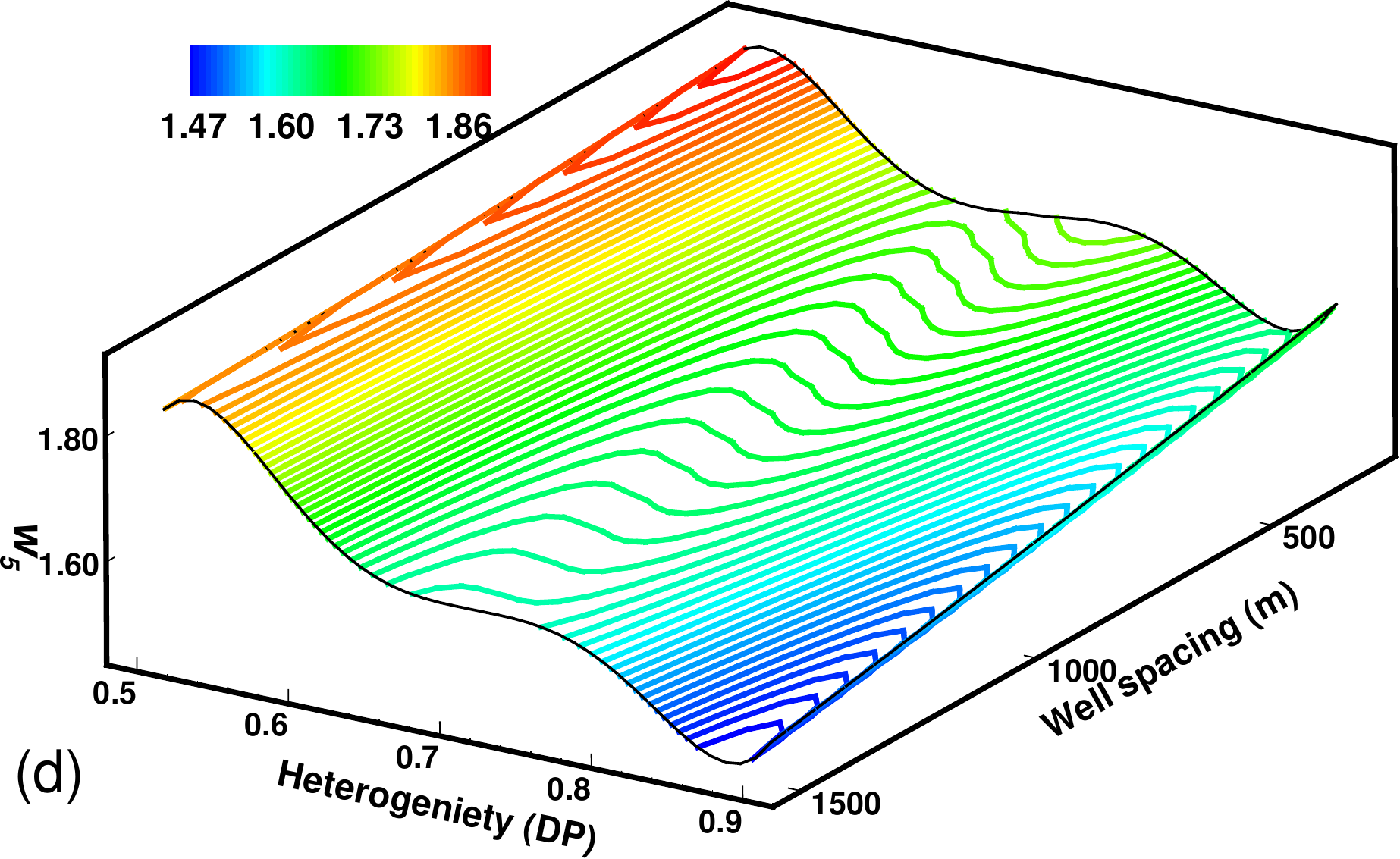} 
\end{minipage}
\caption{Response surfaces of the time-independent  $w_i$ coefficients as functions of the uncertain parameters, where several examples are shown including (a)$w_{1}$, (b)  $w_{2}$, (c)  $w_{4}$, and (d)  $w_{5}$,  versus well spacing $L$ and heterogeneity $V_{DP}$.}
\label{figure15}
\end{figure}

\begin{figure}[h!]
\centering
\begin{minipage}[b]{0.45\linewidth}
\centering
\includegraphics[scale=0.45]{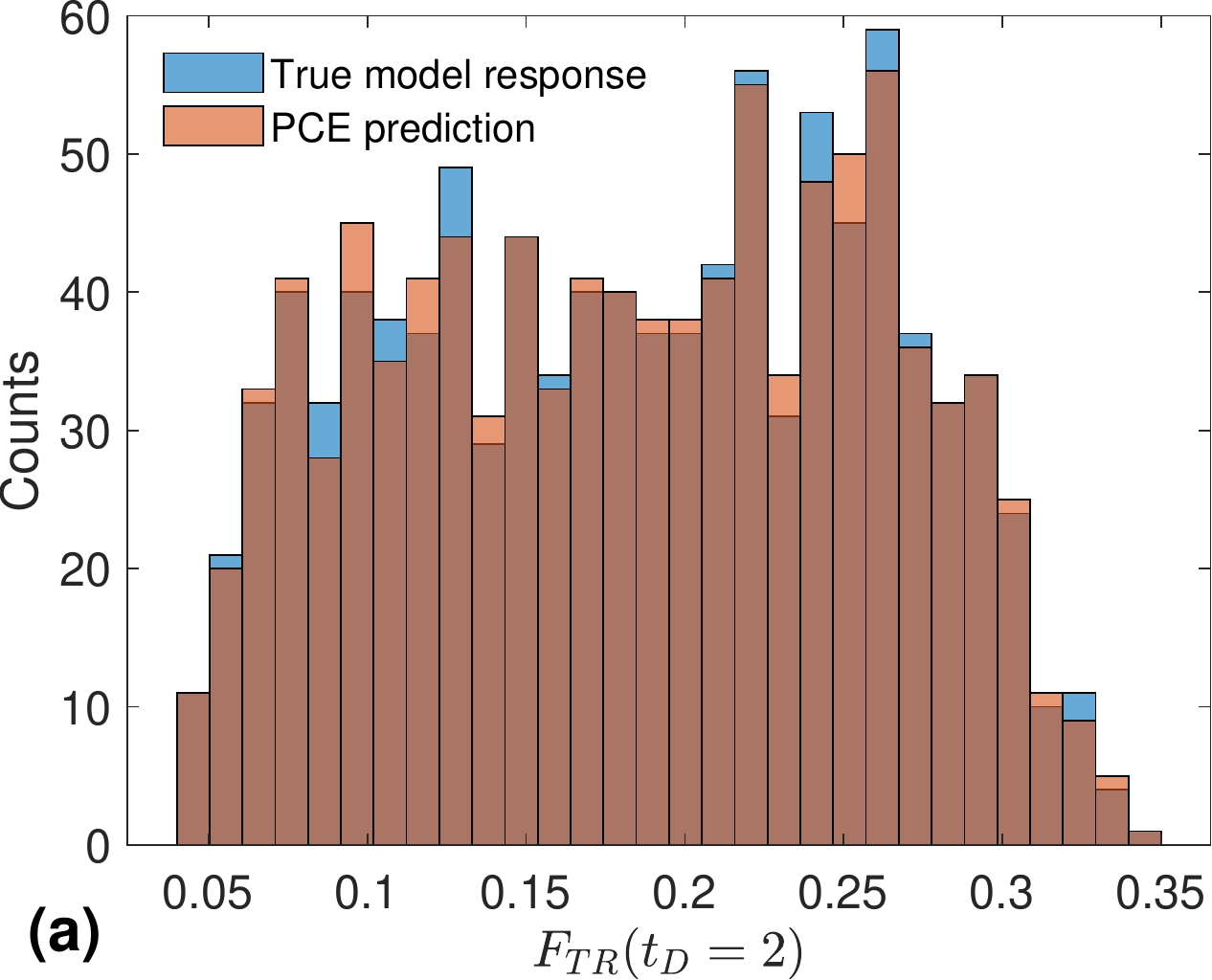}
\end{minipage}
\hspace{0.5cm}
\begin{minipage}[b]{0.45\linewidth}
\centering
\includegraphics[scale=0.45]{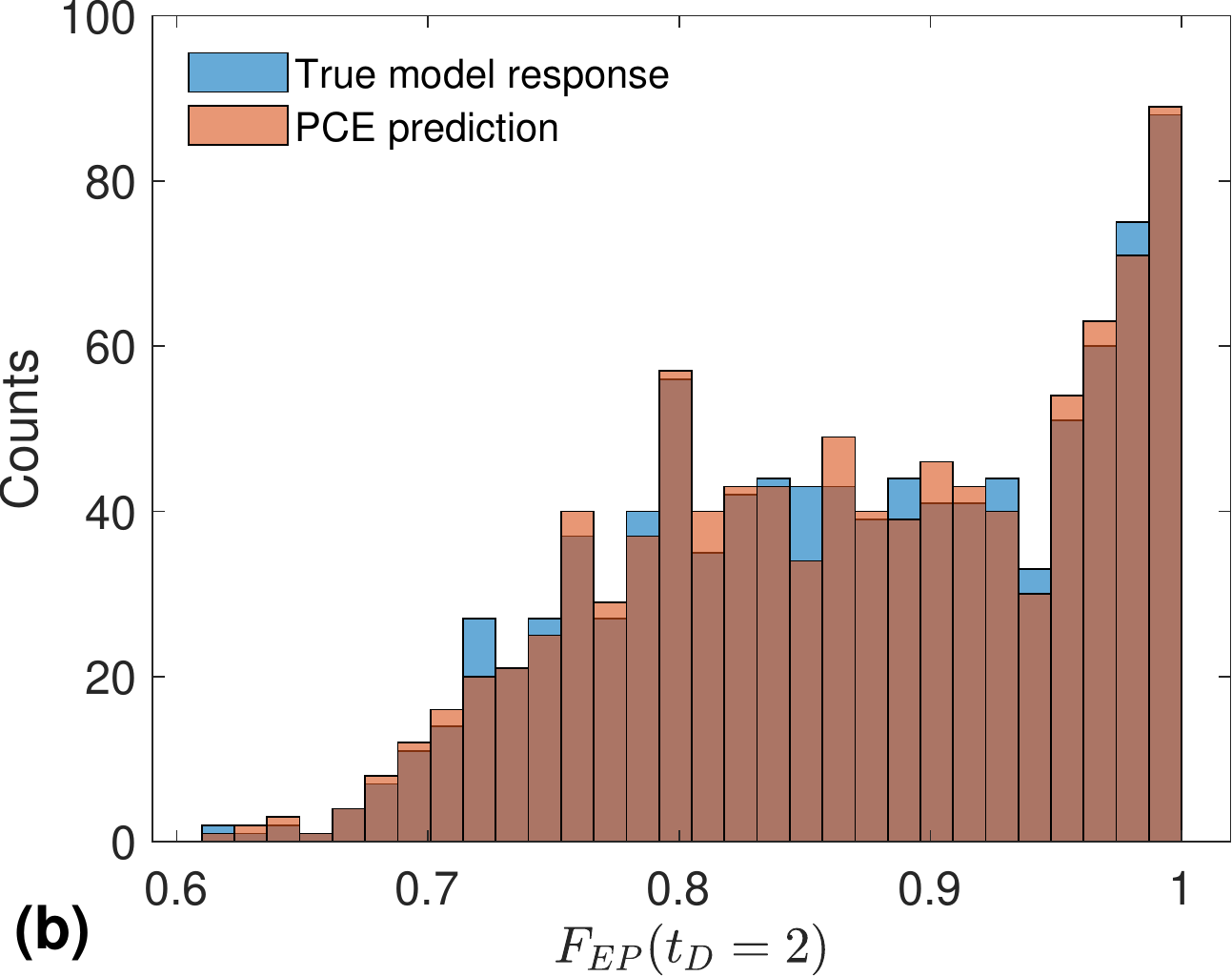}
\end{minipage}
\caption{Prediction accuracy  of time-dependent objective response (PCE prediction) compared to the blind-testing data including 1000 cases (True model) at $t_D=2$  corresponding to  (a) thermal recovery factor ${{F}_{TR}}$, and (b) enthalpy production factor ${{F}_{EP}}$.}
\label{figure12}
\end{figure}

\begin{figure}[h!]
\centering
\begin{minipage}[b]{0.49\linewidth}
\centering
\includegraphics[scale=0.32]{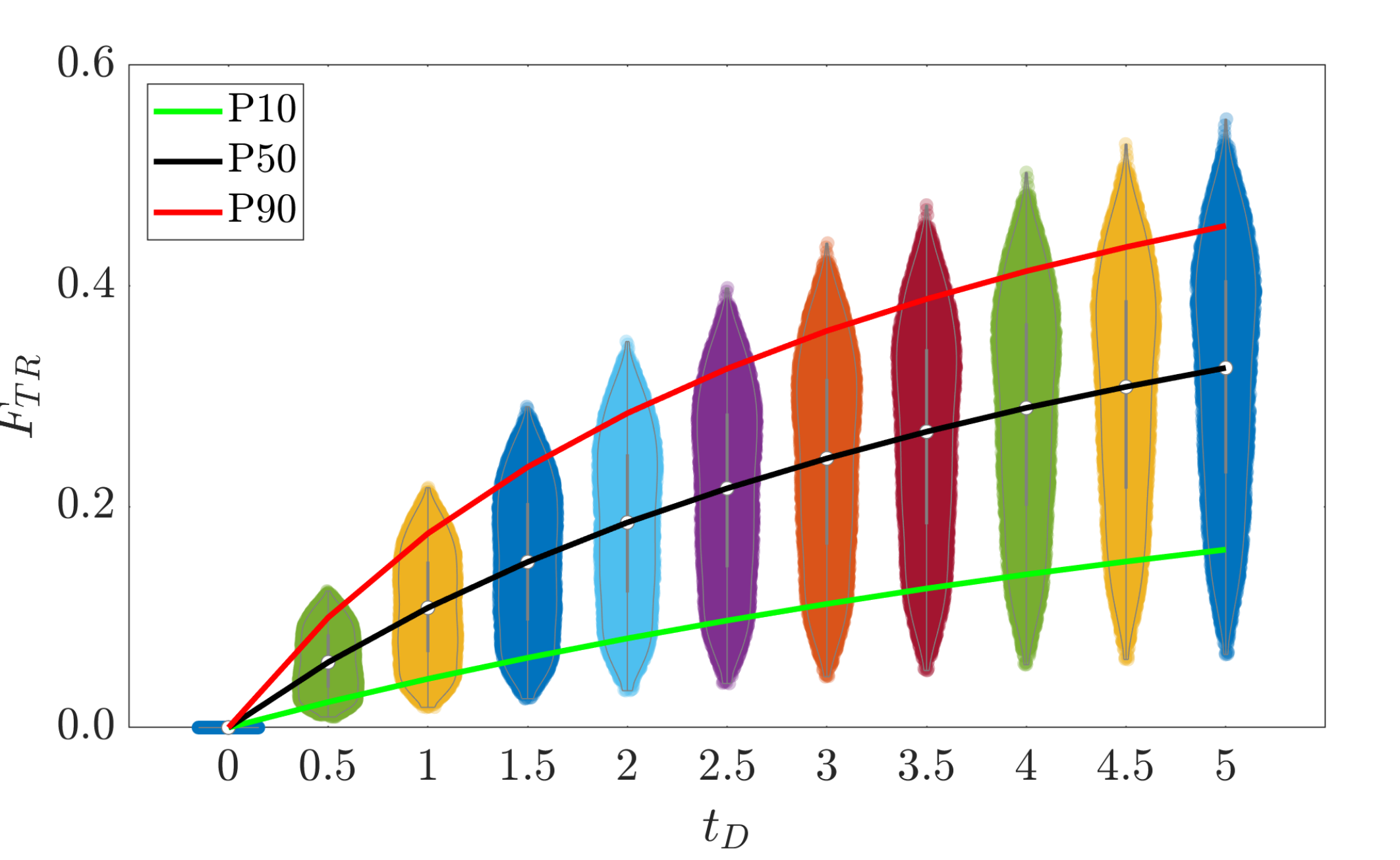}
\end{minipage}
\begin{minipage}[b]{0.49\linewidth}
\centering
\includegraphics[scale=0.32]{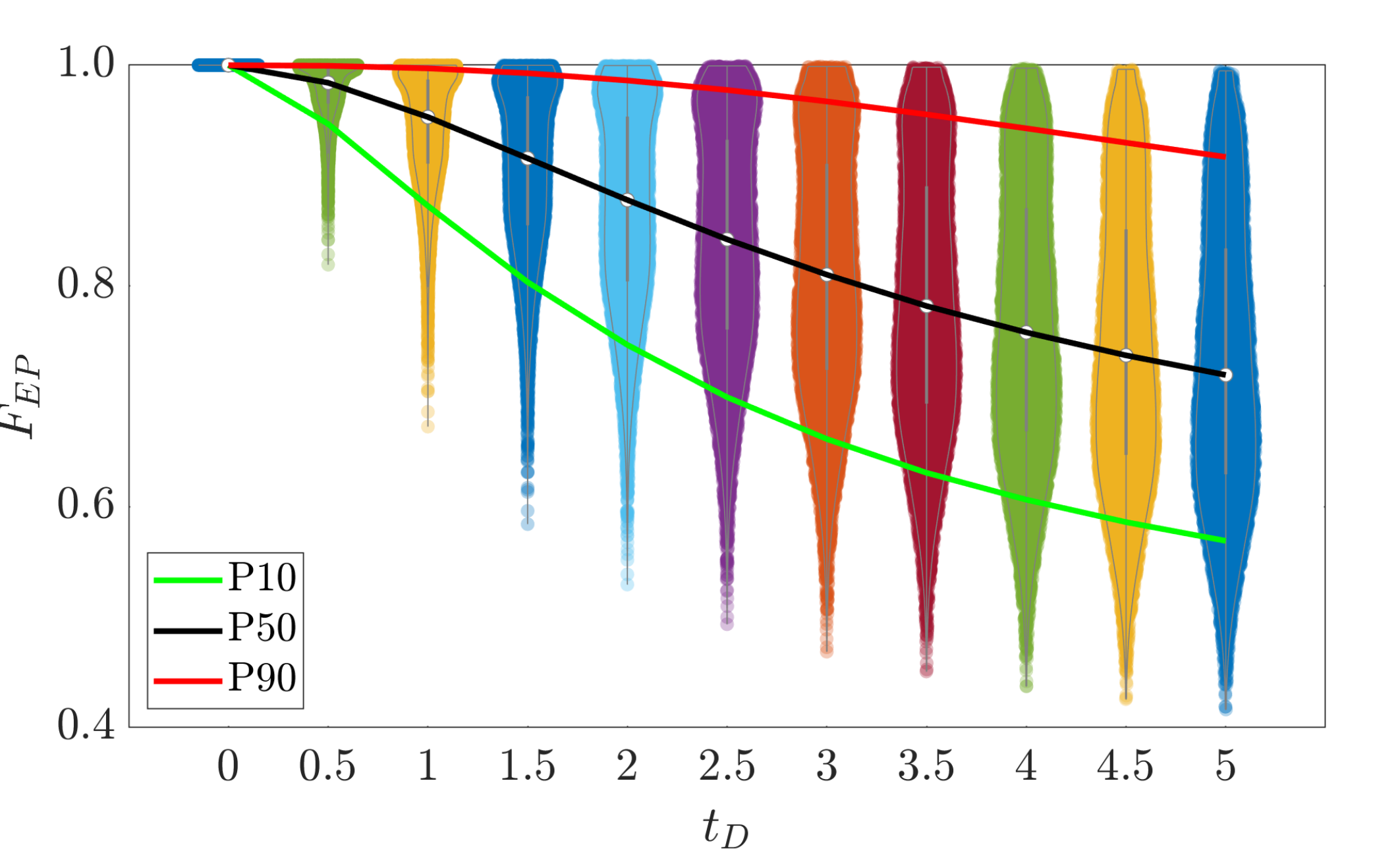}
\end{minipage}
\caption{Uncertainty propagation results for the time-dependent objective function $F_{TR}$ (left) and $F_{EP}$ (right) using Monte Carlo simulations applied to the low-fidelity models.}
\label{figure13}
\end{figure}

\begin{figure}
\begin{minipage}[b]{0.45\linewidth}
\centering
\includegraphics[scale=0.45]{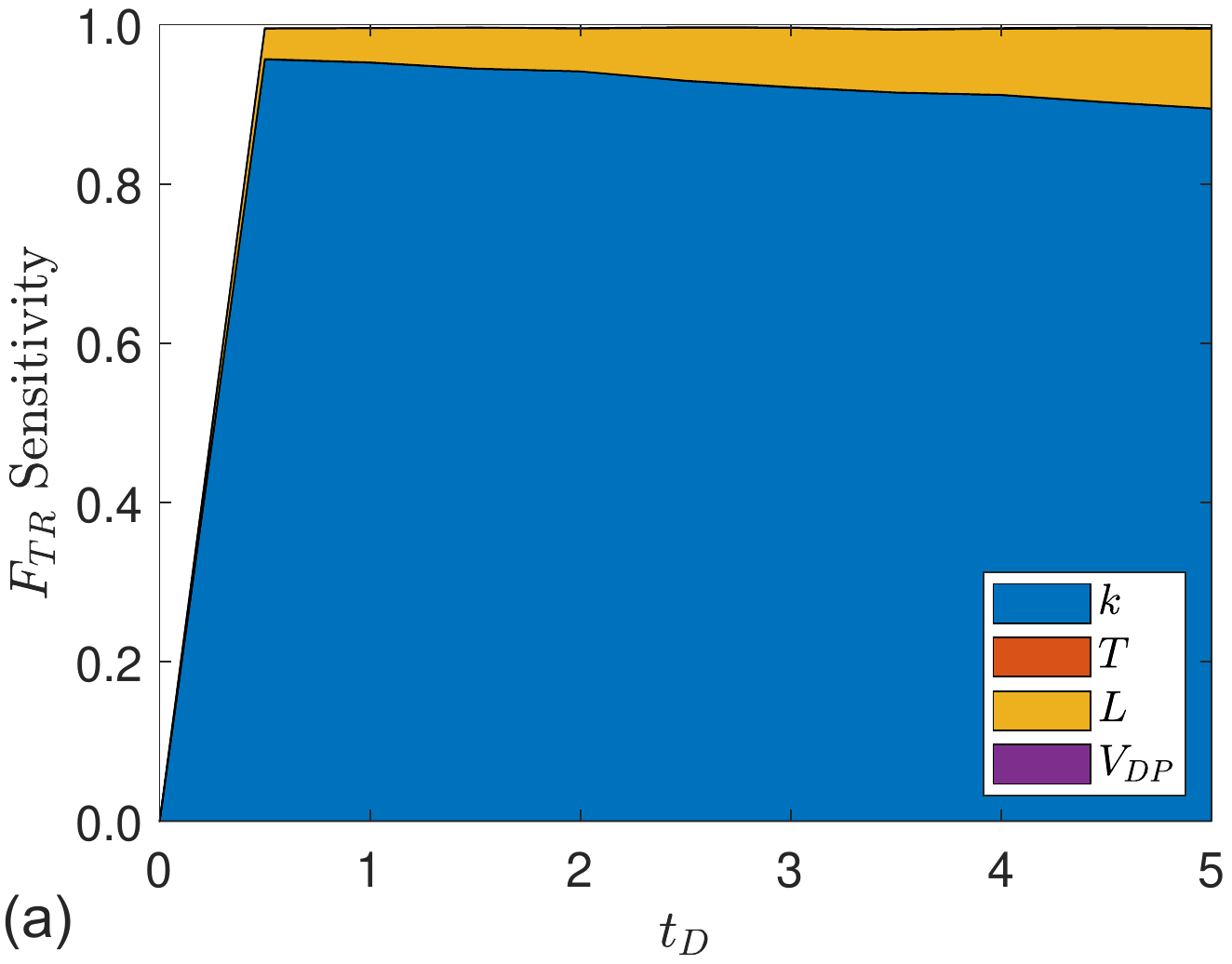}
\end{minipage}
\hspace{0.5cm}
\begin{minipage}[b]{0.45\linewidth}
\centering
\includegraphics[scale=0.45]{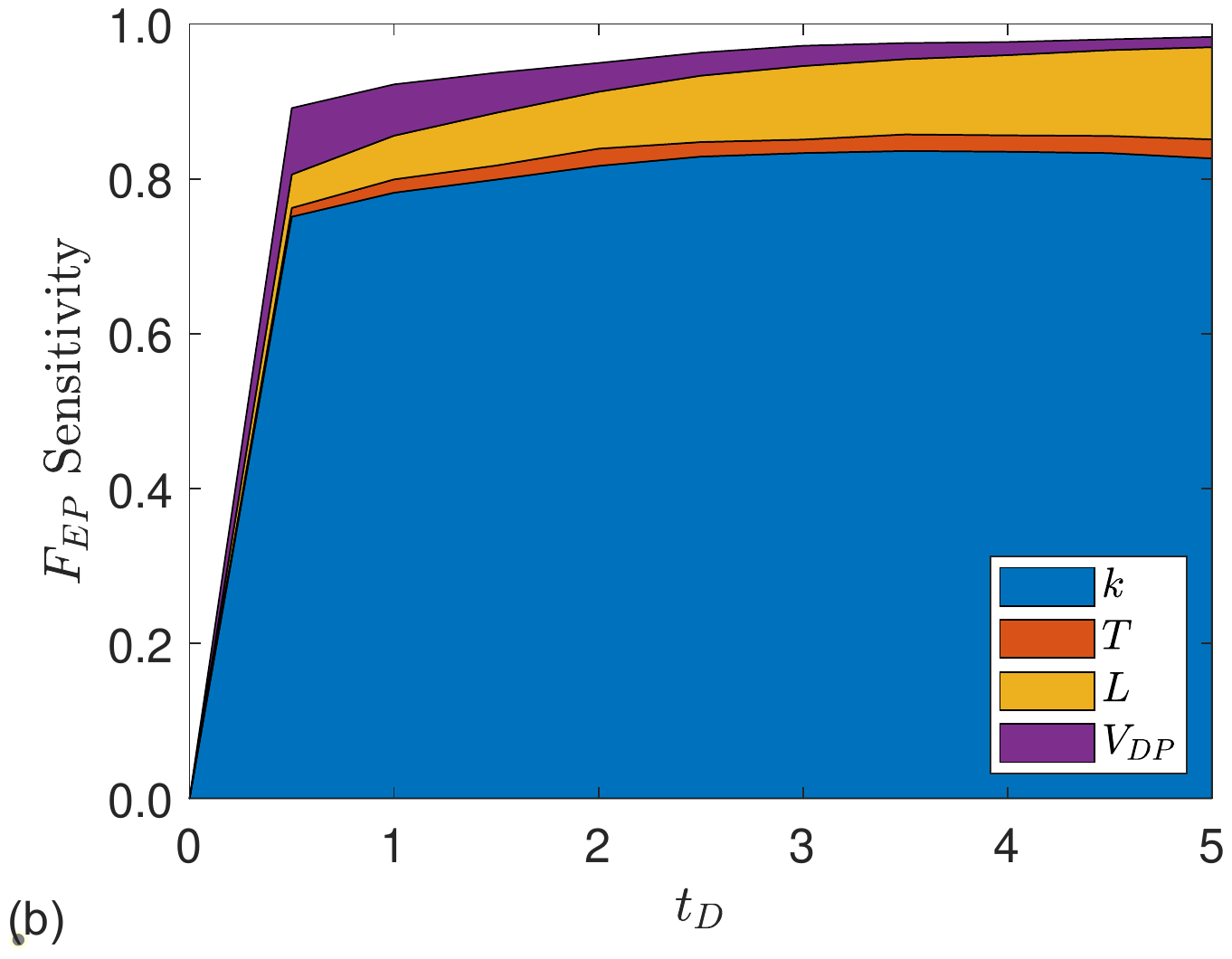}
\end{minipage}
\caption{First-order ANCOVA sensitivity indices of the low-fidelity models: $F_{TR}$ (a) and $F_{EP}$ (b).}
\label{figure14}
\end{figure}

The quantified  uncertainty and design parameters are the  permeability ($k$), temperature ($T$), well spacing ($L$), heterogeneity ($V_{DP}$), and thermal conductivity ($\sigma$). An efficient 2-level folded Plackett-Burman DoE is initially used to determine  the significant parameters (feature selection). Reservoir porosity and production rate are considered as  dependent parameters on  the reservoir permeability to avoid physical inconsistency in the sampling  \cite{Carman1997,Cardona2020}. 
The feature selection process, used to identify the significant parameters, evaluates the sensitivity of the objective functions, represented by the fitting  coefficients $\omega_{i,i=1,...,5}$. The design requires 13 simulation cases  including the endpoints and the center point of the variation intervals of the permeability $k$, temperature $T$, well spacing $L$, heterogeneity $V_{DP}$, and thermal conductivity $\sigma$. Some of the simulation cases from the design and the corresponding fitted low-fidelity model are shown in Figure \ref{figure10}. Figure \ref{figure9} shows that the p-values  for $k$, $T$, $L$, and $V_{DP}$ are above 0.05, while  the one for $\sigma$ is below 0.05, indicating insignificance of its variation on the objective functions. Therefore, $\sigma$ was not included as an uncertainty parameter in the construction of the low-fidelity models for $F_{TR}$ and $F_{EP}$.  Instead, the value of $\sigma$ is fixed at its P50. 
\\
\\
After eliminating the insignificant parameters, a Latin Hypercube design is used to construct the responses of the time-independent coefficients $w_i$ using 4 groups of parameters. These coefficients are then used to assemble the objective functions $F_{TR}$ and $F_{EP}$.  Convergence analysis was conducted to select the order of  approximation of the polynomial chaos expansions (PCE), as well as the size of the design.  Increasing the order of PCE can improve the accuracy of the fitting of the training sample. However, it may cause over-fitting, leading to poor model predictability.   Figure \ref{figuren1}a shows that as the order of PCE increasing, the accuracy of reproducing the training set increases, while  predictivity  may decrease. The predictability of the model, measured by the blind-test samples including 1000 cases, shows that a PCE of order 3 is optimum, and higher-order leads to overfitting.  On the other hand, the size of the training sample in the Latin Hypercube design should be high enough to guarantee  the desired accuracy in the blind test. Figure \ref{figuren1}b shows no significant improvement in the accuracy of the design with more than 150 cases, which was the selected size. Parity plots showing the accuracy of the proxy model in reproducing the training set and in predicting the blind testing set are shown in Figure \ref{figure11}.  The performance  of other designs, including Full Factorial, Box Behken, Taguchi, Central Composite, and D-Optimal are discussed in Appendix A. 
To avoid  over-fitting from the training set, emphasis should be given to predictability using blind testing. The obtained training accuracy ($R^2$) for the low-fidelity models for the 150 cases is above 0.99, while the predictability accuracy ($Q^2$) for 1000 blind cases is above 0.92.  The low-fidelity models provide the PCE-based proxy to estimate the time-independent  $w_i$ as a function of the well spacing $L$, permeability $k$,  heterogeneity $V_{DP}$, and temperature $T$. Figure \ref{figure15} shows examples of the behavior of four coefficients, $w_{1}$,  $w_{2}$,  $w_{4}$, and $w_{5}$, as a function of $L$ and $V_{DP}$. The approximated fitting coefficients are then used in Eqs. \ref{equation19} and \ref{equation20} to calculate the time-dependent objective functions, $F_{TR}$ and $F_{EP}$. The quality of the predictions of $F_{TR}$ and $F_{EP}$  compared to the physics-based simulation results at $t_{D}=2$ are shown in Figure \ref{figure12}. The predictability of the model other times showed similar behavior.
\\
The pdf-s for $F_{TR}$ and $F_{EP}$ are shown versus time in Figure \ref{figure13}. The predictions are generated by performing Monte Carlo simulations using the low-fidelity models with 10,000 samples. The mean prediction of $F_{TR}$ has the square-root-of-time behavior, similar to the trend in diffusion problem \cite{Patzek2013}. At $t_{D}=1$, the percentiles P10, P50, and P90 of $F_{TR}$  are 0.04, 0.11, and 0.18, respectively, and the corresponding $F_{EP}$'s are 0.87, 0.95, and 0.997.  
The ANCOVA sensitivity result (Figure \ref{figure14}) shows that permeability and well spacing influence both the thermal recovery factor and enthalpy production factor at all times. For the thermal recovery factor, both permeability and well spacing stay dominant until the end of the process. There is a slight change in sensitivity values after the breakthrough time, but there is no change in the ranking.  For the enthalpy production factor, permeability stays dominant at all times. However,  there is a change in parameter ranking after the breakthrough time. Dykstra-Parson coefficient and well spacing switch places after fluid breakthrough, which indicates a change in recovery mechanisms after the breakthrough time from “heterogeneity-influenced” to “well-spacing-influenced”. 
We show the  response surfaces of the $w_i$ coefficients  versus several uncertainty parameters in Figure \ref{figure15}. The surfaces reflect the  variability of the coefficients with respect to the uncertainty parameters which show quasi-linear to non-linear behaviors. To capture the significant uncertainty variables,  we emphasize the use of ANCOVA in conducting the sensitivity analysis.
\begin{figure}[h!]
\begin{minipage}[b]{0.5\linewidth}
\centering
\includegraphics[scale=0.35]{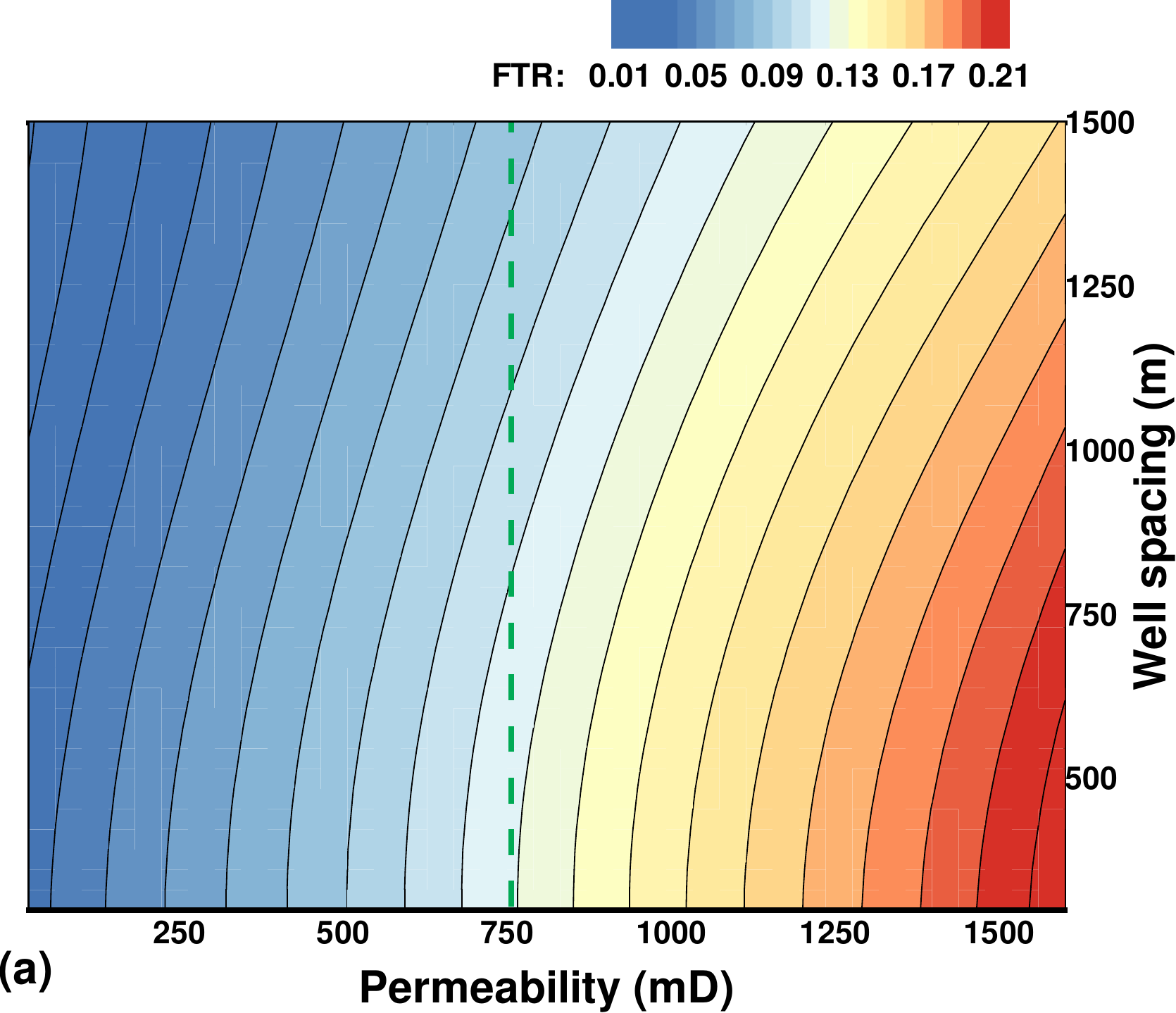}
\end{minipage}
\begin{minipage}[b]{0.5\linewidth}
\centering
\includegraphics[scale=0.35]{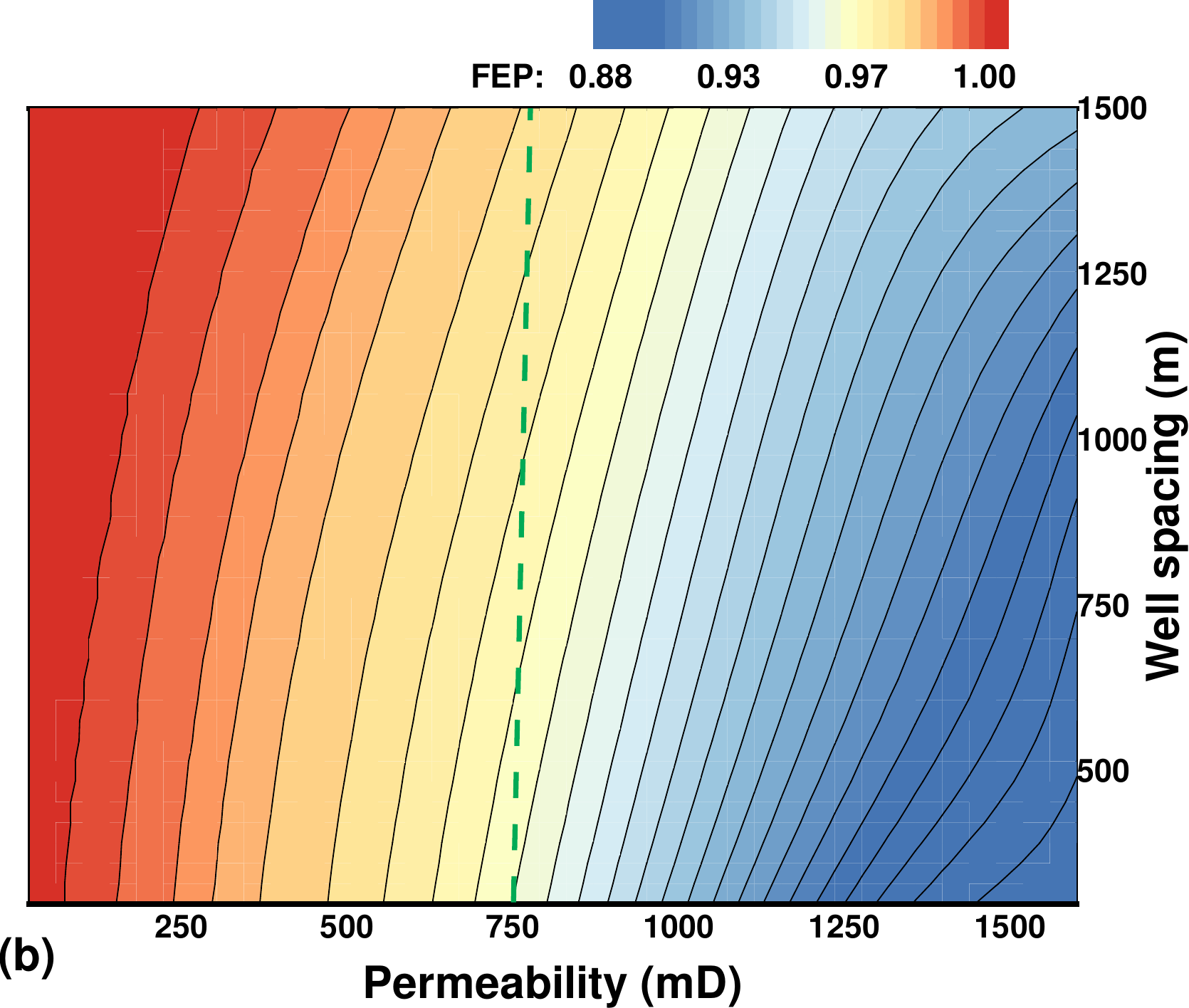}
\end{minipage}
\\~\\
\begin{minipage}[b]{0.5\linewidth}
\centering
\includegraphics[scale=0.35]{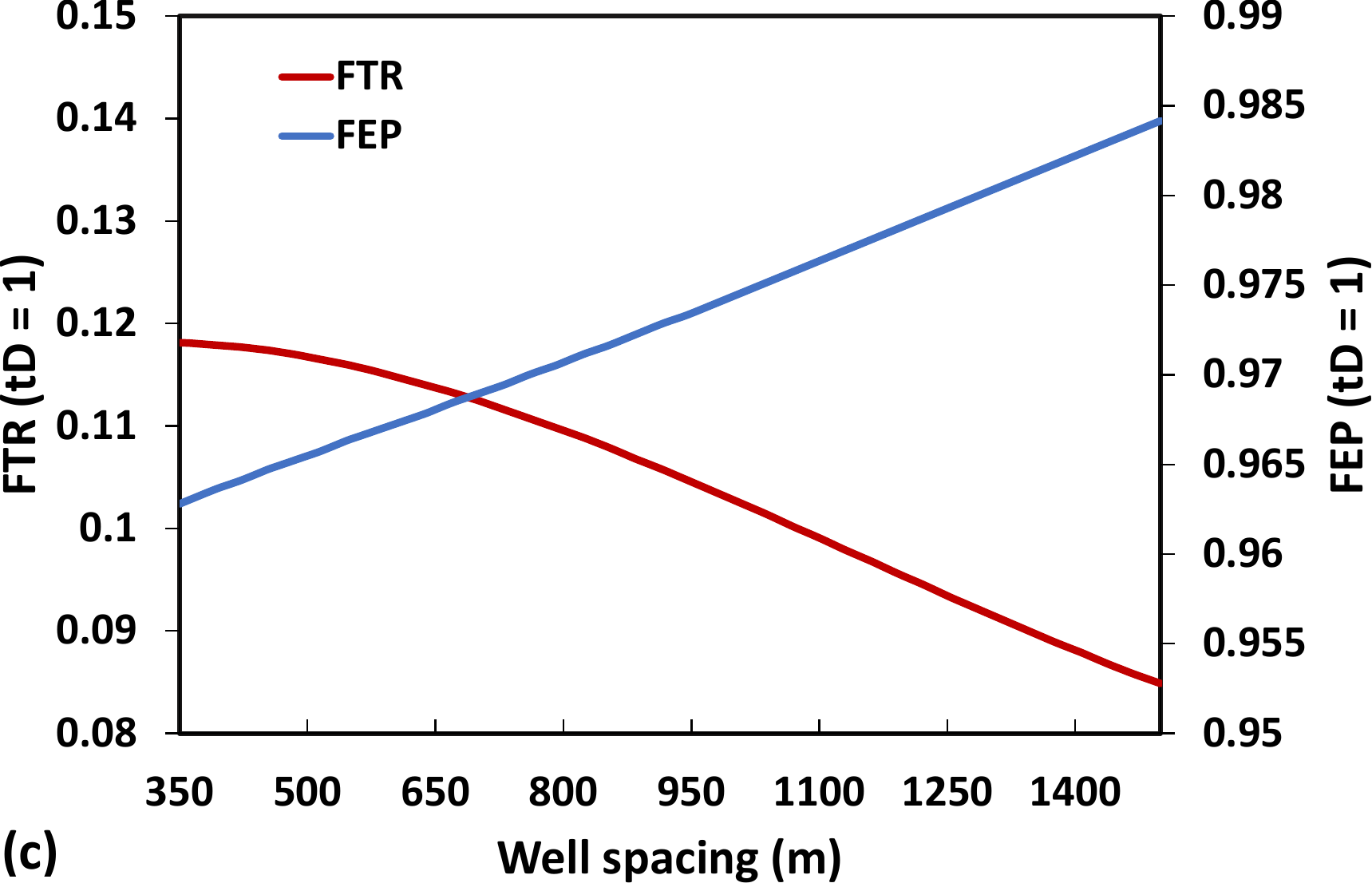}
\end{minipage}
\begin{minipage}[b]{0.5\linewidth}
\centering
\includegraphics[scale=0.35]{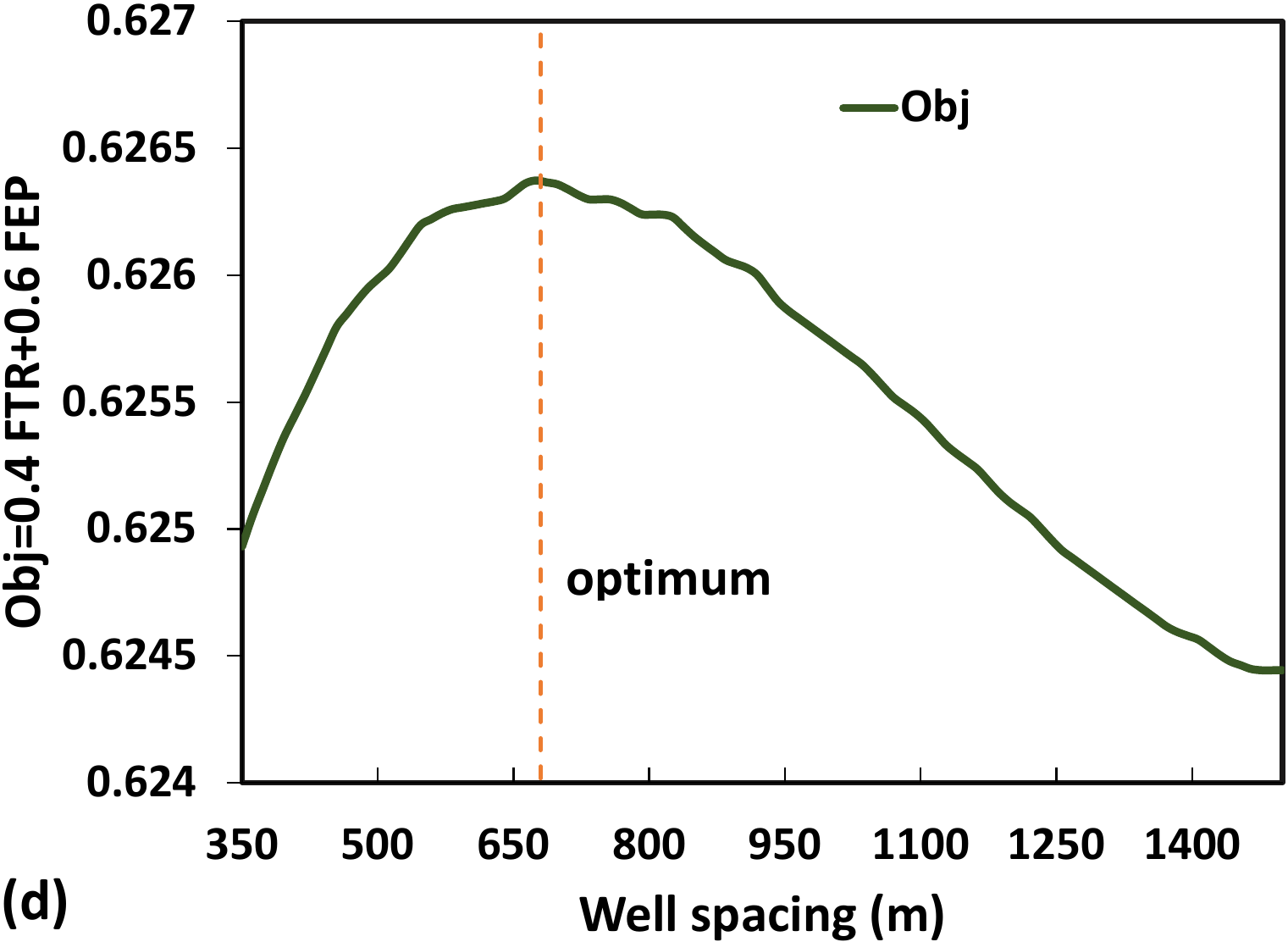}
\end{minipage}
\caption{Distribution of  $F_{TR}$ (a) and $F_{EP}$ (b) at $t_D =1$ as a function for reservoir permeability  and well spacing showing opposite trends. Extractions of $F_{TR}$ and $F_{EP}$ trends versus well spacing at permeability $k=750mD$ is shown in (c). A multi-objective optimization given in Eq. \ref{eq:obj} shows an optimum  well spacing at $L=655m$ (d).}
\label{figure16}
\end{figure} 
We demonstrate the use of our workflow to analyze the optimum injector-producer spacing ($L$) for cases with different  reservoir permeabilities ($k$), as shown in Figure \ref{figure16}. The parameters $k$ and $L$ are selected based on the ANCOVA result. Figure \ref{figure16}a and b show the distributions of $F_{TR}(t_{D}=1)$ and $F_{EP}(t_{D}=1)$ in terms of the variations in $k$ and $L$. Since the optimization should be conducted on both functions, $F_{TR}(t_{D}=1)$ and $F_{EP}(t_{D}=1)$,  simultaneously, the optimum well spacing  should maximize both $F_{TR}$ and  $F_{EP}$, which can be captured by maximizing an objective function $Obj$, such that:
\begin{equation}\label{eq:obj}
 Obj= \text{w}_1 F_{TR} + \text{w}_2  F_{EP},
\end{equation}
where  $\text{w}_1 $ and $\text{w}_2$ are weighting factors relative to the field conditions. For instance,  selecting   $\text{w}_1 =0.4$ and $\text{w}_2=0.6$ for a reservoir case with $k = 750 mD$ ($V_{DP} = 0.55$), the corresponding optimum spacing is $L=655 m$ (see Fig.\ref{figure16}c and d). The optimization problem becomes multidimensional when including other variables, such as reservoir heterogeneity ($V_{DP}$).  Figure \ref{figure18_3d}, shows the distribution of the  objective function in 3D versus the uncertainty variable $k$, and $V_{DP}$, and the design variable $L$. The selected objective function is case dependent, which reflects the desired maximization of  $F_{TR}$ and  $F_{EP}$.  In this case, the optimum well spacing corresponds to maximum value "Obj" as indicated in Fig. \ref{figure18_3d}.
\begin{figure}[h!]
\centering
\includegraphics[scale=0.5]{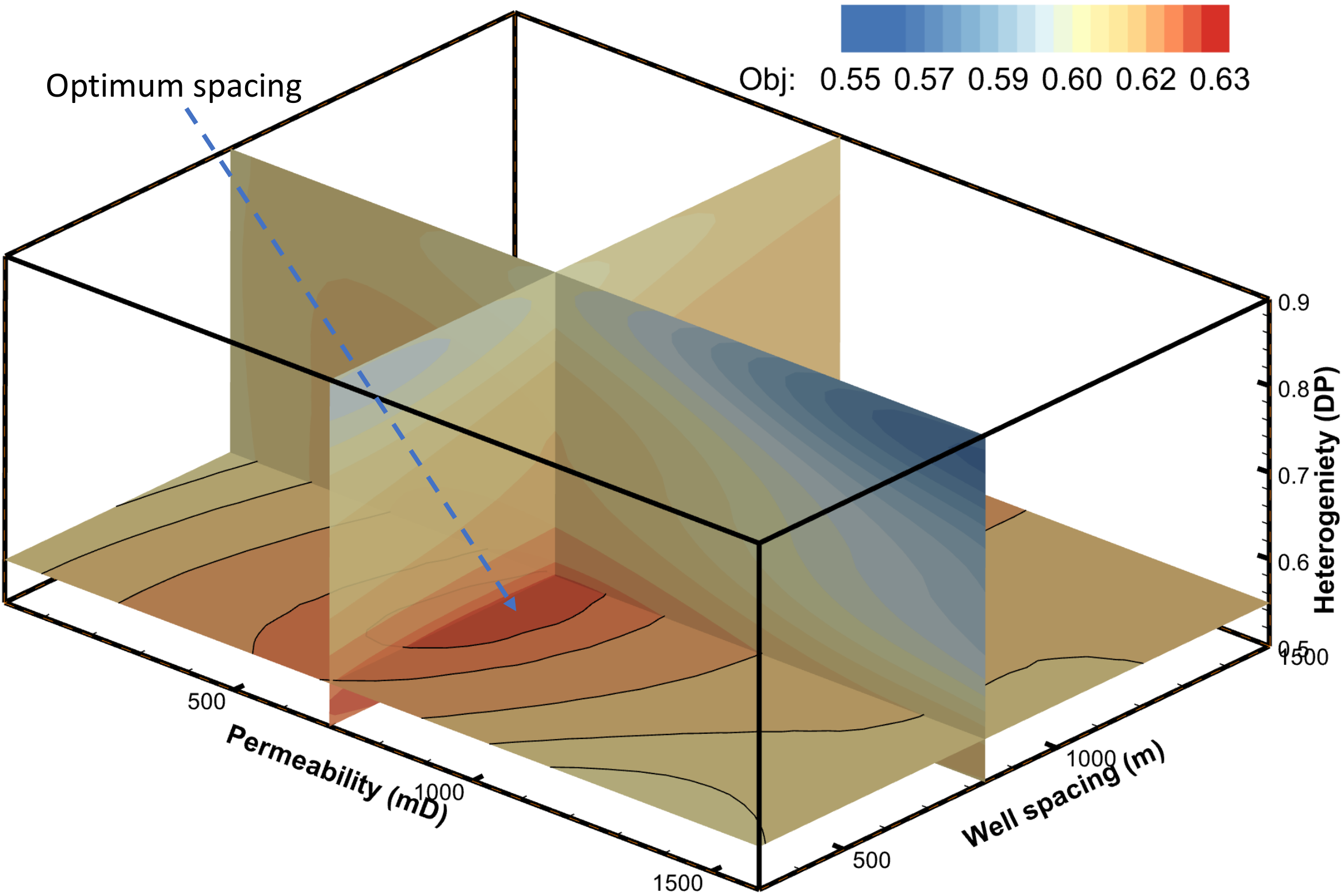}
\caption{A 3D heat map showing an objective function (Eq. \ref{eq:obj} ) versus reservoir permeability, heterogeneity, and well spacing. The indicated zone corresponds to the optimum well spacing corresponding to a specific reservoir's permeability and heterogeneity condition.}  
\label{figure18_3d}
\end{figure} 
~\\
To demonstrate the  applicability of the optimization approach, we applied the model to various geothermal fields considering their temperature, permeability, and heterogeneity ranges \cite{Zarrouk2014,RIVERADIAZ2016105,KAMILA2021101970}.  The estimated optimum well-spacing  between injectors and producers is compared  to the actual average spacing  in the fields, as shown in Fig. \ref{Fig19_spacing}.  The predictions are in reasonable agreement with the actual data. However, several exceptions showed significant discrepancies, which could be related to various reasons. For instance, water re-injection in Dieng field (Java) has experienced mineral deposition at the injection wells, resulting in injectivity loss where the current injection to production volume ratio is about 68\% \cite{KAMILA2021101970}. The actual injection-production well spacing is around 1500m, and with a permeability of 0.1-30 md, temperature range 240-300$\,^\circ$C \cite{Hino2013}, the calculated  well spacing is around 400m. Momotombo field ( \cite{RIVERADIAZ2016105}), on the other hand, has full re-injection capacity to wastewater with a well spacing of 350m, resulting in enthalpy decline \cite{KAMILA2021101970,Kaspereit2016}. With a permeability range of 1-600md and  temperature 180-200 $\,^\circ$C  (\cite{Porras2005}), the corresponding estimated optimum spacing is around 500m, indicating a potentially too tight well spacing in the current well development configuration.

\begin{figure}[h!]
\centering
\includegraphics[scale=0.6]{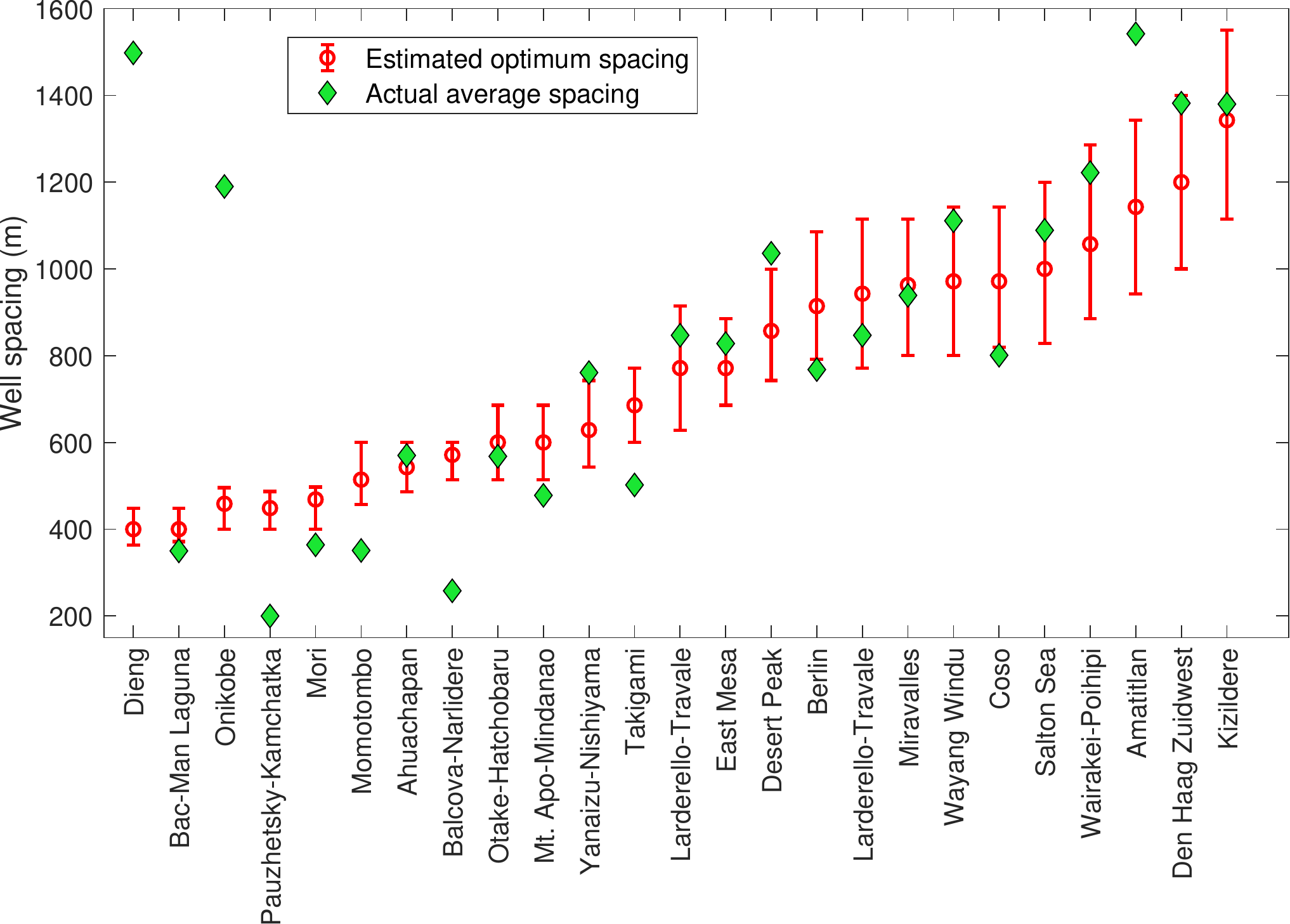}
\caption{Computed ranges of optimum well spacings for various geothermal fields and compared the to actual  average spacing in the fields .}  
\label{Fig19_spacing}
\end{figure} 

\section{Discussion}
The proposed simulation approach utilizes reservoir conditions compiled from worldwide geothermal field data, which may serve as a quick analog to assess field production efficiency. The flow behavior in geothermal reservoir under re-injection can be described by two dimensionless number: Peclet and Rayleigh \cite{Prasad1988,Holzbecher1998,Saar2011,Nield2013}. The Peclet number describes the significance of convective versus conduction heat transfer, while the Rayleigh number reflects the ratio between buoyant-driven and viscous-driven flow  \cite{Horne1974}.\\
The feature selection process   using folded Plackett Burman shows the insignificance of thermal conductivity, as an uncertainty parameter. The  Peclet number within our realizations is in the range of 5 - 110. The reported Peclet number for tracer injection in geothermal reservoirs is 3 - 80 \cite{Lecain2000}. It denotes that in most geothermal re-injection process, the dominant heat transfer process is through convection.
The dimensionless time term reflects the injected/produced volume relative to the reservoir’s pore volume. Therefore, the practical ranges for injection are around  $t_{D}=1$. The obtained thermal recovery factor at  $t_{D}=1$ is within 0.05 and 0.18, with the most likely case at 0.11, which is in agreement with the average worldwide thermal recovery of 0.13 - 0.14 \cite{Zarrouk2014}. \\
The ANCOVA enables to identify the parameters that govern the physical phenomenon at different times, where the permeability, rate, and well spacing are identified as the most significant parameters over time. This result is supported by the obtained Rayleigh number in our realizations within the range of 1 - 440. The reported Rayleigh number for geothermal reservoirs is less than 500 \cite{Donaldson1962,Donaldson1970,Horne1974,Sorey1978}. Furthermore, in a buoyant-dominated setting, permeability, rate, and well spacing are to trigger the convection \cite{Lai1991}. Therefore, our results are in  agreement with  field observations, which indicate that flow in a geothermal reservoir during re-injection is gravity-dominated. 
The workflow provides a novel insight for constructing continuous space-time low-fidelity models for geothermal-related problems. It is useful to accelerate complex and expensive computations \cite{He2013,Tang2020} and perform efficient optimization and history matching \cite{Marzouk2007,Aliyev2017,JagalurMohan2018,Schulte2020}. However, this model does not capture the effect of detailed mechanisms such as injectivity loss from mineral precipitation or other field specifics such as the economic aspects, including the minimum operational production rates and enthalpy.
\section{Conclusions} 
In this work, we present a new workflow to develop time-continuous, multi-objectives uncertainty quantification of geothermal production under re-injection. We identify the parameters and their ranges using a  database reflecting multiple analogs from literature. The feature selection step is to select the significant parameters utilizing a simple design. The time-continuous response is constructed using a nested-function approach and optimized under multi-objectives setting. We also propose the use of ANCOVA indices to perform global sensitivity analysis with the Design of Experiment (DoE). 
Emphasis should be given to  assess the proxy predictability $Q^{2}$, and not to solely rely on training accuracy $R^{2}$ to avoid overfitting. Improved predictability is achieved using space-filling design with an optimum number of realizations.

The feature selection  results showed that thermal conductivity is insignificant to the  re-injection process. The heat transfer process is dominated by convection (driven by fluid flow) with a high value of Peclet number. The thermal recovery and enthalpy production efficiency are significantly influenced by permeability, rate, porosity, and well spacing. 
As the flow is gravity-dominated, the Rayleigh number is relatively high. Monte Carlo simulations show that the most likely recoverable heat under re-injection is 11\%. Such a reservoir is likely to operate within a 95\% enthalpy production factor, which is in agreement  with the reported geothermal recoveries under injection. The optimization workflow was applied to several field cases, which provided insights about the estimated well spacing versus the actual well configurations,  and therefore highlight opportunities for field optimization. The proposed approach could also be used for screening of new field developments, whose initial parameters will have the biggest impact on how much would lowering uncertainty improve optimization of recovery.
\section*{Aknowledgments}
We would like to thank CMG Ltd. for providing the STARS academic license, KAUST for the support, and UQLab for the software's license.

\section*{Declaration of Competing Interest}
The authors declare that they have no known competing financial interests or personal relationships that could have appeared to influence the work reported in this paper.

\appendix

\section{Sensitivity analysis}
We discuss the effect of 1) different 2-level DoE's on feature selection, 2) different 3-level DoE on the construction of the low-fidelity models and uncertainty propagation, and 3) the effect of the number of different realizations  on the accuracy of  Latin Hypercube DoE. 
\subsection{Sensitivity of different 2-level DoEs}
\begin{figure}[h!]
\begin{minipage}[b]{0.45\linewidth}
\centering
\includegraphics[scale=0.45]{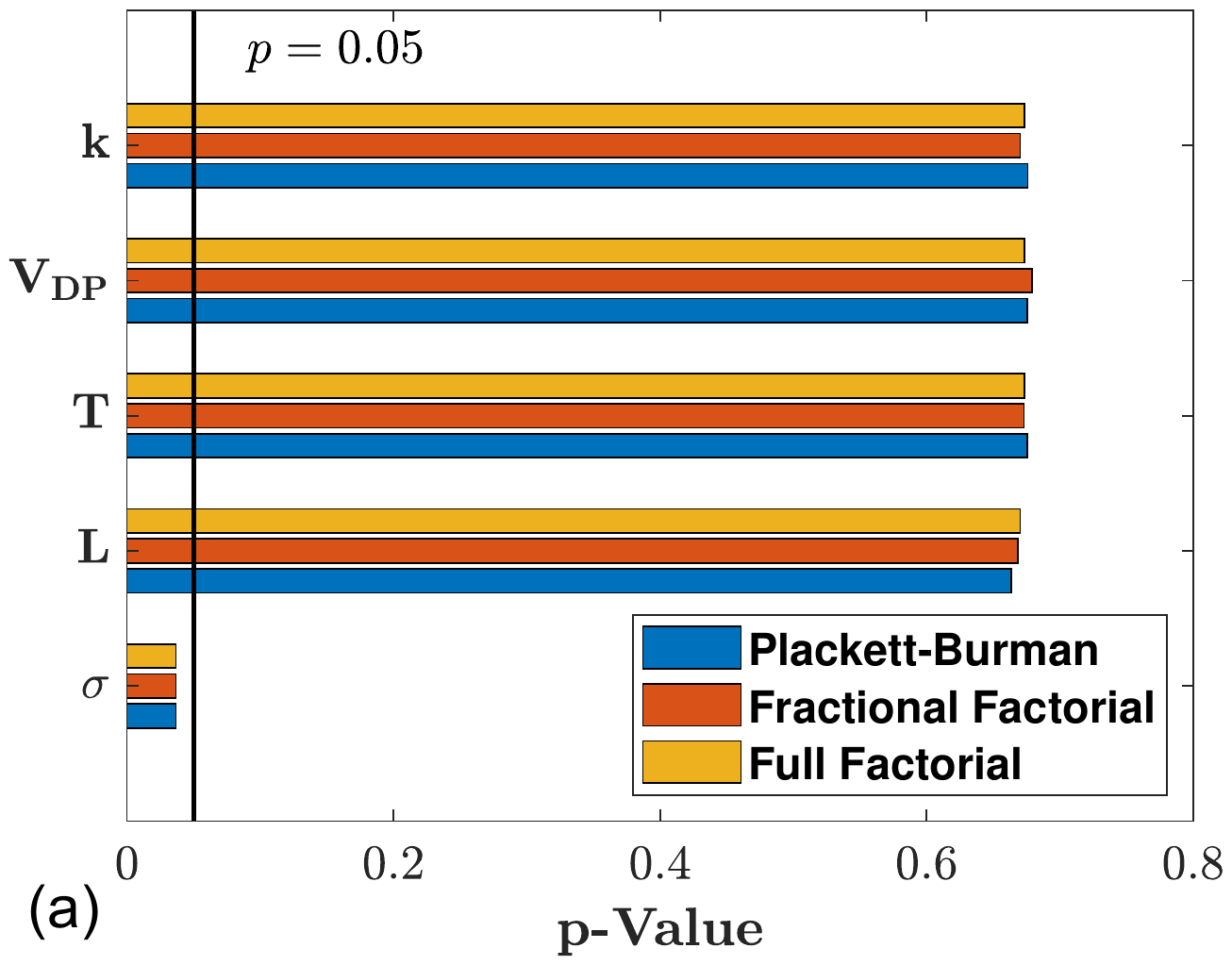}
\end{minipage}
\hspace{1cm}
\begin{minipage}[b]{0.45\linewidth}
\centering
\includegraphics[scale=0.45]{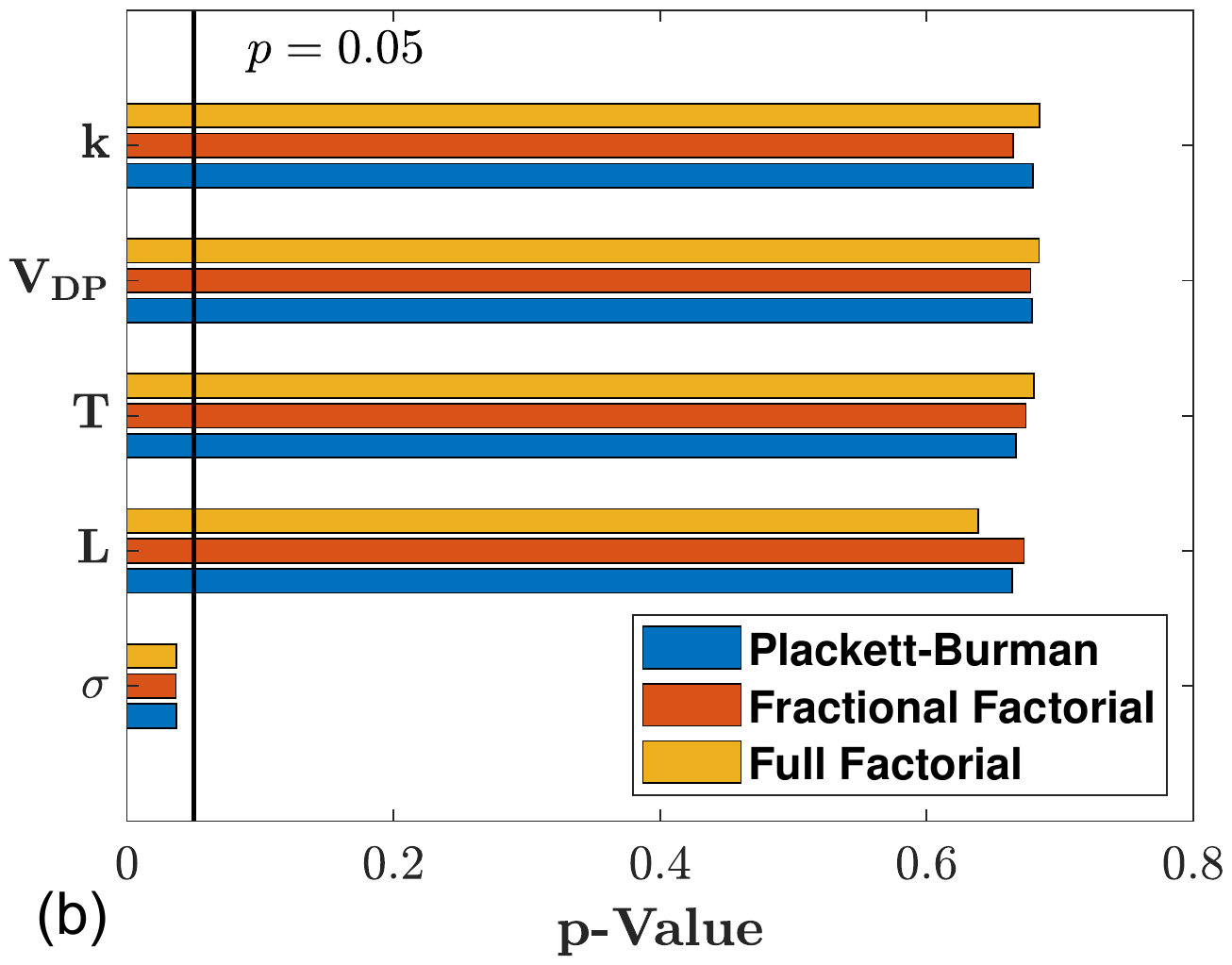}
\end{minipage}
\caption{Selection of significant variables relative to $w_{1}$ (a) and $w_{2}$ (b), using the Plackett-Berman DoE, and compared with other designs for confirmation. The black line denotes a cut-off of 0.05 probability (95\% confidence). The other  ( $w_{3}$, $w_{4}$ , and $w_{5}$) showed similar behavior and therefore not shown. }
\label{figure17}
\end{figure}

We use 2-level full factorial design, 2-level fractional factorial design, and folded Plackett-Burman design to investigate the process accuracy for feature selection. These designs inherently assume uniform probability for every realization \cite{Li2011}, and use first-order regression. Therefore, we only quantify the effect of the linear terms associated with the first-order sensitivity. 
The full factorial design samples the low and high values within the range of uncertainty of each variable, following the permutation rule. For $d$ variable, the total number of cases is $N_s=2^d$. 
The fractional factorial design performs a similar sampling as the full factorial design but considers a resolution optimization to reduce the total number cases. 
On the other hand, the folded Plackett-Burman design samples some of the low and high sides of each variable, such that the total number of cases is $N_s=4i_{, i=1,2,3,...}$. The corresponding maximum number of variables for each case is $d=4i-1$. Figure \ref{figure17} shows consistent ranking of the  assessed variables, which confirms the robustness of the folded Plackett-Burman design. This design is recommended for variable ranking and  feature selection. It requires a smaller number of cases than the full factorial design with comparable accuracy.      
\subsection{Sensitivity of different 3-level DoEs}
\begin{figure}[h!]
\centering
\begin{minipage}[b]{0.45\linewidth}
\centering
\includegraphics[scale=0.45]{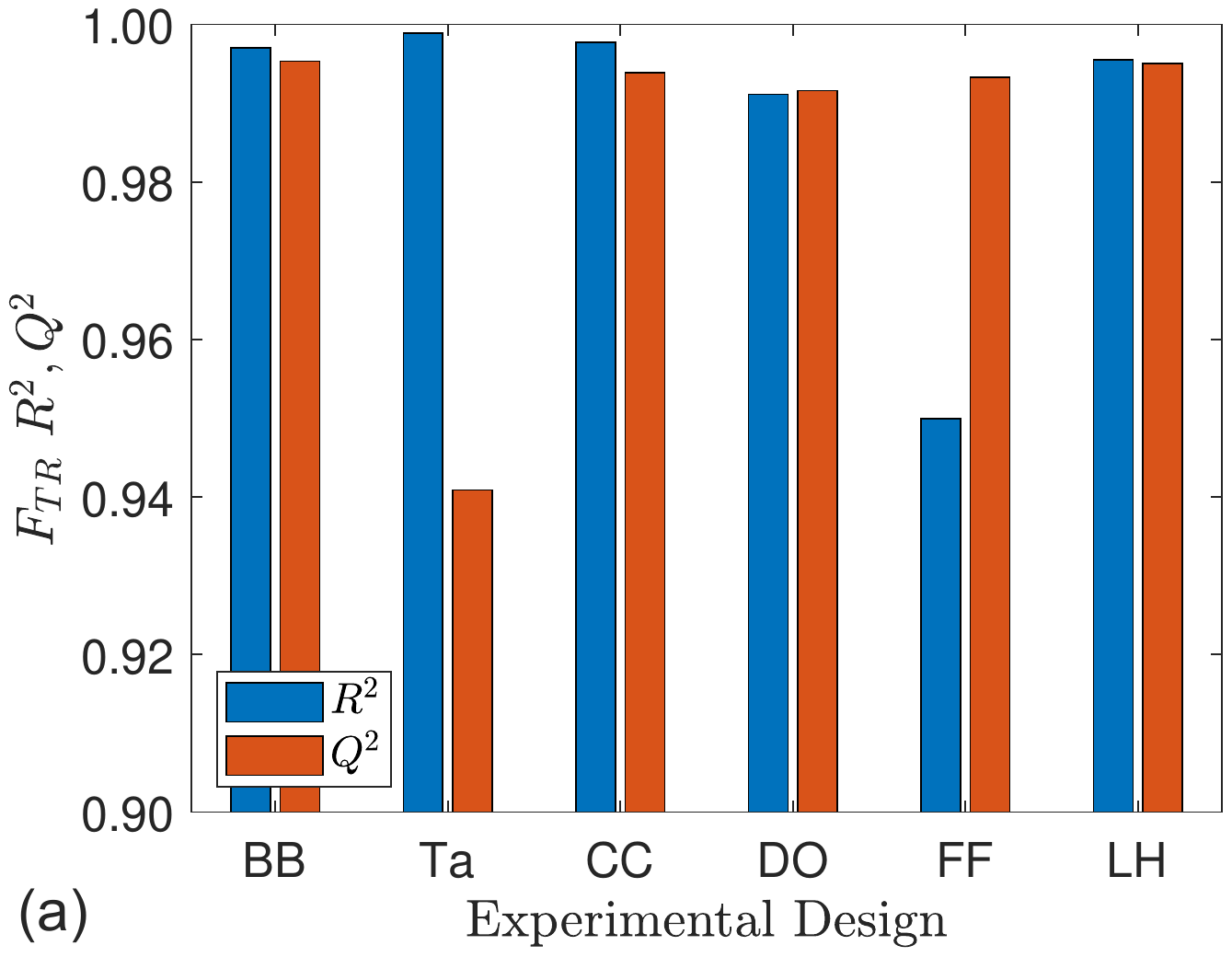}
\end{minipage}
\hspace{0.5cm}
\begin{minipage}[b]{0.45\linewidth}
\centering
\includegraphics[scale=0.45]{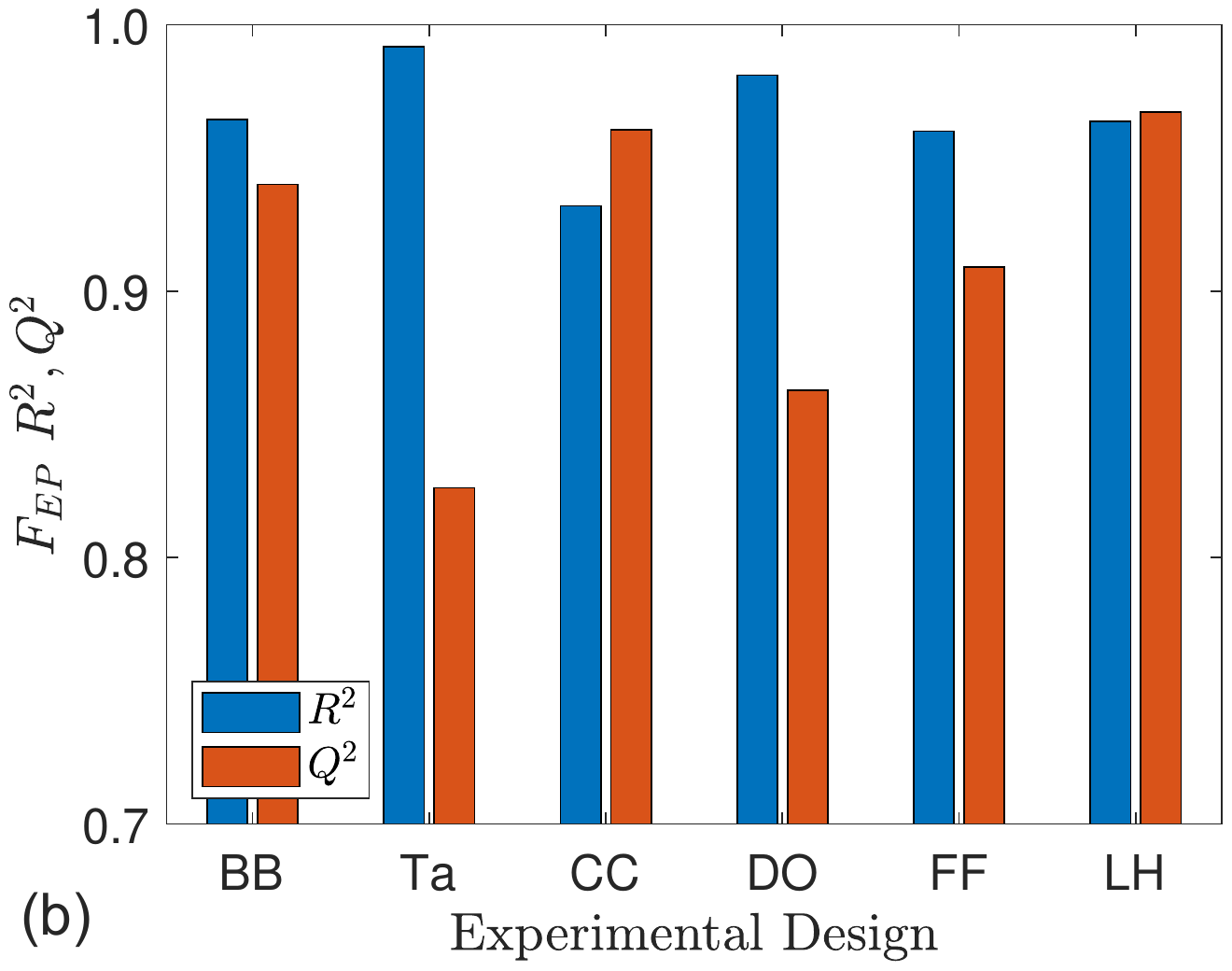}
\end{minipage}
\caption{Accuracy of different DoE evaluated by average $R^{2}$ and  $Q^{2}$ obtained for the thermal recovery factor $F_{TR}$ (a) and enthalpy production factor $F_{EP}$ (b). The corresponding symbols:  FF is Full Factorial, CC is Central Composite, BB is Box Behken, DO is D-Optimal, LH is Latin Hypercube, and Ta is Taguchi. The number of realizations in the different designs are: BB=27,  Ta=28, CC=36, DO=74, FF=81, LH = 1000.}
\label{figure18}
\end{figure}

We use 3-level Full Factorial, Central Composite, Box-Behken, D-Optimal, Latin Hypercube, and Taguchi designs to investigate the effect of sampling and training on the reliability of the low-fidelity models, including predictability  and uncertainty propagation. All the designs assume uniform probability for every realization \cite{Li2011}. 
The 3-level Full Factorial design requires  ${{N}_{s}}={{3}^{d}}$ cases. The Central Composite design provides the same sample points as in Full Factorial design, added with the center point and augmented with a group of axial points. The axial points represent extreme values to allow estimation of the curvature. The total number of cases becomes $
{{N}_{s}}=( 2\times {{3}^{d}} )+2$. 
The additional two cases  correspond to the center and the axial points. Box-Behken design includes the midpoint of the edges and the center point, while the D-Optimal design includes sample points by optimizing a D-optimality criterion.  
The Latin Hypercube design creates realizations by maximizing the distance among the samples, while the
Taguchi design utilizes special orthogonal arrays to generate a minimal number of realizations to capture full information of all variables. 
The summary of the total number of realizations, training accuracy, and predictability for the coefficient low-fidelity models is shown in Figure \ref{figure18}. 
Achieving high predictability is more important than high training accuracy since high predictability enables accurate interpolation for the prediction. For instance, Taguchi design showed the highest training accuracy but produced the lowest predictability. It cannot provide accurate predictions for unknown points.  Figure \ref{figure18} shows that higher predictability $Q^{2}$ is achieved by increasing the total number of realizations under random and space-filling conditions. Randomization is critical to tackle unknown non-linearity within the model and correlation among parameters, while space-fillingness to obtain an equal response from the whole region \cite{Santner2003}. Full Factorial, D-Optimal, and Taguchi design sample extreme/corner points. Therefore, they have low predictability compared to the other random and space-filling designs. Full Factorial has more realizations than Taguchi and D-Optimal, which gives higher predictability. Central composite design is more space-filling than Box-Behken design since it is Full Factorial design equipped with axial points \cite{Ansari2017}. Therefore, Central Composite design gives higher predictability than Box-Behken design. Latin Hypercube has both random and space-filling properties, therefore, it gives the highest predictability.
\subsection{Sensitivity of the size of Latin Hypercube design}
\begin{figure}[h!]
\begin{minipage}[b]{0.45\linewidth}
\centering
\includegraphics[scale=0.5]{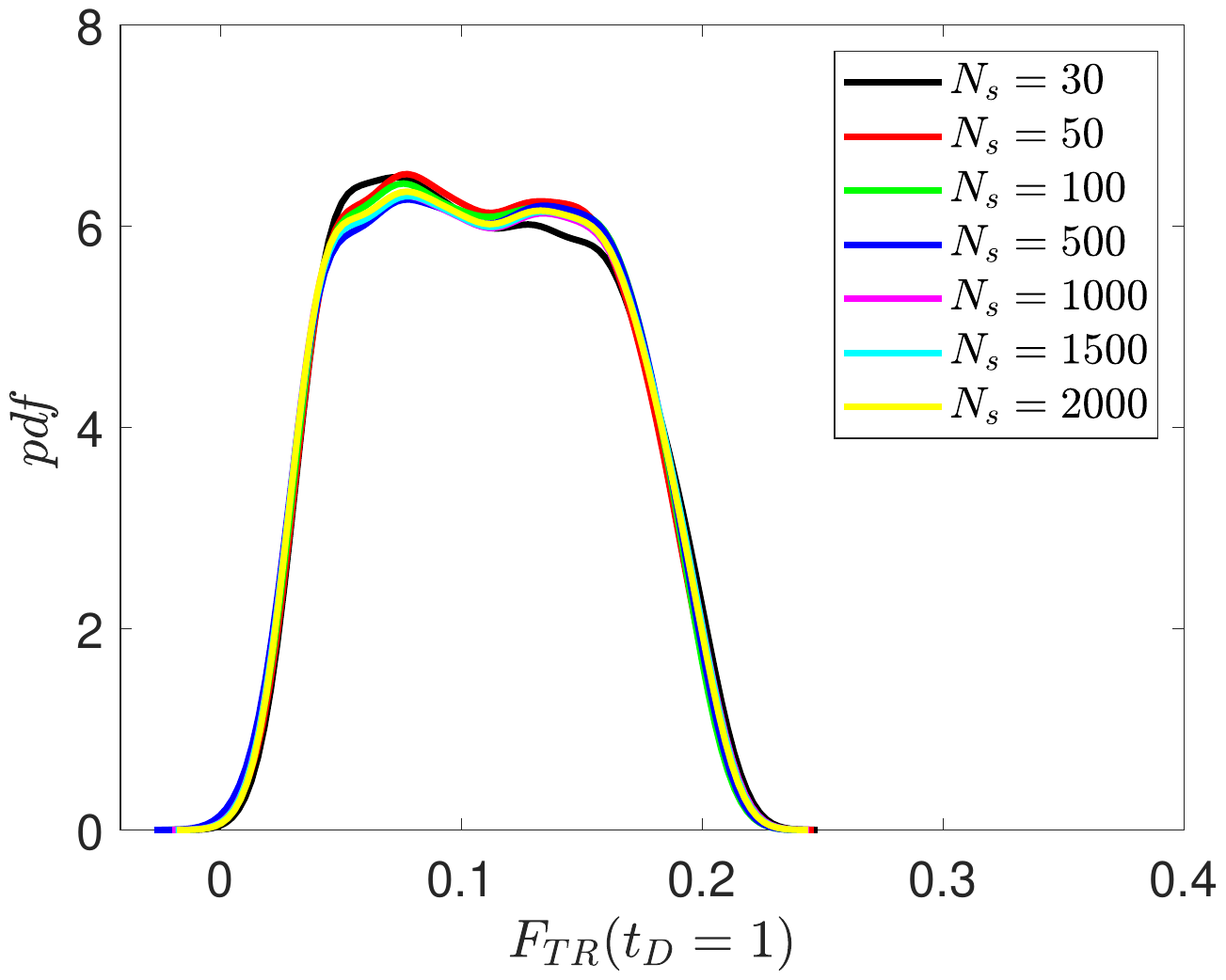}
\begin{center}
\hspace{1cm}
\textbf{(a)}
\end{center}  
\end{minipage}
\hspace{0.5cm}
\begin{minipage}[b]{0.45\linewidth}
\centering
\includegraphics[scale=0.5]{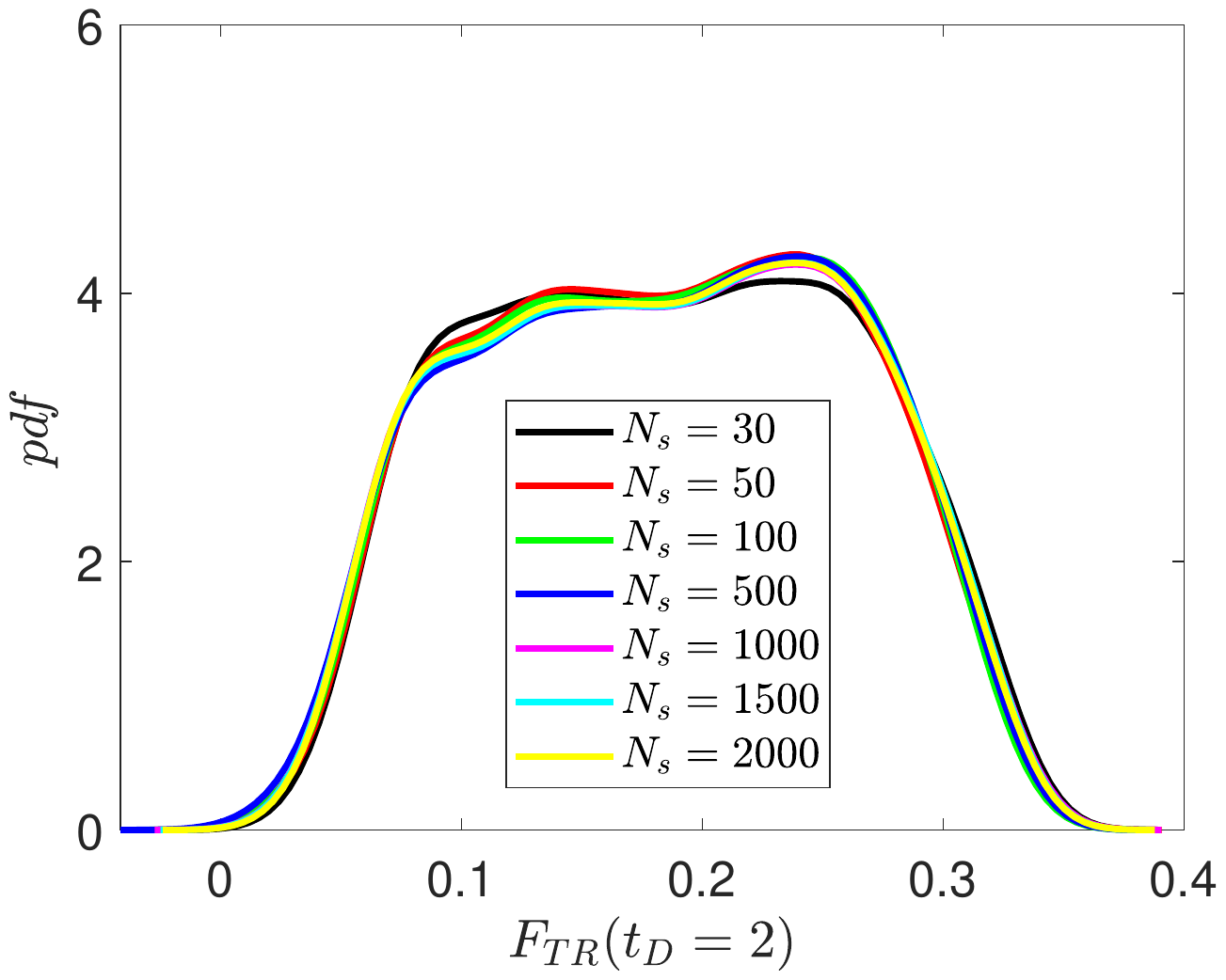}
\begin{center}
\hspace{1cm}
\textbf{(b)}
\end{center}  
\end{minipage}
\vspace{2ex}

\begin{minipage}[b]{0.45\linewidth}
\centering
\includegraphics[scale=0.5]{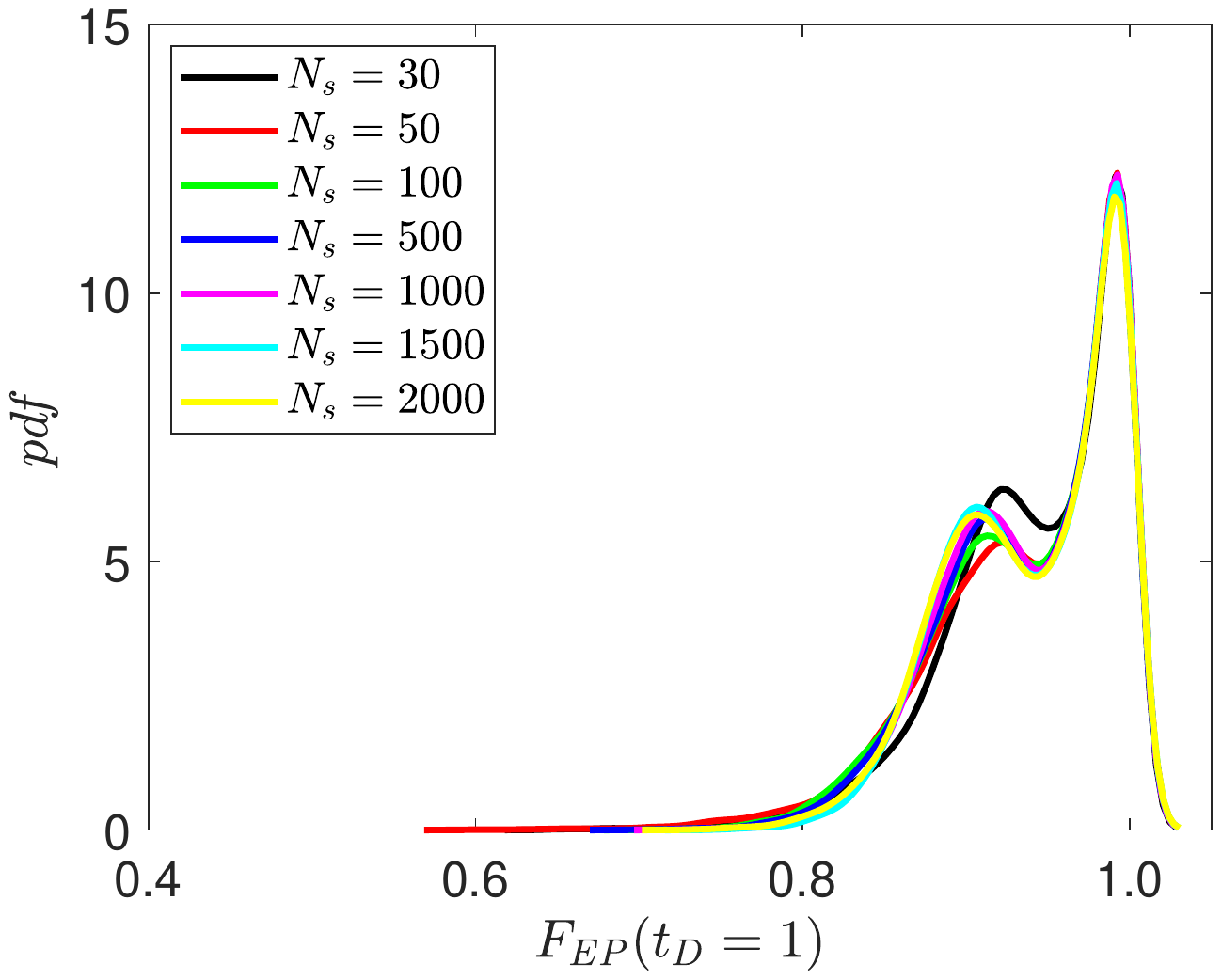}
\begin{center}
\hspace{1cm}
\textbf{(c)}
\end{center}  
\end{minipage}
\hspace{0.5cm}
\begin{minipage}[b]{0.45\linewidth}
\centering
\includegraphics[scale=0.5]{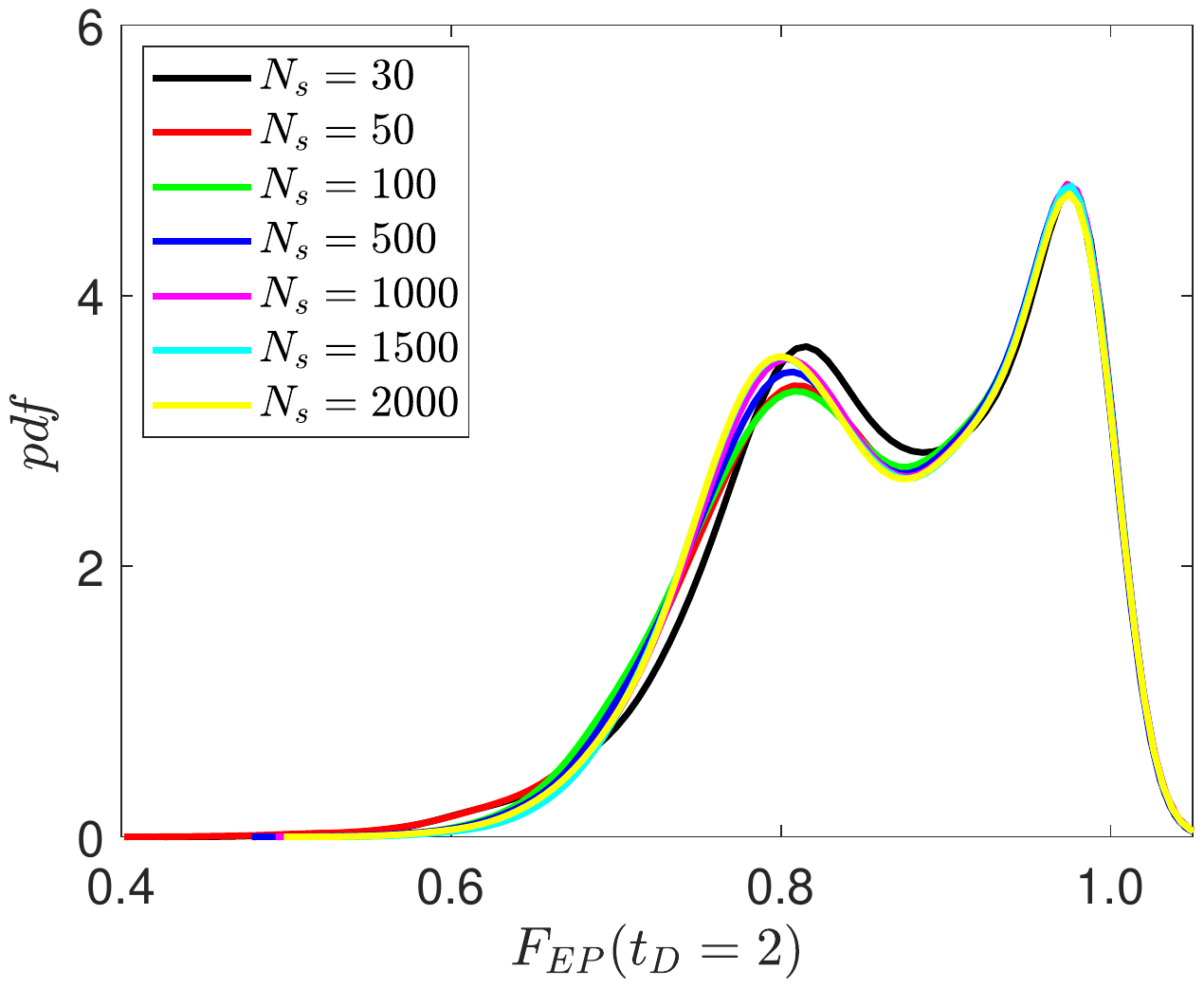}
\begin{center}
\hspace{1cm}
\textbf{(d)}
\end{center}  
\end{minipage}
\vspace{2ex}  
  
\caption{Monte Carlo results using different number of realizations on Latin Hypercube design: ${{F}_{TR}}\left( {{t}_{D}}=1 \right)$ (a), ${{F}_{TR}}\left( {{t}_{D}}=2 \right)$ (b), ${{F}_{EP}}\left( {{t}_{D}}=1 \right)$ (c), and ${{F}_{EP}}\left( {{t}_{D}}=2 \right)$ (d). The plots are produced by applying Kernel smoothing.}
\label{figure21}
\end{figure}

We perform sensitivity of a different number of realizations to the Latin Hypercube design. The aim is to find the optimum condition which gives high predictability and small realizations. The comparison result for  Monte Carlo simulations is shown in Figure \ref{figure21}. 
Change in the number of realizations gives a slightly different distribution in Monte Carlo results, as shown in Figure \ref{figure21}. All tested cases with different number of realizations produced consistent parameter ranking, which demonstrates the robustness of the method. 

\bibliographystyle{unsrt}

\end{document}